%% file: Database_Paper.tex
\documentclass[journal]{IEEEtran}

\usepackage[pdftex]{graphicx}
\usepackage{amsmath}
\usepackage{array}
\usepackage{url}
\usepackage{multirow,makecell}
\usepackage{color,soul}

\begin{document}

\title{Deep Neural Networks for Blind Image Quality Assessment: Addressing the Data Challenge}

\author{Shahrukh~Athar,~\IEEEmembership{Member,~IEEE,}
        Zhongling~Wang,~\IEEEmembership{Student Member,~IEEE,}
        and~Zhou~Wang,~\IEEEmembership{Fellow,~IEEE}
\thanks{The authors are with the Department of Electrical and Computer Engineering, University of Waterloo, Waterloo, ON, N2L 3G1, Canada (e-mail: \{shahrukh.athar, zhongling.wang, zhou.wang\}@uwaterloo.ca).}}%

\maketitle

\begin{abstract}
The enormous space and diversity of natural images is usually represented by a few small-scale human-rated image quality assessment (IQA) datasets. This casts great challenges to deep neural network (DNN) based blind IQA (BIQA), which requires large-scale training data that is representative of the natural image distribution. It is extremely difficult to create human-rated IQA datasets composed of millions of images due to constraints of subjective testing. While a number of efforts have focused on design innovations to enhance the performance of DNN based BIQA, attempts to address the scarcity of labeled IQA data remain surprisingly missing. To address this data challenge, we construct so far the largest IQA database, namely Waterloo Exploration-II, which contains 3,570 pristine reference and around 3.45 million singly and multiply distorted images. Since subjective testing for such a large dataset is nearly impossible, we develop a novel mechanism that synthetically assigns perceptual quality labels to the distorted images. We construct a DNN-based BIQA model called EONSS\footnote{Preliminary work and partial results of Section \ref{sec:EONSS}, related to EONSS, were presented at \cite{eonss_iciar}.}, train it on Waterloo Exploration-II, and test it on nine subject-rated IQA datasets, without any retraining or fine-tuning. The results show that with a straightforward DNN architecture, EONSS is able to outperform the very state-of-the-art in BIQA, both in terms of quality prediction performance and execution speed. This study strongly supports the view that the quantity and quality of meaningfully annotated training data, rather than a sophisticated network architecture or training strategy, is the dominating factor that determines the performance of DNN-based BIQA models. (Note: Since this is an ongoing project, the final versions of Waterloo Exploration-II database, quality annotations, and EONSS, will be made publicly available in the future when it culminates.)
\end{abstract}


\section{Introduction}
\label{sec:intro}

\IEEEPARstart{A}{dvances} in technology have enabled increasing and affordable connectivity, and the development of a multitude of mobile devices, leading to a well-connected world. An increasingly large proportion of the global population now accesses visual content through the Internet for various purposes such as communication, entertainment, education, sports, social media sharing, and so on. It is projected that by 2022 the annual global IP traffic will reach 4.8 zettabytes per year, with videos constituting the vast majority of this traffic at an expected 82\% \cite{cisco}. Visual content undergoes a number of distortions during the processes of acquisition, storage, transmission under bandwidth constraints, and display, any of which can degrade its perceived quality. Given the important role that such content now plays in our lives, perceptual image and video quality assessment has become a fundamental problem that is pivotal for the design, optimization and evaluation of various image and video processing algorithms and systems. Image quality assessment (IQA) can be classified into \textit{subjective} and \textit{objective} quality assessment (QA). In subjective QA, humans are tasked to rate the visual quality of content. Since humans are the ultimate receivers of visual content, subjective QA is regarded as the most reliable way to quantify its perceptual quality. However, subjective QA is time consuming, expensive, cannot be embedded in algorithms for optimization purposes, and cannot be deployed in a large-scale and real-time manner. To address these issues, the goal of objective QA is to automatically predict the perceptual quality of visual content as perceived by humans, where subject-rated IQA datasets are used for testing, and at times for training. Objective IQA algorithms can be categorized into three major frameworks \cite{iqa_book,rrnrSPM}: 1) Full-Reference (FR) IQA methods require complete access to the pristine or reference version of a distorted image to evaluate its quality; 2) Reduced-Reference (RR) IQA methods require partial access to the reference image through certain extracted features; 3) No-Reference (NR) or Blind IQA (BIQA) methods evaluate the quality of a distorted image in the absence of its reference version. 

In the last two decades, significant progress has been made in the development of FR IQA algorithms. State-of-the-art FR methods (such as but not limited to \cite{fr_iwssim,fr_fsim,fr_vsi,fr_dss}) are training-free and their predictions correlate well with human perception of quality while evaluating images afflicted with common distortion types. This is evident when they are tested on a wide variety of subject-rated datasets \cite{eval_fr_icip12,eval_fr_icip15,eval_survey_latest,eval_Waterloo}. However, the practical application of FR methods remains limited because in real-world media delivery systems, access to pristine reference images is either extremely rare or altogether nonexistent especially at the end-user level. In such practical scenarios, NR IQA or BIQA is the only feasible option. While a lot of work has also been done on the development of BIQA methods, significant room for improvement exists to further enhance their performance \cite{eval_Waterloo,eval_gmad}. This is understandable as BIQA is a much more difficult task owing to lack of access to the reference image.

While a few recent BIQA methods are either training-free \cite{md_sisblim_db_mdid2013,nr_lpsi} or require training only to learn a universal model of pristine images \cite{nr_niqe,nr_ilniqe}, a large number of BIQA methods try to alleviate the constraints posed by lack of access to the reference image by employing machine learning algorithms where training is done against human-annotated distorted content. Most training-based BIQA methods extract features from the distorted image and use standard regression methods such as support vector regression (SVR) \cite{SVRbook,libsvm} in conjunction with subject-rated data to learn a quality model. These features are either domain knowledge based handcrafted features \cite{nr_biqi,nr_diivine,nr_bliinds2,nr_brisque,nr_gmlog,nr_nferm,md_gwhglbp,nr_nrsl,nr_friquee_jrnl} or learned features \cite{nr_cornia,nr_hosa}. We recently showed \cite{eval_Waterloo} that BIQA models that employ learned features (such as \cite{nr_cornia,nr_hosa,nr_dipiq}) offer better general-purpose performance compared to those that use handcrafted features. Since data-driven end-to-end optimized deep neural networks (DNN) combine the tasks of learning goal-oriented features and regression, they have great potential to outperform the traditional two-stage approach where feature extraction and regression are optimized independently, but not jointly. A prerequisite requirement for using such deep networks is to have an adequately large set of training data. This is because such networks have hundreds of thousands, if not millions of parameters, and an insufficient amount of training data leads to overfitting, thereby degrading the generalizability of the trained model to unseen data. While a number of DNN based BIQA models have been proposed recently \cite{eval_dnn_latest}, they all encounter a significant hurdle: \textit{the lack of large-scale perceptually annotated training data} \cite{eval_dnn_latest,spm_dnn}. Since obtaining large-scale subject-rated data (millions of quality annotated images) is difficult, if not impossible, contemporary DNN based BIQA models focus on data augmentation techniques to enhance the size of the small-scale annotated IQA data that is available \cite{eval_dnn_latest,spm_dnn}. However, even with data augmentation, the size of the training data remains limited, and such augmentation techniques lead to their own issues. When tested on unseen data, the performance of DNN based BIQA models remains inadequate \cite{eval_Waterloo}.

In this work, we focus on the fundamental problem plaguing the development of high performance DNN based BIQA models, i.e., the lack of large-scale annotated training data. Through the development of a very large-scale synthetically annotated IQA dataset, we show that the performance of a DNN model with a simple architecture, when tested on wide-ranging subject-rated unseen data, can be elevated so much so that it not only outperforms recent DNN based BIQA models but also the very state-of-the-art in BIQA. The rest of the paper is organized as follows. The data challenge in the area of IQA is discussed in Section \ref{sec:dataCH} along with the contributions of this work. Section \ref{sec:DB_Const} discusses the construction of a very large-scale IQA dataset. Section \ref{sec:SGT} presents a novel technique to synthetically annotate this dataset with perceptual quality ratings and evaluates its performance. Section \ref{sec:EONSS} discusses the construction of a simple DNN based IQA model that trains on the newly developed dataset along with extensive performance evaluation of this model on subject-rated IQA datasets in a variety of scenarios to demonstrate the effectiveness of our approach. Section \ref{sec:Conc} concludes this work.

\section{The Data Challenge}
\label{sec:dataCH}

\subsection{DNNs in Visual Recognition: A Case Study}
\label{subsec:DNNsVisRec}

The application of DNNs has led to tremendous progress in the area of visual recognition, thus it is important to ascertain the reasons for this success. Like machine learning based BIQA, methods in visual recognition are composed of two important components, the model and the data used to train the model. Although a lot of work has been done on the modeling component of the visual recognition task, around ten years ago researchers started to focus on the data component of this task \cite{db_Tiny,db_ImageNet}. While the Tiny Image dataset \cite{db_Tiny} has around 80 million loosely labeled low-resolution images, the ImageNet database \cite{db_ImageNet,db_ImageNet_source} is composed of more than 14 million more precisely labeled higher resolution images and has led to many breakthroughs in visual recognition. The images in ImageNet populate close to 22,000 \cite{db_ImageNet_source} synonym sets (synsets) of the WordNet \cite{WordNet_Miller,WordNet_Fellbaum} hierarchy with an average of 650 images per synset. For the image classification task, ImageNet first obtains images by crawling the Internet through synset specific search queries to several image search engines. Next, it engages human subjects, through the online crowdsourcing platform Amazon Mechanical Turk \cite{amazonMT}, to verify that images have been associated with the correct labels regardless of any distortions that may be present. Although it is not an easy task to ask human subjects to verify the labels associated with millions of images, it is still manageable because: 1) Subjects are not being asked to label an image from scratch, instead they are required to verify if an image contains an object associated with the given label. This simplifies the task. 2) The verification requires binary answers (Yes/No). 3) Each image can be treated as an independent entity. 4) Votes from only a few subjects are sufficient to verify the label of each image. Higher levels of the hierarchy are usually easier to verify and require votes from just a few subjects (much less than five), while deeper levels of the hierarchy may require votes from more subjects (five to ten) \cite{db_ImageNet}. 5) Viewing conditions and devices do not impact the authenticity of label verification by subjects, and hence crowdsourcing can be conveniently used.

\input{Tables/Table1.tex}

The annual ImageNet Large Scale Visual Recognition Challenge (ILSVRC) \cite{db_ILSVRC} ran from 2010 to 2017 and was composed of subsets of ImageNet images for the purposes of algorithm training, validation, and testing. ILSVRC tasks included image classification, object localization, and object detection. With the availability of a large-scale dataset such as ImageNet along with computationally powerful GPUs coming of age, the stage was set for the development of DNN based visual recognition models.  While the first two years of ILSVRC did not see DNN based entries, a significant turning point was observed in ILSVRC 2012, when a deep convolutional neural network (CNN) based model \cite{NIPS_DNN}, with 60 million parameters, comprehensively won both the classification and localization challenges in terms of top-5 error rate \cite{db_ILSVRC}. The margin with which \cite{NIPS_DNN} outperformed other models in the 2012 challenge had such an impact that submissions to the ILSVRCs in the subsequent years were predominantly deep CNN based. Since 2012, deep CNN based models have won the various ILSVRCs in terms of top-5 error rate for the image classification and object localization tasks, and in terms of mean average precision for the object detection task \cite{db_ILSVRC}. Thus, the development of high performance, generalizable and robust deep CNN based visual recognition models such as \cite{NIPS_DNN,Clarifai_uses_ImageNet,OverFeat_uses_ImageNet,Caffe_uses_ImageNet,VGG_uses_ImageNet,GoogLeNet_uses_ImageNet} has become possible due to the ImageNet database \cite{db_ImageNet} and the ILSVRC \cite{db_ILSVRC}.

Finally, it is pertinent to mention that some other datasets in the area of visual recognition, such as the PASCAL VOC datasets \cite{db_pascal_voc}, Caltech 101 dataset \cite{db_caltech101}, and Caltech 256 dataset \cite{db_caltech256}, that have between 9,000 to 31,000 images in 20 to 256 object classes, are considered small-scale and training DNN models from scratch on these datasets is considered infeasible due to overfitting concerns \cite{VGG_uses_ImageNet}.

\subsection{DNNs in BIQA: The Data Challenge}
\label{subsec:DNNsBIQAch}

\subsubsection{Small Scale of IQA Datasets}
\label{subsubsec:SmallScaleIQAdb}

Compared to annotating datasets for visual recognition tasks such as ImageNet \cite{db_ImageNet}, obtaining subjective ratings of \textit{image quality} is an altogether different and much more complex scenario because: 1) Subjects need to provide their \textit{opinion} of an image's quality, which is a rather abstract concept and requires substantial critical thinking on the subject's part. 2) The quality scale is not binary, instead it either has a number of discrete levels (five or more) or is continuous. 3) A subject's opinion of quality needs to be \textit{calibrated} before the experiment, so that they have a rough idea about the range of quality to expect. While subjects are asked to treat each image independently during the experiment, they still need to provide ratings relative to the quality range introduced to them. 4) To ensure reliability, it is recommended that at least 15 subjects rate each image in the subjective experiment \cite{bt500}. 5) It is suggested that a test session should last no longer than 30-minutes to avoid fatigue effects and that participating subjects be screened for visual acuity and color vision \cite{bt500}. 6) Viewing conditions play a crucial role in the appearance of visual content, and hence on its quality. Therefore, viewing conditions such as display and background luminance, room illumination, observation angle, and viewing distance, play an important role in subjective tests and need to be set according to established norms \cite{bt500}. 

Considering the above-mentioned constraints, it is much more difficult, if not impossible, to carry out subjective testing for IQA datasets composed of millions of images, even with crowdsourcing. To-date IQA datasets consist of hundreds or a few thousands of distorted images. A summary of contemporary subject-rated IQA databases of two-dimensional (2D) natural images is given in Table \ref{table:IQAdbSummary}. IQA datasets are classified either as \textit{simulated} or \textit{authentic} distortion databases, depending upon whether distortions were simulated on a set of pristine reference images or if they were captured directly in the real-world environment, respectively. Simulated distortions datasets can further be classified into either \textit{singly} or \textit{multiply} distorted databases, where each distorted image is afflicted by a single distortion in the former case or by multiple simultaneous distortions in the latter. Among simulated distortion datasets, the multiply distorted ones are more accurate representations of practical content since visual content almost always undergoes multiple distortions in the real-world. From Table \ref{table:IQAdbSummary} it can be seen that the largest singly distorted simulated distortion dataset, the recently released PDAP-HDDS \cite{db_pdaphdds}, has only 12,000 distorted images, while the largest multiply distorted dataset, MDID \cite{db_mdid}, has only 1,600 distorted images. Among authentic distortion databases, the recently released KonIQ-10K \cite{db_koniq10k}, has only 10,073 distorted images. As is evident from recent surveys of DNN based BIQA models \cite{eval_dnn_latest,spm_dnn}, datasets such as LIVE R2 \cite{stateval_db_liveR2}, LIVE MD \cite{db_livemd}, LIVE Challenge \cite{db_livewc}, CSIQ \cite{fr_mad_db_csiq}, and TID2013 \cite{db_tid2013} are used to train such models. The largest dataset among them is the singly distorted database TID2013 \cite{db_tid2013} which has only 3,000 distorted images. From Table \ref{table:IQAdbSummary}, it can be observed that each dataset has only a limited number of images per distortion type. The number of images per distortion type per level of distortion is even smaller. For example, for singly distorted datasets, this is usually equal to the number of pristine images in the dataset. It can also be observed from Table \ref{table:IQAdbSummary} that simulated distortion datasets have a very limited amount of pristine reference content. The PDAP-HDDS database \cite{db_pdaphdds} has 250 reference images, however it is itself an outlier as all other datasets have less than 100 reference images (usually 10 to 30). Since these pristine reference images are \textit{supposed} to be representatives of the enormous space of all possible natural images, contemporary IQA datasets do rather inadequately in terms of overall content variation. These shortcomings of subject-rated IQA datasets create enormous hurdles in the development of generalizable and robust DNN based BIQA methods. Recently two large-scale singly distorted datasets have been constructed. These include the Waterloo Exploration-I database \cite{db_waterlooed}, which has 4,744 reference and 94,880 distorted images, and the KADIS-700k database \cite{db_kadid10k}, which has 140,000 reference and 700,000 distorted images. However, they cannot be used to train DNN based BIQA models as their distorted content has not been annotated with perceptual quality labels. Thus, the issue of large-scale annotated IQA data is an open problem which needs to be resolved.

\subsubsection{Current Strategies to Deal with Lack of Data}
\label{subsubsec:CurrStrat}

Contemporary DNN based BIQA models employ a number of design innovations and data augmentation techniques to deal with the issue of small-scale training data. 

A widely used technique adopted by DNN based BIQA methods to increase the size of the training set is to extract multiple fixed size small patches from each labeled image \cite{nr_cnn,nr_cnn_plusplus,md_cnn_JieFu,nr_cnn_prewitt,nr_biecon_C,nr_biecon_J,nr_cnn_svr,nr_deepIQA_conf,nr_fr_deepIQA,nr_deepBIQ,nr_cnn_pqr,nr_diqa,nr_pccnn}. While a popular patch size is $32 \times 32$ \cite{nr_cnn,nr_cnn_plusplus,md_cnn_JieFu,nr_cnn_prewitt,nr_biecon_C,nr_biecon_J,nr_cnn_svr,nr_deepIQA_conf,nr_fr_deepIQA,nr_pccnn}, larger sized patches have also been employed \cite{nr_deepBIQ,nr_cnn_pqr,nr_diqa}. Since local patch-level quality labels are not available in IQA databases, the global image-level quality score is usually applied to each patch extracted from an image. Due to the influence of distortions and the visual attention property of the human visual system (HVS), some regions of an image might seem perceptually more relevant to a human subject while assigning global quality scores \cite{fr_vis_imp_pool,fr_vis_attn_qa}. An image might be assigned a low quality score, yet patches extracted from it might receive high quality scores when viewed independently. Thus, the assignment of the global quality score to local patches extracted from an image leads to a significant label noise problem. The method in \cite{nr_biecon_C,nr_biecon_J} tries to address this problem by splitting model training into two steps. In the first step, an FR method (FSIM \cite{fr_fsim}) was used to assign a quality score to each local patch, and a CNN was pre-trained using this patch-level data. In the second step, the model was fine-tuned on subject-rated datasets. Similar to \cite{nr_biecon_J}, the method in \cite{nr_diqa} also follows a two-step training process, however, instead of using FR methods to label local patches, it uses the exponent difference function to generate objective error maps for these patches, which are then used as intermediate training targets. The model is fine-tuned on subject-rated datasets. The method in \cite{nr_deepIQA_conf,nr_fr_deepIQA} tries to alleviate the label noise problem by weighted-average patch quality aggregation. It estimates the quality of $32 \times 32$ image patches and also determines the relative weight of each patch to account for its contribution in the global quality of the parent image. Patch weight estimation is carried out by adding a parallel branch to the patch quality regression layers, and the whole network is optimized in an end-to-end manner. In \cite{nr_cnn_prewitt}, an image is segmented and the Prewitt operator is used to generate the gradient map through which patch weights are determined. The quality of each image patch is predicted by a CNN and the global image quality is computed through a weighted average of patch qualities. Even with the adoption of the patch-based data augmentation technique, the volume of training data remains limited given the small-scale of IQA datasets. While a recent method \cite{nr_pccnn} tries to further increase the training data size by using various combinations of distorted and their corresponding reference image patches, to generate patch pairs which are used for CNN training, the overall amount of training data still remains limited. The very small amount of reference content and the label noise issue further impacts the utility of this data augmentation technique.

Some methods \cite{nr_biecon_J,nr_diqa,nr_meon,nr_nima} increase the size of the training data by horizontally flipping the images or image patches, and using the quality label assigned to the parent image. Other kinds of geometric transformations cannot be applied in the area of IQA as they can significantly impact the perceptual quality of an image \cite{spm_dnn}. In addition to horizontal flipping, the method in \cite{nr_meon} creates additional training samples by changing the saturation and contrast of images as long as these changes do not impact their perceptual quality. Given the small-scale of IQA datasets, this data augmentation technique also leads to a limited expansion of training data and suffers from the limited nature of reference content.

Since very large-scale annotated databases are available in the area of visual recognition, some BIQA methods utilize DNNs that have been pre-trained for the visual recognition task. In \cite{nr_deepBIQ}, the Caffe network \cite{Caffe_uses_ImageNet} that has been pre-trained on the ImageNet \cite{db_ImageNet} and Places \cite{db_places} visual recognition databases is used in two ways: 1) As a feature extractor, where SVR and IQA databases are used to map these features to perceived quality scores; 2) As an initialization, where the network is fine-tuned with respect to IQA databases. In \cite{nr_blinder}, the VGG network \cite{VGG_uses_ImageNet}, that has been pre-trained on the ImageNet database \cite{db_ImageNet} for the object recognition task, is used where feature vectors are extracted from different layers of the network and form a multi-level representation of the image. Next, IQA databases are used along with SVR to learn a mapping from each feature vector to a quality score. A global quality score for an image is then computed as the average of quality scores predicted by different network layers. Instead of predicting a single quality score, the method in \cite{nr_cnn_pqr} predicts the quality distribution of a given image using CNNs. The output of the CNN model is in terms of probabilistic quality representation (PQR) vectors that are then mapped to scalar quality scores using SVR. Of the three CNNs used in \cite{nr_cnn_pqr}, two are deep CNNs (AlexNet and ResNet50) that have been pre-trained for the image classification task on the ImageNet database \cite{db_ImageNet}. These deep CNNs are then fine-tuned by using subject-rated IQA databases. Although features extracted from a DNN that has been trained for a particular visual recognition task, such as image classification, are known to be effective generic features for other visual recognition tasks \cite{CNN_Feat_OffShelf,Delv_Deep_ConvNets}, their use in an altogether different area, such as IQA, is open to doubt \cite{eval_dnn_latest}.

Some DNN based BIQA methods adopt a multi-task strategy to deal with the lack of quality-annotated training data. The work in \cite{nr_cnn_plusplus} is a pioneering effort in this direction, where image quality and distortion type are simultaneously estimated. It is demonstrated in \cite{nr_cnn_plusplus} that such a multi-task approach allows for a reduction in the model's learnable parameters without loss in model performance. The method in \cite{nr_meon} uses the multi-task approach of distortion identification and quality prediction, however in a causal manner. It splits these two tasks between two sub-networks such that their early layers are shared. Sub-network 1, which identifies distortion type through a probability vector is fed into sub-network 2, which predicts image quality. Since a large amount of labeled data can be generated for the distortion identification task without the need for human annotations, 840 pristine images are degraded at five distortion levels for different distortion types in \cite{nr_meon} to generate a large amount of training data, which is used to pre-train sub-network 1 along with the shared layers of the overall network. The entire network is subsequently joint optimized using subject-rated data. The method in \cite{nr_dbcnn} utilizes two deep CNNs to separately deal with the scenarios of synthetically (simulated distortions) and authentically distorted images. For synthetic distortions, the distortion type and level information is used for pre-training, where the training set includes 852,891 distorted images that have been obtained by using 9 distortion types to degrade 21,869 pristine images at various distortion levels. For authentic distortions, the CNN VGG-16 \cite{VGG_uses_ImageNet}, that has been pre-trained on the ImageNet database \cite{db_ImageNet} for the image classification task, is used. The feature sets from the two CNNs are transformed into one representation set through bilinear pooling. The entire network is fine-tuned on subject-rated IQA datasets. In these methods, specially \cite{nr_meon,nr_dbcnn}, the use of multi-task learning, where distortion identification is carried out in addition to quality score prediction, has the benefit of enabling the training process to be split into pre-training and joint optimization stages. Labels for distortion identification do not require human annotation, and thus a large amount of data can be generated for the pre-training step. However, such an approach does not take into account the impact of content variations even though distortions of the same type and magnitude can lead to drastically different perceived quality results for two different contents. This is a fundamental limitation of such multi-task learning based approaches for data augmentation.

Since deep models require a large amount of training data due to the high-dimensional nature of images as model inputs, some techniques (such as \cite{nr_dliqa}) use low-dimensional representations of images, by using natural scene statistics (NSS) \cite{NatImgStat_NeuRep} features extracted from the images, as inputs to the model. While this reduces the training size requirements of training data, such models are unable to realize the full potential of DNNs since end-to-end learning is lacking.

\subsection{Contributions of this Work}
\label{subsec:Contributions}

While a lot of efforts have been made to construct DNN based BIQA methods that focus on the modeling part of the problem and try to alleviate the lack of training data by using data augmentation and multi-task learning techniques (as described in Section \ref{subsubsec:CurrStrat}), efforts to address the fundamental problem of lack of large-scale quality-annotated IQA datasets remain surprisingly missing. In this work we focus on addressing this fundamental problem plaguing the development of robust and generalizable DNN based BIQA models. Specifically, we make the following three novel contributions: 1) We construct the largest IQA dataset to-date, called the Waterloo Exploration-II database, which has 3,570 pristine and more than 3.45 million distorted images (including both singly and multiply distorted content). 2) Since annotating so many images through subjective testing is not possible, we devise a novel synthetic quality benchmark generation mechanism that annotates the images with perceptually oriented quality ratings. Our tests on a wide range of available subject-rated IQA datasets show that this mechanism leads to quality annotations that are highly correlated with human perception of quality, and thus they can be used as alternatives to human quality ratings. 3) We develop a DNN based BIQA model called End-to-end Optimized deep neural Network using Synthetic Scores (EONSS) and train it using the Waterloo Exploration-II database. Although EONSS has a simple architecture, we show that when tested across a wide range of available subject-rated IQA datasets, it not only comprehensively outperforms other DNN based BIQA models with more complex architectures that use data augmentation, but also outperforms the very state-of-the-art in BIQA, thereby highlighting the significance of our approach to overcome the data challenge encountered when constructing DNN based BIQA models. The EONSS model and a few results of Section \ref{sec:EONSS} were first presented in our earlier work \cite{eonss_iciar}. However, compared to \cite{eonss_iciar}, this work makes the following new contributions: 1) The construction of the Waterloo Exploration-II database has only been exclusively discussed in this work. 2) The generation mechanism of the novel synthetic quality benchmark used to assign perceptual quality annotations to dataset images has only been exclusively discussed in this work. 3) The target of \cite{eonss_iciar} was to develop a DNN-based BIQA model, EONSS, specifically for multiply distorted images. However, in this work our goal is not to develop one particular DNN-based BIQA model but to use EONSS as a common testing platform to ascertain the impact of data on the performance of DNN-based IQA models. 4) Only three multiply distorted subject-rated datasets were used for model testing in \cite{eonss_iciar}, while we have used nine subject-rated datasets for model testing in this work, which include five singly and four multiply distorted datasets. 5) Section \ref{sec:EONSS} of this work presents an extensive array of tests specifically concerned with analyzing and revealing the impact of training data on the performance of DNN-based IQA models, all of which are missing in \cite{eonss_iciar}.

\section{Waterloo Exploration-II Database Construction}
\label{sec:DB_Const}

\begin{figure*}[t!]
	\centering
	\begin{tabular}{c @{\hspace{-1ex}} c @{\hspace{-1ex}} c @{\hspace{-1ex}} c @{\hspace{-1ex}} c}
		\includegraphics[width=0.2\textwidth]{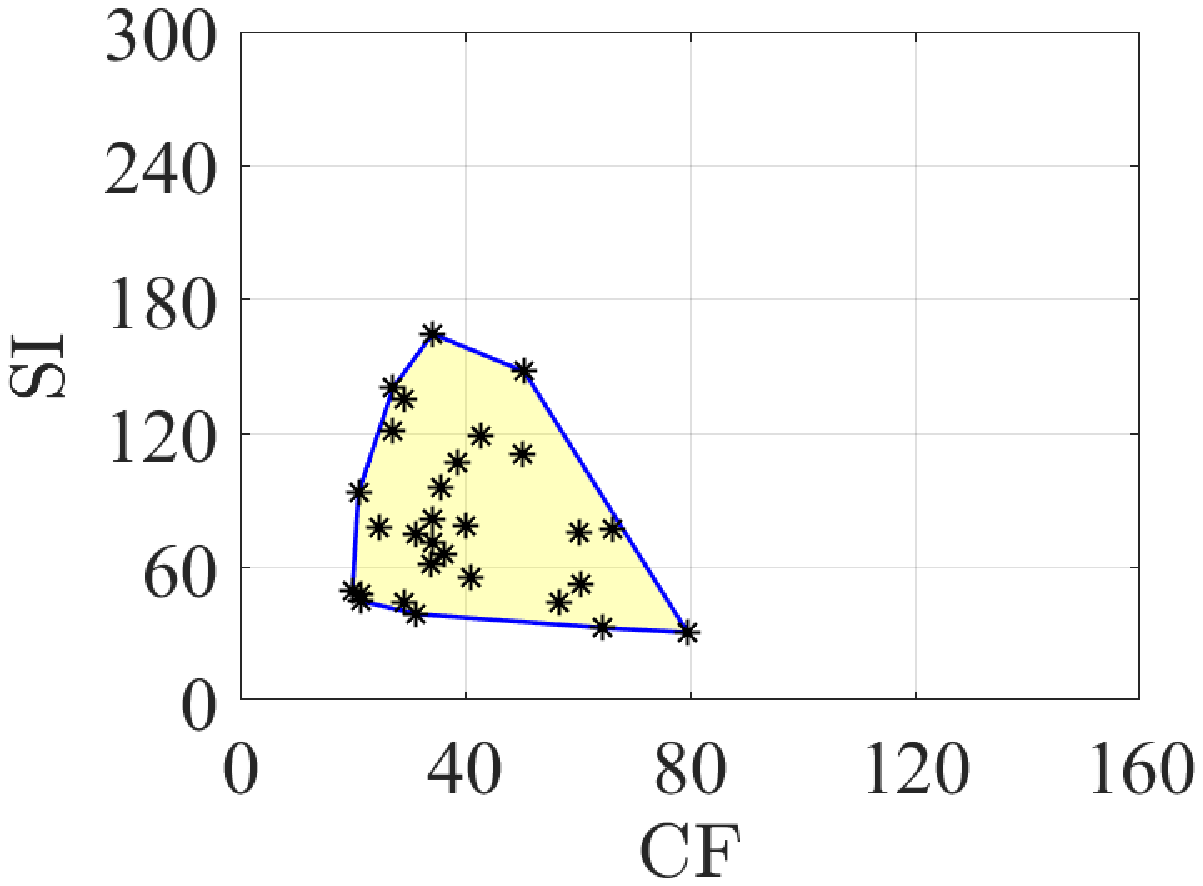} &
		\includegraphics[width=0.2\textwidth]{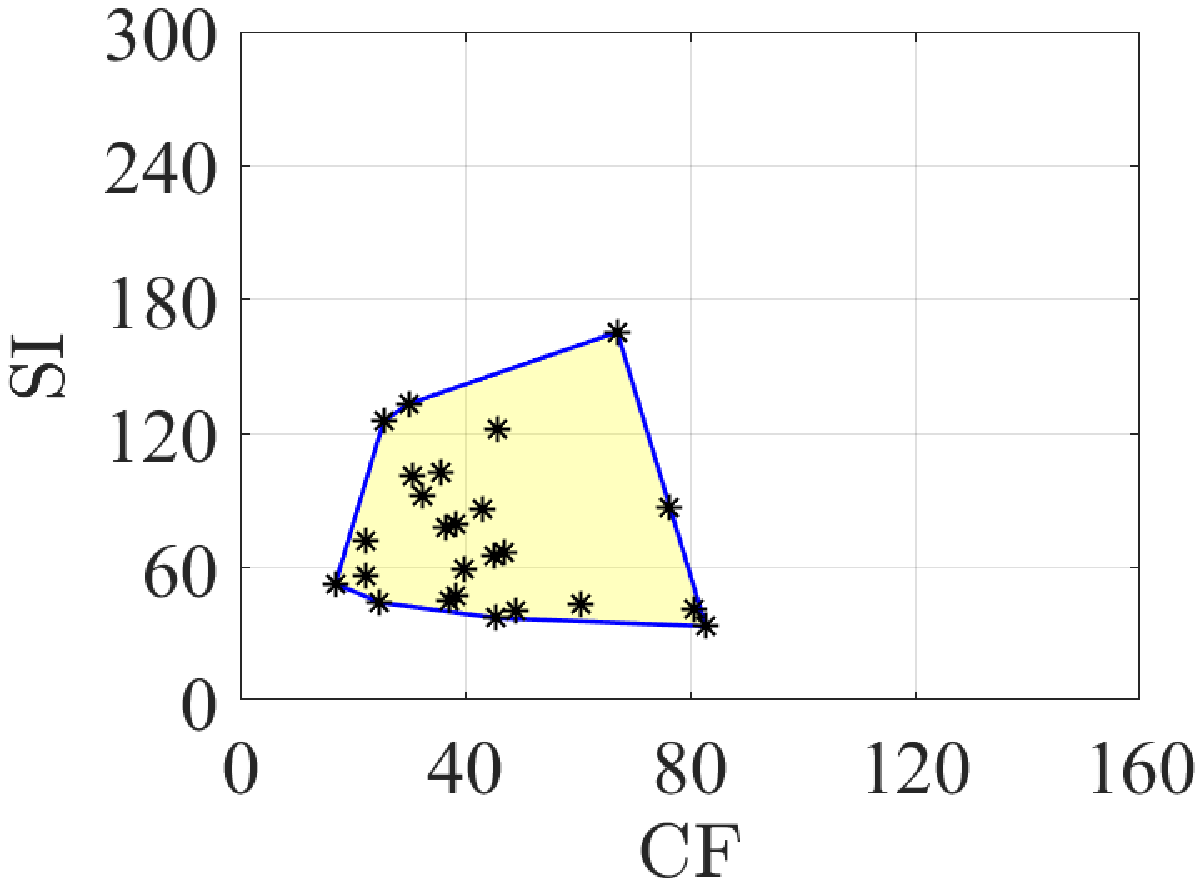} &
		\includegraphics[width=0.2\textwidth]{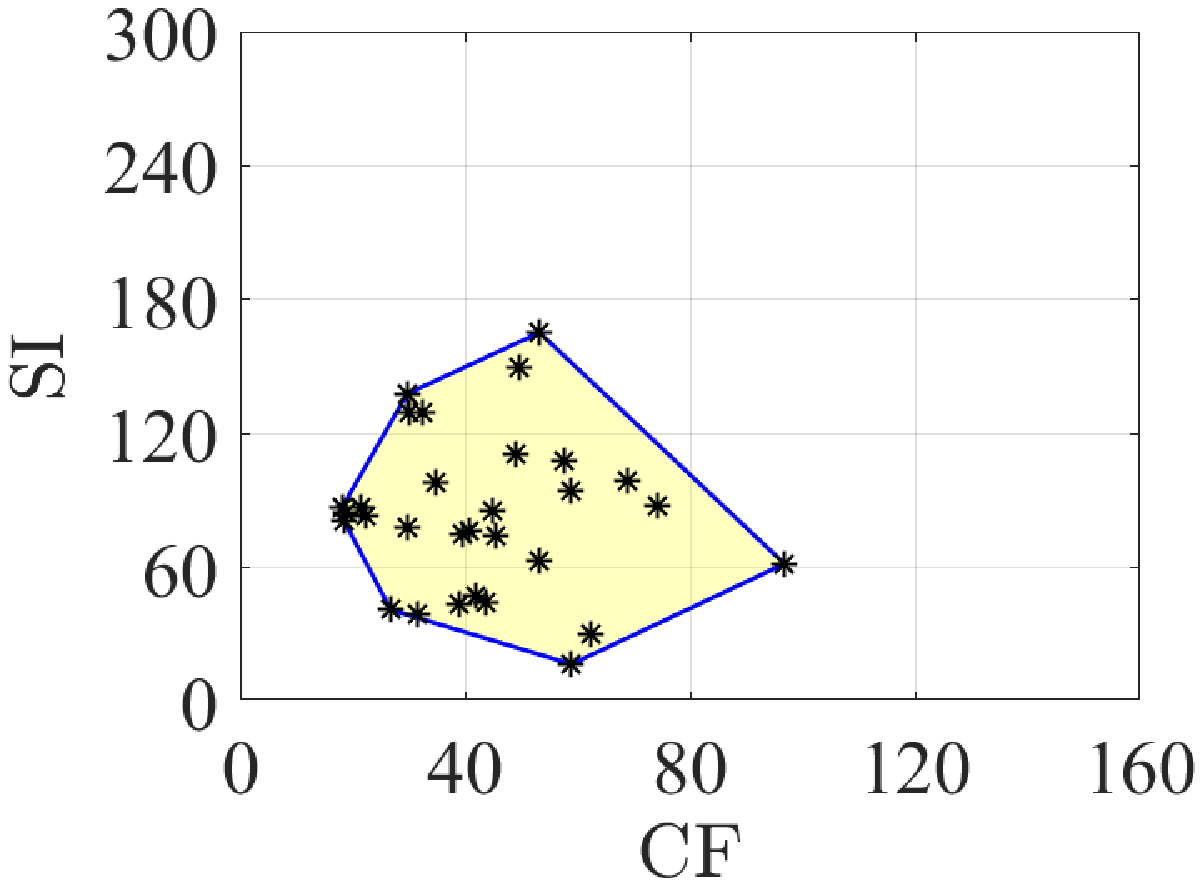} &
		\includegraphics[width=0.2\textwidth]{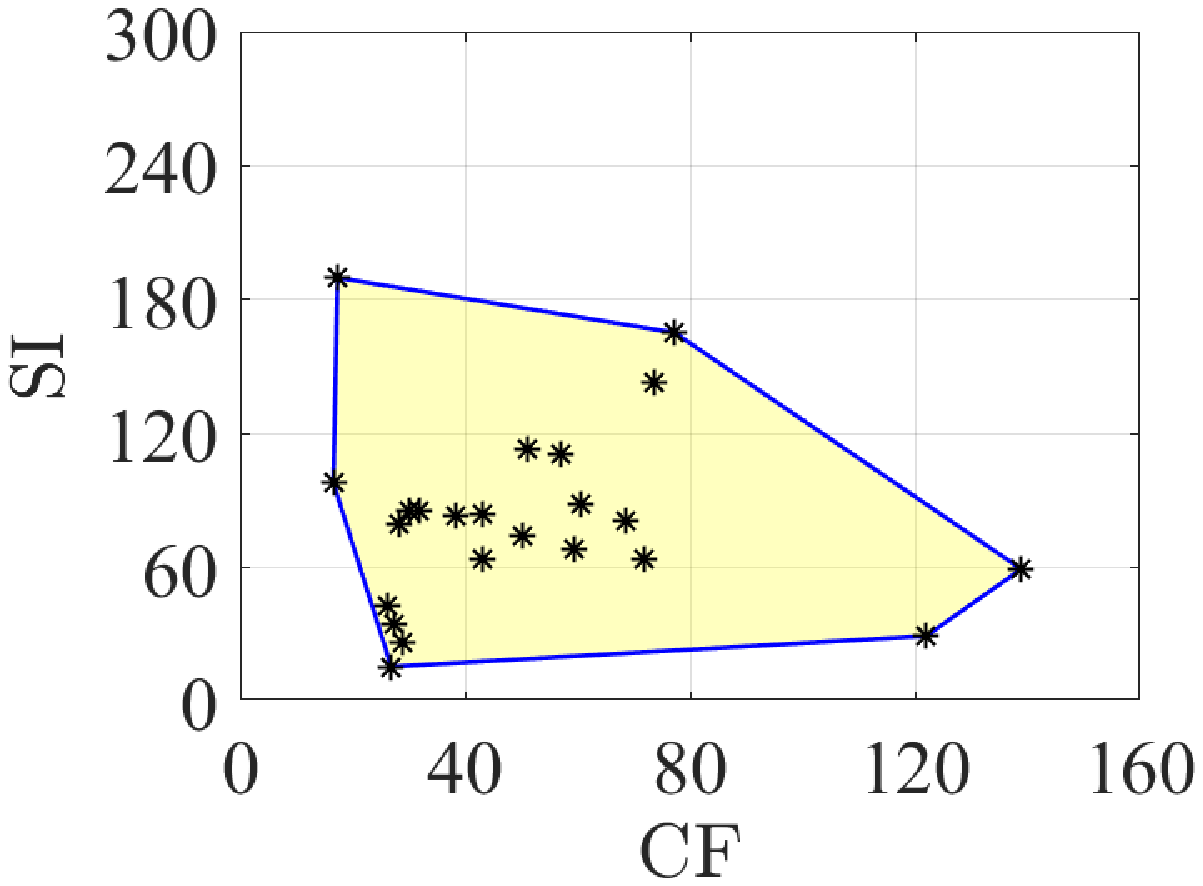} &
		\includegraphics[width=0.2\textwidth]{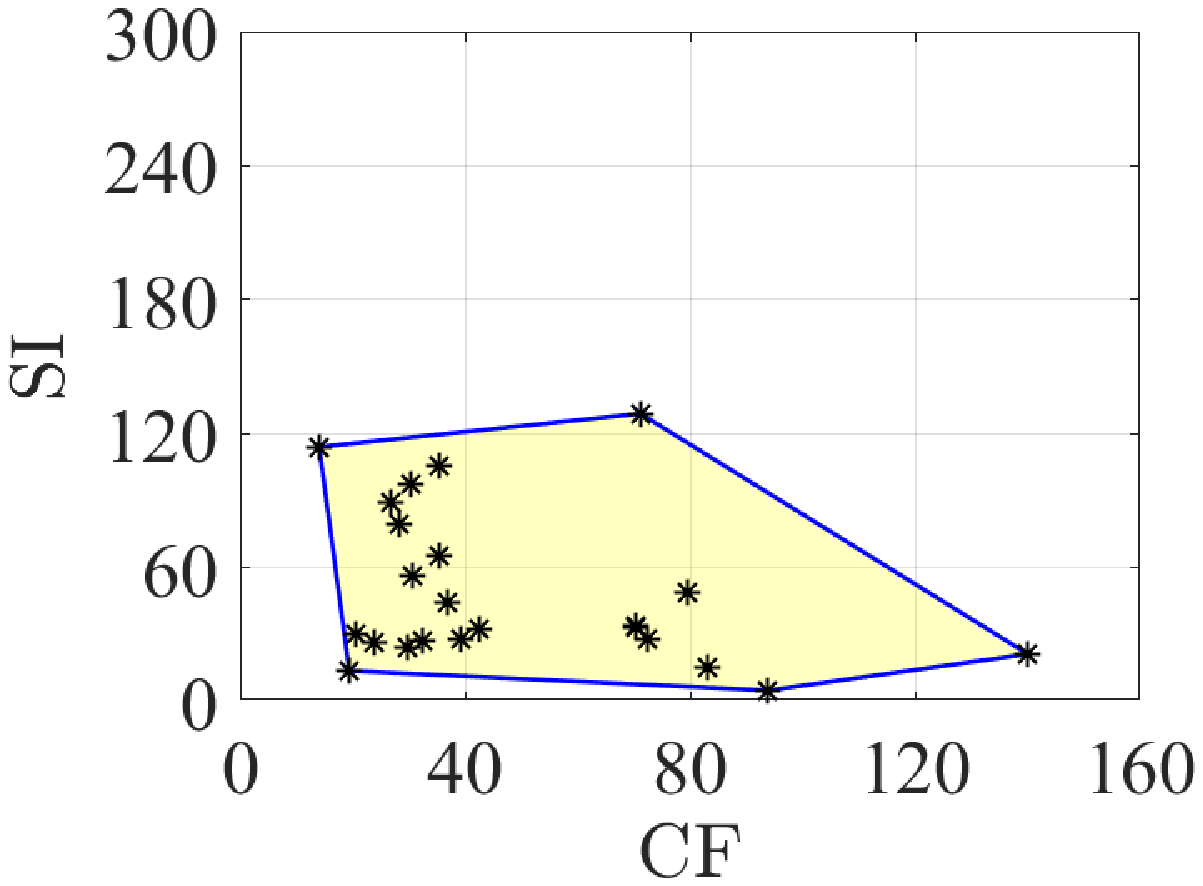} \\
		(a) LIVE R2  & (b) TID2013 & (c) CSIQ & (d) VCLFER & (e) CIDIQ  \\[6pt]
	\end{tabular}
	\begin{tabular}{c @{\hspace{-1ex}} c @{\hspace{-1ex}} c @{\hspace{-1ex}} c @{\hspace{-1ex}} c}
		\includegraphics[width=0.2\textwidth]{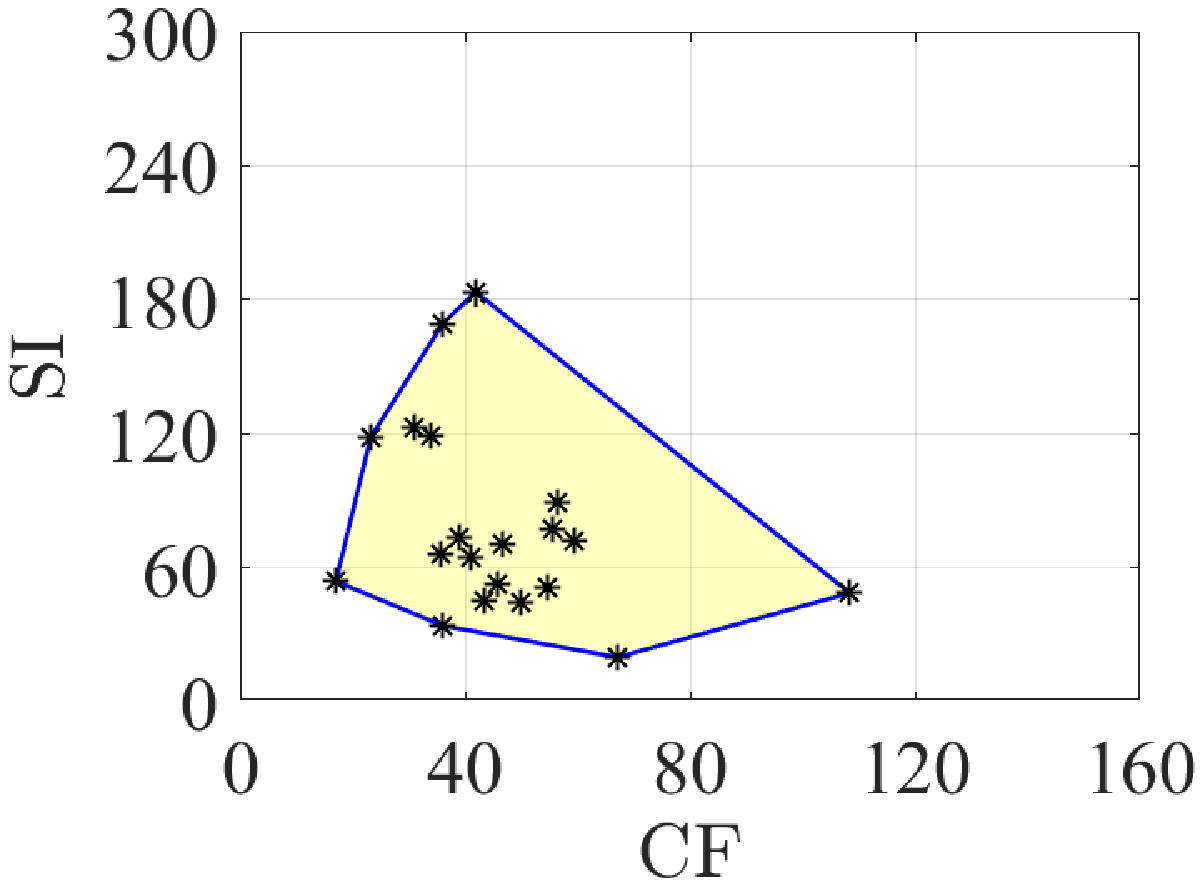} &
		\includegraphics[width=0.2\textwidth]{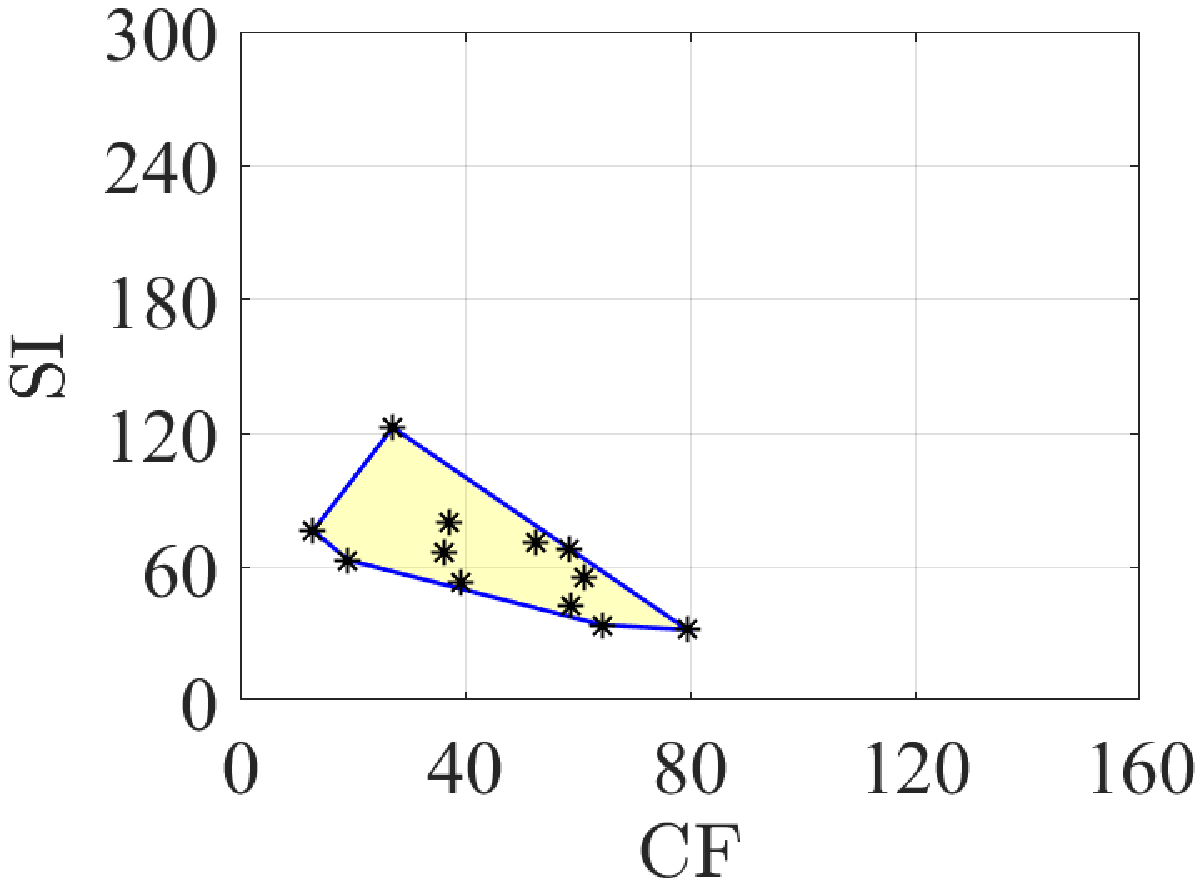} &
		\includegraphics[width=0.2\textwidth]{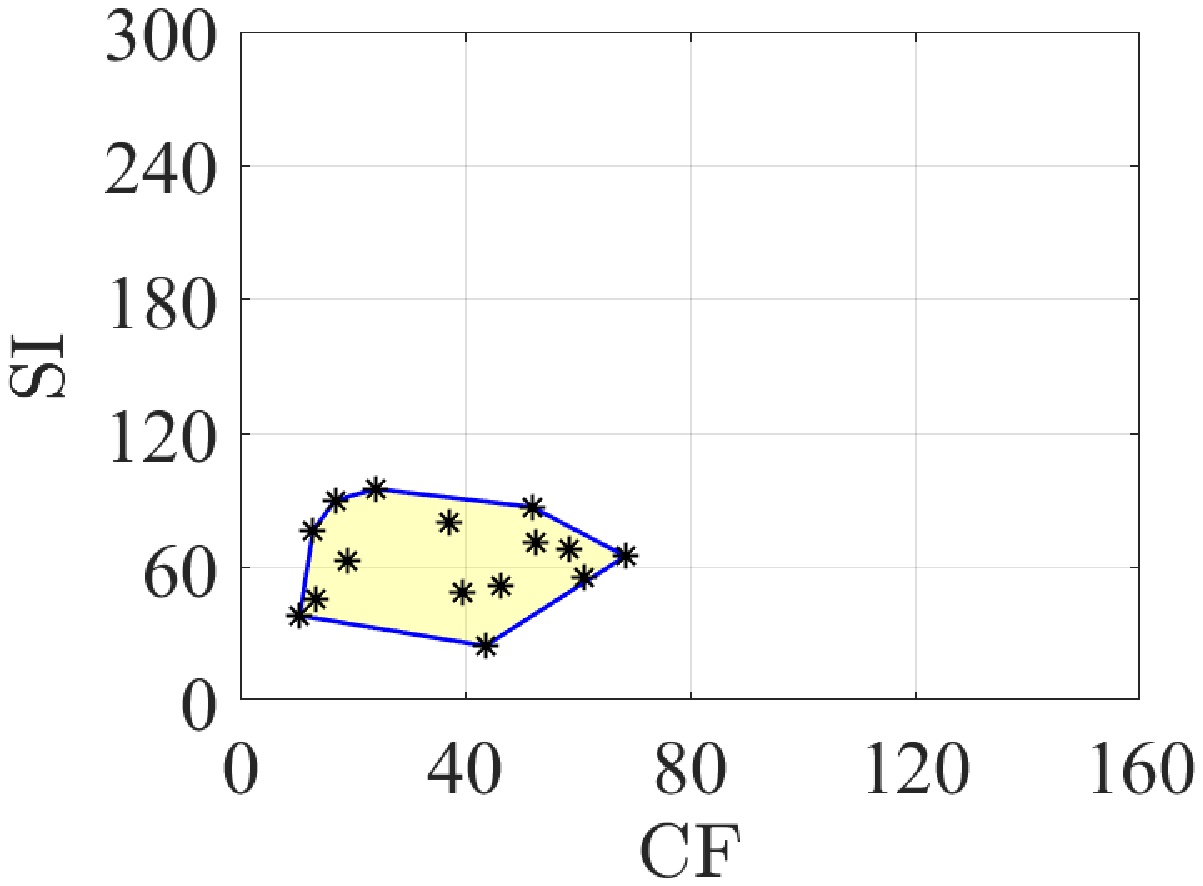} &
		\includegraphics[width=0.2\textwidth]{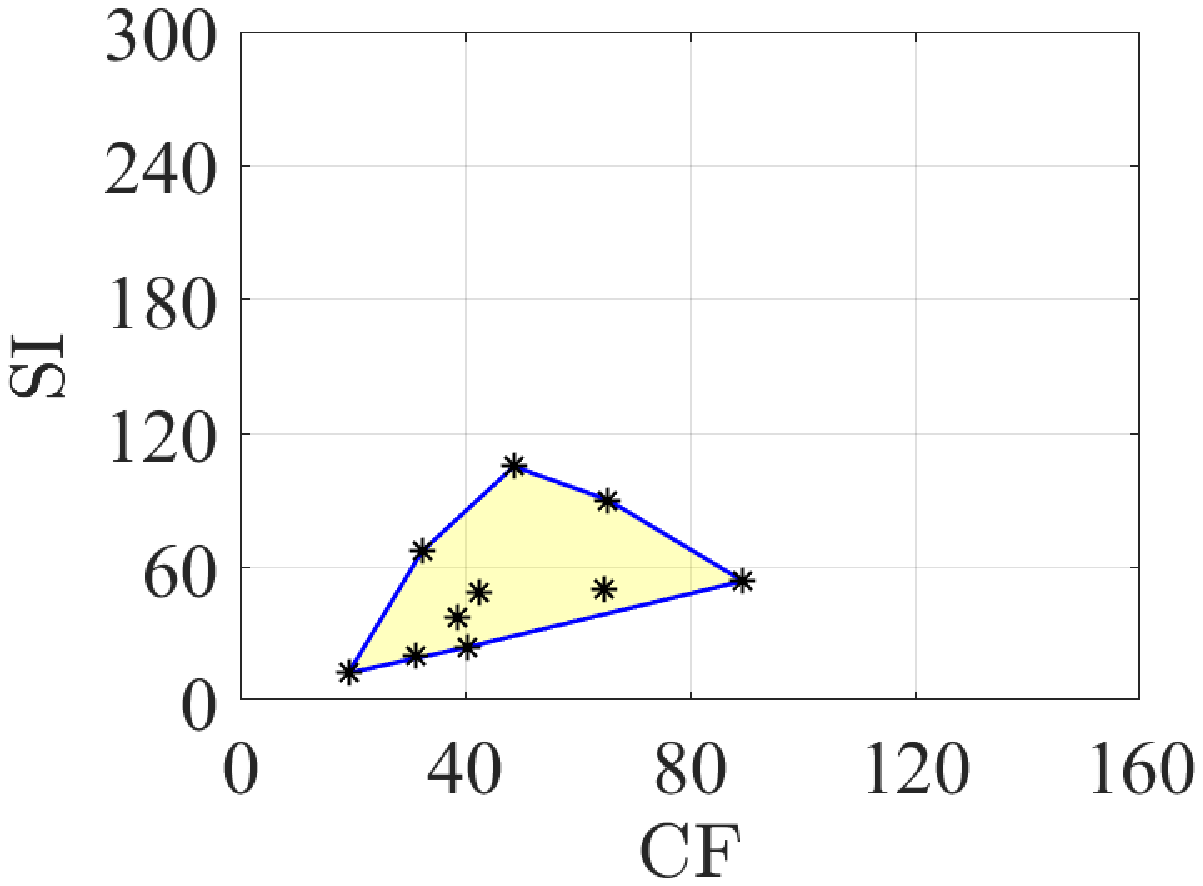} &
		\includegraphics[width=0.2\textwidth]{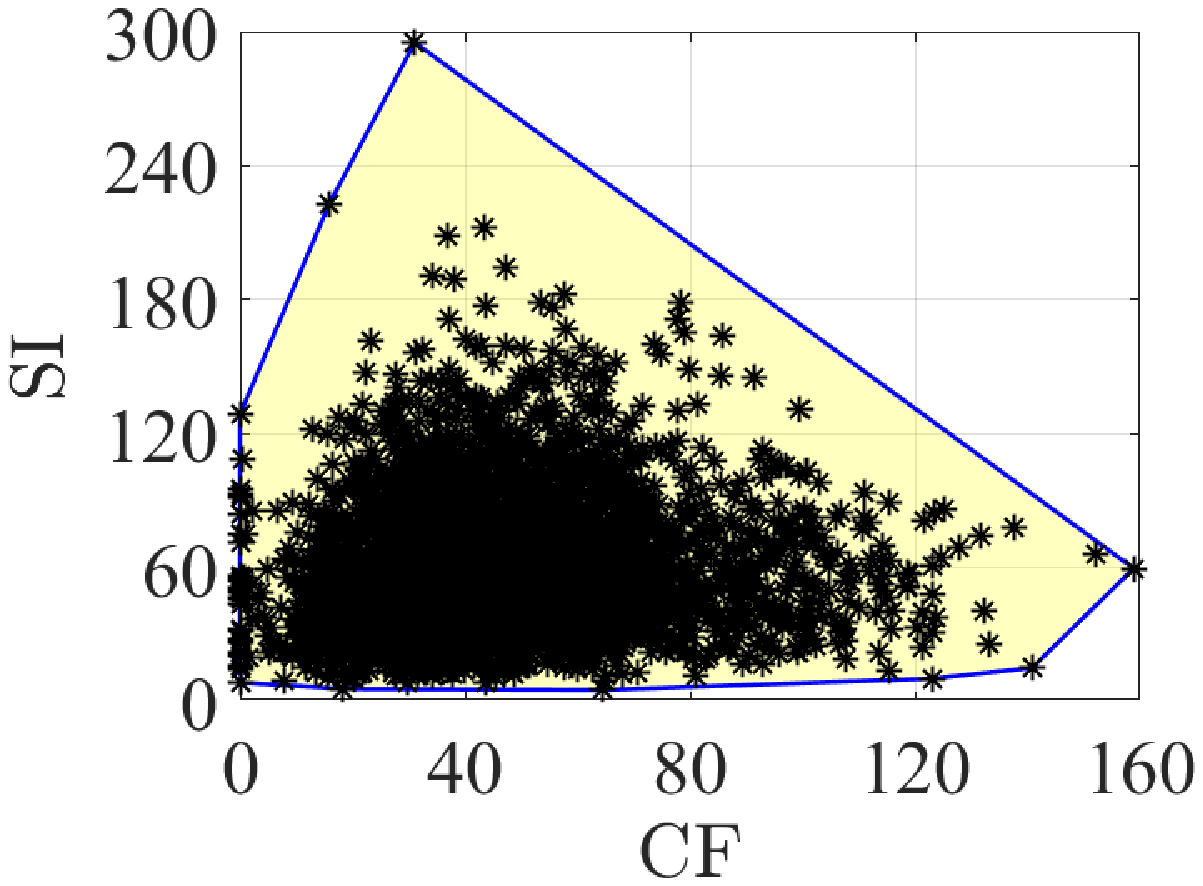} \\
		(f) MDID  & (g) MDID2013 & (h) LIVE MD & (i) MDIVL & (j) Waterloo Exp-II  \\[6pt]
	\end{tabular}
	\vspace{-3mm}
	\caption{SI versus CF plots of the reference images of Waterloo Exploration-II database (j) and nine subject-rated databases (a-i). The blue outer boundary marks the convex hull in each case.}
	\label{fig:SICF_WaterlooMD}
	\vspace{-3mm}
\end{figure*}

The Waterloo Exploration-II database\footnote{Since this paper presents a preliminary version of our work, the version of the Waterloo Exploration-II database presented here is not final and may change in the future.} is a simulated distortions dataset which starts with a set of pristine reference images and simulates both singly and multiply distorted images at various distortion levels.

\subsection{Reference Content}
\label{subsec:Ref_Cont}

The \textit{pristine} or \textit{reference} content in an IQA dataset is representative of the enormous space of all possible natural images. As is evident from Table \ref{table:IQAdbSummary}, contemporary IQA datasets have only a small number of reference images, which can be regarded as a very sparse and inadequate representation of this enormous space. To ensure wide coverage of image content, we include 3,570 reference images in Waterloo Exploration-II, which we take from the Waterloo Exploration-I database \cite{db_waterlooed}. To ensure images that are representatives of what humans see in their daily lives, the creators of Waterloo Exploration-I \cite{db_waterlooed} use 196 keywords to search the Internet for images which broadly belong to seven categories (human, animal, plant, landscape, cityscape, still-life, and transportation) and obtain an initial set of 200,000 images. Next, they manually view each image and remove images that have any visible distortions, leading to a filtered set of 7,000 images. Finally, they carry out another round of filtering where they view each remaining image by zooming in multiple times and remove any images with visible compression distortions leaving 4,744 high quality natural images that become the reference image set of Waterloo Exploration-I \cite{db_waterlooed}. For Waterloo Exploration-II, we start with the 7,000 images that were obtained after the first round of filtering while obtaining Waterloo Exploration-I reference images. Using a similar manual procedure of viewing each of these images multiple times and zooming in, we carry out another round of filtering and select 3,570 pristine quality images as our reference image set.

While the usual practice is to describe the variety of reference content by using subjective terms, a few quantitative descriptors have also been used to describe such content, such as image spatial information (SI) which is indicative of edge energy in an image, and colorfulness (CF) which represents the variety and intensity of colors in an image \cite{db_winkler_analysis}. The two-dimensional SI versus CF space has been used to represent and compare the diversity of source content in different IQA databases \cite{db_winkler_analysis}. Three different SI measures were compared in \cite{db_winkler_SI} and it was found that a mean based SI measure ($\text{SI}_{Mean}$) has the highest correlation with compression based image complexity measures. We use $\text{SI}_{Mean}$ \cite{db_winkler_SI} and a computationally efficient CF measure \cite{db_colorfulness} to evaluate the diversity of the reference image content. The SI versus CF scatter plot of the Waterloo Exploration-II database is shown in Fig. \ref{fig:SICF_WaterlooMD}(j), which suggests a comprehensively improved content representation in terms of both diversity and density in comparison with nine well-known IQA datasets as depicted in Fig. \ref{fig:SICF_WaterlooMD}(a-i).

\subsection{Distorted Content}
\label{subsec:Dist_Cont}

An ideal simulated distortions IQA dataset should be diverse in terms of distortion types and levels. The goal is to simulate varying degrees of distortions so that the perceptual quality scale is uniformly sampled, which ensures that objective IQA methods are tested (and trained) across the quality spectrum. Distortions also need to be realistic, and thus multiply distorted content takes precedence over singly distorted images. Existing simulated distortion datasets, as summarized in Table \ref{table:IQAdbSummary}, have the following shortcomings apart from their small-scale nature: 1) Most of them are singly distorted. 2) Usually 4 to 6 distortion levels per distortion type are used, which does not allow for a dense sampling of the perceptual quality scale. 3) Existing multiply distorted datasets usually have 3 to 4 distortion levels per distortion type per stage, leading to a sparse multiply distorted image set which is inadequate for learning how two or more different (or same) distortions interact with each other. 4) Distorted content in most IQA datasets is not uniformly distributed across the quality spectrum as demonstrated in \cite{eval_Waterloo}, which is because fixed distortion parameters are used to generate each distortion type. While this is a convenient approach, it does not adapt to the impact of content variations on the perceptual appearance of distortions. Since it is known that many objective IQA methods find it more difficult to evaluate better quality images compared to lower quality ones \cite{db_tid2013}, effective representation of the entire quality scale, specially the higher quality region is necessary. To address the above-mentioned shortcomings of existing IQA datasets, we generate the distorted content of the Waterloo Exploration-II database in the following manner.

\subsubsection{Content Adaptive Distortion Parameters}
\label{subsubsec:ContAdThresh}

\input{Tables/Table2.tex}
\input{Tables/Table3.tex}

To ensure uniform coverage of the entire quality spectrum, we use content adaptive distortion parameters instead of fixed ones. For its first version, we choose the following four base distortions for the Waterloo Exploration-II database: 1) Gaussian white noise, 2) Gaussian blur, 3) JPEG compression, and 4) JPEG2000 compression. We use one of the most advanced FR quality-of-experience (QoE) measures called SSIMplus \cite{fr_ssimplus}, to identify distortion parameters that correspond to a particular level of distortion for each reference image. SSIMplus predicts the quality of images on a scale of 0-100 where 100 corresponds to the best quality while 0 corresponds to the worst quality. A significant advantage of using SSIMplus is that its quality scale was calibrated to be linear with respect to perceptual quality, which means that the loss of quality associated with $x$ SSIMplus points has the same perceptual significance regardless of the starting point on the quality scale, allowing for a division of the quality scale into uniformly spaced intervals. To densely sample the quality spectrum, we choose to have 17 distortion levels for the four base distortion types. These distortion levels, their target SSIMplus scores, and quality categories are depicted in Table \ref{table:SSIMplus_DistLevels}, where it can be seen that we do not go below the SSIMplus score of 20 as the resulting images are severely distorted and may not make a useful contribution to the dataset. For each reference image of the Waterloo Exploration-II database, we use different distortion parameters to create 15,000, 10,000, 101, and 20,000 distorted images for the base distortions of Gaussian white noise, Gaussian blur, JPEG compression, and JPEG2000 compression, respectively. Finally, distortion parameters for each base distortion that lead to SSIMplus scores closest to the target scores of the 17 distortion levels (see Table \ref{table:SSIMplus_DistLevels}) are selected for subsequent database generation. Thus, each of the 3,570 reference images in the Waterloo Exploration-II database has its own set of distortion parameters for each of the four base distortion types.

\subsubsection{Dense Singly and Multiply Distorted Content}
\label{subsubsec:SingMultCont}

To better mimic real-world distortions, we construct the Waterloo Exploration-II database in two stages to include both singly and multiply distorted content, with emphasis on the latter. Table \ref{table:WaterlooMD_Comp} outlines the composition of the database. Stage-1 contains singly distorted images belonging to three distortion types: 1) Gaussian white noise, 2) Gaussian blur, and 3) JPEG compression. Images for each of these single distortion types are obtained by distorting the reference images using their respective content adaptive distortion parameters belonging to Levels 1 to 11, as depicted in Table \ref{table:SSIMplus_DistLevels}. Thus, the 11 Stage-1 distortion levels correspond to the SSIMplus \cite{fr_ssimplus} quality range of 50 to 100, which is representative of \textit{fair} to \textit{excellent} perceptual quality. We restrict ourselves to the top half of the perceptual quality spectrum as distorted images in the earlier part of the media distribution pipelines are expected to be in this quality range. This leads to 39,270 singly distorted images for each single distortion category for a total of 117,810 singly distorted images. Stage-2 contains multiply distorted images belonging to five distortion combinations: 1) Gaussian white noise followed by JPEG compression (Noise-JPEG), 2) Gaussian white noise followed by JPEG2000 compression (Noise-JP2K), 3) Gaussian blur followed by JPEG compression (Blur-JPEG), 4) Gaussian blur followed by Gaussian white noise (Blur-Noise), and 5) JPEG compression followed by JPEG compression (JPEG-JPEG). Stage-2 multiply distorted images are obtained by starting from the respective Stage-1 singly distorted images and distorting them by using the content adaptive distortion parameters of the parent reference image belonging to Levels 1 to 17, as depicted in Table \ref{table:SSIMplus_DistLevels}, where it can be seen that this covers SSIMplus quality range of 20 to 100, which is representative of \textit{bad} to \textit{excellent} perceptual quality. Thus, the first distortion in multiply distorted images belongs to the fair to excellent quality range and the subsequent distortion belongs to the entire meaningful quality spectrum (bad to excellent). Each of the five multiple distortion combinations has 667,590 images for a total of 3,337,950 multiply distorted images. Overall, Waterloo Exploration-II has 3,455,760 distorted images, which we annotate with synthetic perceptual quality ratings (explained in Section \ref{sec:SGT}), making it by far the largest annotated dataset in IQA.

\subsubsection{Distorted Content Analysis}
\label{subsubsec:DistContAnalysis}

\begin{figure}[!t]
	\centering
	\includegraphics[width=2.5in]{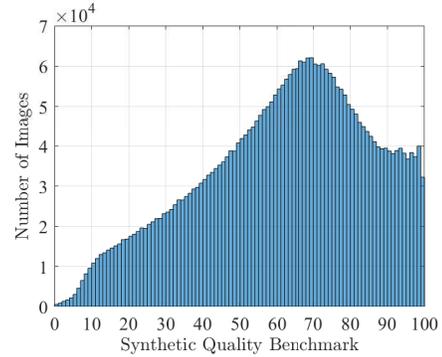}
	\vspace{-2mm}
	\caption{SQB histogram of the Waterloo Exploration-II database.}
	\label{fig:SGThist_WaterlooE2}
	\vspace{-4mm}
\end{figure}

To observe how well the Waterloo Exploration-II database covers the perceptual quality spectrum, we plot the synthetic quality benchmark (SQB) histogram of the dataset in Fig. \ref{fig:SGThist_WaterlooE2}. The generation of these synthetic quality labels will be explained in Section \ref{sec:SGT}. The SQB has a quality range of 0 to 100, where 100 is representative of the best quality while 0 represents the worst quality. It can be seen from Fig. \ref{fig:SGThist_WaterlooE2} that the Waterloo Exploration-II database has more than at least 10,000 annotated images for each integer quality value above 10, thereby ensuring adequate representation of each quality value. It can also be seen that the quality range of 50 to 100 has the most images, which ensures that the higher quality range, which is difficult to assess for objective IQA methods \cite{db_tid2013}, is adequately represented.

\begin{figure}[!t]
	\centering
	\includegraphics[width=2.5in]{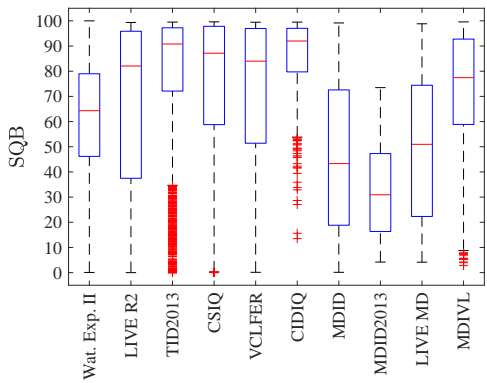}
	\vspace{-2mm}
	\caption{SQB box plot of Waterloo Exploration-II (Wat. Exp. II) in comparison with nine well-known IQA datasets. The top and bottom edges of the blue boxes represent the 75th and 25th percentiles, respectively. The red line represents the median (50th percentile). The top and bottom black lines represent extreme data points. The outliers are represented by red + symbols.}
	\label{fig:SGT_BoxPlot}
	\vspace{-4mm}
\end{figure}

To see how well the Waterloo Exploration-II database covers the quality spectrum in comparison with other well-known IQA datasets, we compute the SQB values of all the distorted images in Waterloo Exploration-II and nine well-known IQA datasets, and provide the corresponding boxplots in Fig. \ref{fig:SGT_BoxPlot}, where the range of distortions in different databases can be directly compared. By observing these boxplots, it becomes clear that while most contemporary IQA datasets tend to favor either the higher or lower end of the quality spectrum, Waterloo Exploration-II offers a better spread and more balanced coverage, with the highest concentration at the practically most common mid-to-high quality range.

\section{Synthetic Quality Benchmark}
\label{sec:SGT}

\subsection{Background and Extensive Review}
\label{subsec:PerfSurvey}

As discussed in Section \ref{subsubsec:SmallScaleIQAdb} it is not possible to annotate large-scale IQA datasets through human observers, and thus assignment of perceptual quality annotations through alternative means is necessary. Given that the area of FR IQA has matured quite well \cite{eval_Waterloo}, one possible alternative is to replace subjective ratings with scores from reliable FR methods. In fact, a number of works in IQA literature have already taken this route. During its training phase, the BIQA method QAC \cite{nr_qac} uses the FR method FSIM \cite{fr_fsim} to annotate image patches with quality scores based on which subsequent grouping is done. The BLISS framework \cite{fusion_bliss} proposes a way to convert opinion-aware BIQA methods into opinion-unaware ones. It first fuses five FR methods (FSIM \cite{fr_fsim}, FSIMc \cite{fr_fsim}, GMSD \cite{fr_gmsd}, IWSSIM \cite{fr_iwssim}, and VIF \cite{fr_vif}) to generate synthetic scores for a dataset composed of 100 reference and 3,200 distorted images. It then uses these synthetic scores to retrain a BIQA method CORNIA \cite{nr_cornia} which was previously trained through subjective ratings. The BIQA method dipIQ \cite{nr_dipiq} uses three FR methods (GMSD \cite{fr_gmsd}, MSSSIM \cite{fr_msssim}, and VIF \cite{fr_vif}) to generate 80-million quality-discriminable image pairs (DIPs) from a dataset that has 840 reference and 16,800 distorted images, which are then used to learn a blind quality model. A recent BIQA method MUSIQUE \cite{md_musique} uses the FR method VIF \cite{fr_vif} in its training stage to find a relationship between estimated distortion parameters and VIF quality scores. The BIQA method BIECON \cite{nr_biecon_C,nr_biecon_J} uses FR methods (FSIMc \cite{fr_fsim}, GMSD \cite{fr_gmsd}, SSIM \cite{fr_ssim}, and VSI \cite{fr_vsi}) to derive local scores of $32 \times 32$ patches and then pre-trains a CNN using these patches with corresponding FR scores. The model is then fine-tuned on a subject-rated dataset. In \cite{md_waterloo17}, the FR method MSSSIM \cite{fr_msssim} is used to annotate four large-scale databases of singly and multiply distorted images, the largest of which is composed of around 2 million images. 

While the above-mentioned works demonstrate that FR scores may be used in place of subjective ratings, their choices of FR methods are rather ad hoc and deeper justification and analysis are lacking. The following questions arise when using FR scores for annotating large-scale IQA datasets as alternatives to subjective ratings: 1) Which FR method or methods can be reliably used? 2) Can fused FR methods, which combine the results of multiple FR methods, offer any further advantages over individual ones? To comprehensively answer these questions, we carried out the largest performance evaluation study to-date in IQA literature in \cite{eval_Waterloo}, as a prerequisite requirement of this work. In this study, we compared the performance of 43 FR and seven fused FR methods (22 versions) on nine subject-rated IQA databases (five singly and four multiply distorted) to ensure the diversity of test data. Our results from \cite{eval_Waterloo} indicate the following: 1) Among individual FR methods, the structural similarity based methods IWSSIM \cite{fr_iwssim}, FSIMc \cite{fr_fsim}, VSI \cite{fr_vsi}, and DSS \cite{fr_dss}, outperform others. 2) However, the performance of even the best individual FR methods varies, at times widely, across different IQA datasets, a point which has earlier been noted in \cite{db_dataset_impact}. This puts into question the robustness of individual FR methods, especially when using them as alternatives to human annotations. 3) Among FR fusion methods, learning based fusion methods such as MMF \cite{fusion_mmf_tip} and CNNM \cite{fusion_cnnm}, and empirical fusion methods such as HFSIMc \cite{fusion_hfsim}, CM3 \cite{md_cm3cm4}, and CM4 \cite{md_cm3cm4}, are outperformed by the best individual FR methods and thus do not offer any further advantages. 4) However, the FR fusion method which we called RRF \cite{fusion_rrf} based Adjusted Scores (RAS) in \cite{eval_Waterloo}, is found to outperform not only the other fusion based methods, but more importantly, the best individual FR methods. In the literature of IQA, RAS was originally proposed as part of the BLISS framework in \cite{fusion_bliss} and uses a rank aggregation based fusion strategy \cite{fusion_rrf}, but no deeper analysis or empirical justification was provided about which FR methods to fuse. 5) The performance of RAS is found to be more stable across different IQA datasets relative to individual FR methods. Thus, the training-free rank aggregation based fusion strategy \cite{fusion_rrf} is a strong candidate for synthetically annotating large-scale IQA datasets.

\subsection{Synthetic Quality Benchmark Generation}
\label{subsec:SGT_Gen}

\begin{figure*}[!t]
	\centering
	\includegraphics[width=\textwidth]{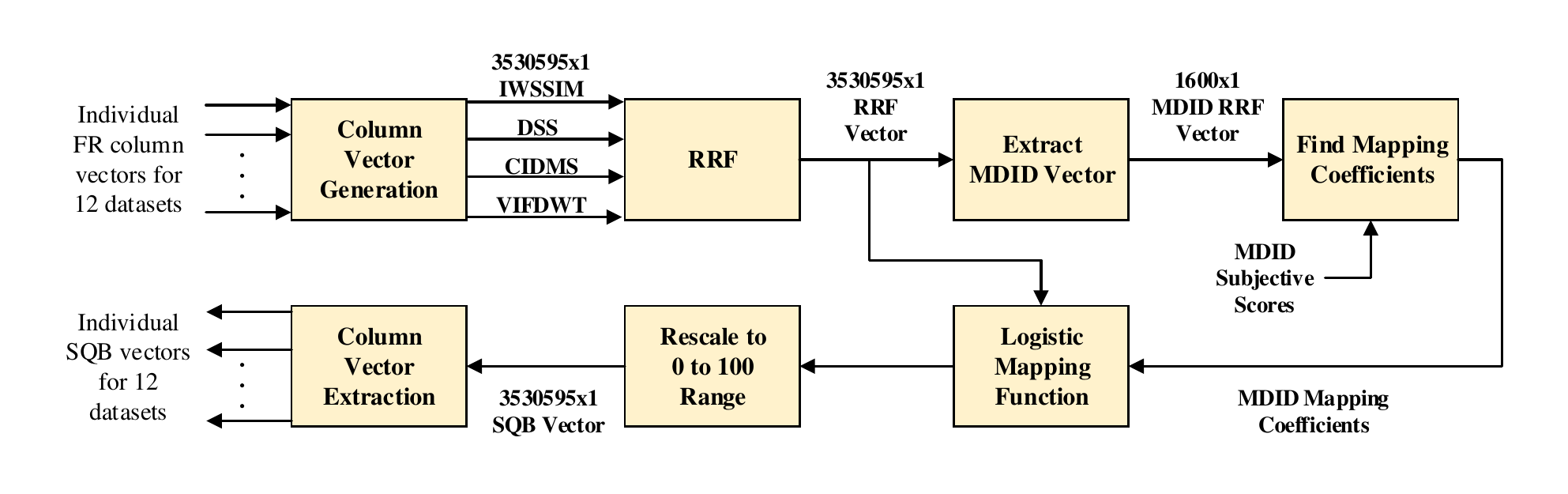}
	\vspace{-8mm}
	\caption{Synthetic Quality Benchmark (SQB) generation procedure.}
	\label{fig:SGT_Gen}
	\vspace{-4mm}
\end{figure*}

At the core of the rank aggregation based fusion strategy is the training-free Reciprocal Rank Fusion (RRF) algorithm \cite{fusion_rrf}, which was first developed to combine document rankings from multiple information retrieval systems in an unsupervised manner. For a given set of test images and their associated quality scores as assigned by different FR IQA methods, a consensus ranking can be obtained in terms of RRF as follows \cite{fusion_rrf}:

\begin{equation}
\centering
{RRF_{score}(I_i) = \sum_{j=1}^{J} \frac{1}{k + r_j(i)}
	\label{eq:RRF}}
\end{equation}

\noindent where $J$ is the number of FR methods being fused, $r_j(i)$ is the rank given by the \textit{j-}th FR method to the image $I_i$, $RRF_{score}(I_i)$ is the RRF score of image $I_i$, and $k$ is a stabilization constant. RRF was first used in IQA as part of the BLISS framework \cite{fusion_bliss}, which replaces human opinion scores with synthetic quality scores that act as ground truth data to train BIQA methods. The BLISS framework \cite{fusion_bliss} produces synthetic quality scores in two steps for a given set of images: 1) Generation of a consensus ranking score through RRF \cite{fusion_rrf}, and 2) Since the ranking score is a measure of quality relative to other images and cannot be considered an independent quality measure, the scores of a base FR method are adjusted based on the consensus ranking, which then act as synthetic quality scores. The latter step is required because the BLISS framework operates in the absence of subject-rated datasets. The choice of FR methods to combine in \cite{fusion_bliss} is ad hoc. To test the rank aggregation based fusion of FR methods more thoroughly, we performed an exhaustive search by testing 737,280 different combinations of 2 to 15 FR methods, and finalized 13 versions of RAS in \cite{eval_Waterloo}. Among them, RAS6 was found to be the top performer and will be used as a basis for the generation of our synthetic annotations, explained next.

To generate the \textit{Synthetic Quality Benchmark} (SQB) for the very large-scale Waterloo Exploration-II database, we use RRF \cite{fusion_rrf} to fuse the same four FR methods as in RAS6 \cite{eval_Waterloo}, i.e., IWSSIM \cite{fr_iwssim}, DSS \cite{fr_dss}, CID\_MS \cite{fr_cid}, and VIF\_DWT \cite{fr_vifdwt_ssimdwt}. However, unlike RAS (BLISS framework \cite{fusion_bliss}), we do not adjust the score of a base FR method to generate the synthetic quality scores. Instead, we use the novel framework shown in Fig. \ref{fig:SGT_Gen} to generate the SQB. First, we acquire the scores of the above-mentioned four FR methods for 12 databases mentioned in Table \ref{table:SGT_ColVec}. These include nine subject-rated datasets (serial number 4 to 12) which have been described earlier in Table \ref{table:IQAdbSummary}, and three large datasets which include Waterloo Exploration-II, and two other datasets called DR-IQA V1 and DR-IQA V2, which we have developed by following the same design process as Waterloo Exploration-II. The purpose of these latter two datasets is to provide training and validation data for machine learning tools that do not require as much data as DNN models, like SVR-based model development, and we use them in our work on degraded reference (DR) IQA which will be published separately. For each dataset, we obtain scores for each FR method in terms of database-wide column vectors, which are all then concatenated into one large column vector of size 3530595$\times$1 for each FR method, as depicted in Table \ref{table:SGT_ColVec}. Next, RRF is used to fuse the four large column vectors, through Equation \ref{eq:RRF}, resulting in an RRF vector which contains the consensus ranking. Since the RRF process involves sorting the constituents of individual vectors being fused, this results in punctuating the scores of the three large databases without subject ratings with scores of the nine databases that do have human annotations of quality. This is done without subjects rating the images of the three large databases, and meanwhile allows us to evaluate the performance of SQB compared to actual human annotations in Section \ref{subsubsec:SGTPerfEval}. The RRF vector is normalized as follows:

\begin{equation}
\centering
{RRF_{normalized} = \frac{RRF_{score}}{max(RRF_{score})}
	\label{eq:RRFnorm}}
\end{equation}

\input{Tables/Table4.tex}

As discussed earlier, the outcome of the RRF step leads to a quality rating for an image relative to other images. To be considered independently, the ratings from the consensus RRF ranking can be mapped to a subjective quality scale by using a subset of RRF scores for which subjective quality scores are available. Since the MDID database \cite{db_mdid} has uniform representation from different parts of the quality spectrum \cite{eval_Waterloo}, we choose it to learn this mapping through a five-parameter modified logistic function \cite{stateval_db_liveR2}:

\begin{equation}
\centering
{S(R) = \beta_1\Bigg[\frac{1}{2} - \frac{1}{1+e^{\{\beta_2(R-\beta_3)\}}}\Bigg] + \beta_4 R +\beta_5
	\label{eq:RRFnlmapping}}
\end{equation}

\noindent where $R$ denotes the RRF scores of the MDID database that have been extracted from the overall normalized RRF vector, $S$ denotes the predicted MDID subjective quality scores, and $\beta_1$, $\beta_2$, $\beta_3$, $\beta_4$, and $\beta_5$ are mapping coefficients that are found numerically to maximize the correlation between MDID subjective quality scores and its RRF scores. Since the MDID RRF scores punctuate the overall RRF vector in a regular manner, the MDID mapping coefficients are then used to map the entire RRF vector to the MDID subjective quality scale (0 to 8) by again using Equation \ref{eq:RRFnlmapping}. These quality scores, denoted by $Q$, are then rescaled to the 0 to 100 range as:

\begin{equation}
\centering
{SQB = 100 \times \frac{Q - min(Q)}{max(Q - min(Q))}
	\label{eq:SGT}}
\end{equation}

Equation \ref{eq:SGT} results in the concatenated \textit{Synthetic Quality Benchmark} (SQB) vector for all databases involved, and ensures that the rescaling process does not disturb the distribution of the quality scores. Finally, the SQB vectors for individual databases are extracted from the overall SQB vector. 

\begin{figure}[!t]
	\centering
	\includegraphics[width=2.5in]{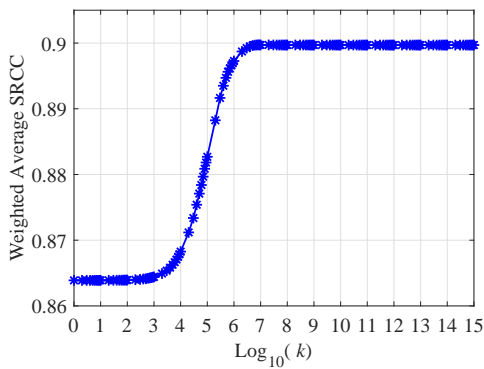}
	\vspace{-2mm}
	\caption{Weighted average SRCC of SQB with respect to subjective scores of the nine subject-rated databases (see Table \ref{table:SGT_ColVec}) for different values of the RRF constant \textit{k}.}
	\label{fig:RRFgammaSel}
	\vspace{-4mm}
\end{figure}

It is mentioned in \cite{fusion_rrf} that the constant $k$ in Equation \ref{eq:RRF} counters the impact of high rankings by outliers and its value was set at 60 through a pilot investigation. This value of the constant $k$ was also used in the BLISS framework \cite{fusion_bliss}. While the value of $k$ may not be critical when the number of data points is small, we believe that it takes on a more significant role when the number of objects to be ranked is large. In our case, with around 3.53 million images, a small value of $k=60$ leads to weights assigned to rank 1 and to rank 3,530,595 that differ by several orders of magnitude. Thus, some ranks are more favored than others and a level playing field is absent. We believe that in order to rank a large number of objects, the value of the constant $k$ should be proportionately higher. To test this hypothesis, we carry out an empirical study where the value of $k$ was progressively increased in terms of order of magnitude, and the overall SQB vector was recomputed each time. The SQB scores for each of the nine subject-rated databases mentioned in Table \ref{table:SGT_ColVec} were extracted for each value of $k$. For each database we compute the Spearman Rank-order Correlation Coefficient (SRCC) of the SQB with respect to the respective subjective scores. The weighted average SRCC values for the nine subject-rated databases for various values of $k$ are depicted in Fig. \ref{fig:RRFgammaSel}, where it can be seen that our hypothesis is indeed correct. Given that we have around 3.53 million images for which RRF is being computed, the weighted average SRCC starts increasing as $k$ goes beyond $10^4$ and keeps on increasing until $k$ attains a value of around $10^7$ beyond which it remains constant. We believe that further increase of the value of $k$ does not lead to further SRCC gain as the weights assigned to all the ranks remain within the same order of magnitude. Through our empirical investigation, we have found that for 3.53 million images, the weighted average SRCC does not increase beyond $k=8 \times 10^6$, and hence this value of $k$ has been used in this work.

\subsection{SQB Performance}
\label{subsec:SGT_Perf}

\subsubsection{Databases and Methods used for Comparison}
\label{subsubsec:DBmethods}

To comprehensively test the perceptual quality prediction performance of SQB, we test it on the nine subject-rated IQA datasets that were made part of the SQB computation process as discussed earlier. Five of these datasets include singly distorted content and include LIVE R2 \cite{stateval_db_liveR2}, TID2013 \cite{db_tid2013}, CSIQ \cite{fr_mad_db_csiq}, VCLFER \cite{db_vclfer}, and CIDIQ \cite{db_cidiq}, while four contain multiply distorted content and include MDID \cite{db_mdid}, MDID2013 \cite{md_sisblim_db_mdid2013}, LIVE MD \cite{db_livemd}, and MDIVL \cite{db_mdivl}. It should be noted that CIDIQ \cite{db_cidiq} contains subject-ratings at two viewing distances (50 cm and 100 cm). We shall refer to these two sets of subjective data as CIDIQ50 and CIDIQ100, respectively. The main features of these databases are given in Table \ref{table:IQAdbSummary}, while more details can be found in \cite{eval_Waterloo} or in their respective papers.

For a thorough comparison, we tested the performance of other state-of-the-art methods, including two fused and 14 individual FR methods, on the above-mentioned datasets. The fused FR methods include the rank aggregation based fusion method RAS6 which was the overall top performer in our earlier performance evaluation study \cite{eval_Waterloo} and uses the approach described in \cite{fusion_bliss}. We also include the learning based fusion method MMF \cite{fusion_mmf_tip} in our comparison. The 14 individual FR methods are top performers in \cite{eval_Waterloo} and belong to three state-of-the-art FR design philosophies. Among them, eight methods belong to the \textit{structural similarity} based design philosophy and include CID\_MS \cite{fr_cid}, DSS \cite{fr_dss}, ESSIM \cite{fr_essim}, FSIMc \cite{fr_fsim}, GMSD \cite{fr_gmsd}, IWSSIM \cite{fr_iwssim}, MCSD \cite{fr_mcsd}, and VSI \cite{fr_vsi}, four are \textit{natural scene statistics} (NSS) based and include QASD \cite{fr_qasd}, SFF \cite{fr_sff}, VIF \cite{fr_vif}, and VIF\_DWT \cite{fr_vifdwt_ssimdwt}, and two belong to the \textit{mixed strategy} based design philosophy and include DVICOM \cite{fr_dvicom} and MAD \cite{fr_mad_db_csiq}. We also include the \textit{error based} method PSNR for legacy purposes.

\subsubsection{Evaluation Criteria}
\label{subsubsec:EvalCrit}

Five evaluation criteria are used to analyze the performance of methods under test: 1) For assessing \textit{prediction accuracy}, we use the Pearson Linear Correlation Coefficient (PLCC) \cite{vqeg_report}. The scores generated by objective IQA methods are usually not linear with respect to subjective ratings. Thus, a nonlinear mapping step is required before the computation of PLCC. To do this, we adopt the five-parameter modified logistic function used in \cite{stateval_db_liveR2} and given in Equation \ref{eq:RRFnlmapping}. PLCC is then computed at the database-level between the subjective scores and the objective scores after passing through the nonlinear mapping step. 2) We assess \textit{prediction monotonicity} by using the Spearman Rank-order Correlation Coefficient (SRCC) \cite{vqeg_report}. SRCC is a non-parametric rank-order based correlation measure. It does not require the preceding nonlinear mapping step. The absolute value of both PLCC and SRCC lies in the 0 to 1 range. A better objective IQA method should have higher PLCC and SRCC values with respect to subjective scores, where a value of 1 would indicate perfect perceptual performance. 3) Since we are using nine different IQA datasets for performance evaluation (ten if the two viewing distances of CIDIQ are considered separately) and PLCC values are at the individual database-level, trying to make conclusions about the overall performance becomes cumbersome and a measure of aggregate performance is required. We provide this measure for \textit{overall prediction accuracy} by calculating the weighted average (WA) PLCC value for each IQA method across all datasets (as in \cite{fr_iwssim,eval_Waterloo}). The total number of distorted images in a dataset define the weight assigned to it in the weighted average computation. The CIDIQ database \cite{db_cidiq} is considered twice in this calculation due to its evaluations being at two viewing distances. 4) Similarly \textit{overall prediction monotonicity} is determined through WA SRCC. 5) Finally we perform statistical significance testing (hypothesis testing) to draw statistically sound and generalizable inferences about the performance of an IQA method compared to another. We carried out these tests on the prediction residuals of different methods for each database. These residuals were obtained by first mapping the IQA method outcomes to subjective scores by using the nonlinear mapping approach described above for PLCC calculation, and then subtracting these predictions from the actual subjective scores. We use the one-sided (left-tailed) two-sample \textit{F}-test \cite{statTest_fTest_ref} to statistically compare the performance of two IQA methods with each other at the 5\% significance level (95\% confidence) for each of the IQA databases. By carrying out this test twice, with the order of the methods reversed, we were able to determine if the method performance was statistically indistinguishable or if one method performed better than another. Since these tests assume the Gaussianity of residuals, we used a simple kurtosis based check (as in \cite{stateval_db_liveR2,eval_Waterloo}), where Gaussianity is assumed if the kurtosis of the residuals is between 2 and 4. The databases and evaluation criteria used in this work are the same as in our earlier work \cite{eval_Waterloo}, thus results in this paper can be directly compared with other methods discussed there. For a more detailed description of these evaluation criteria, readers can refer to \cite{eval_Waterloo}.

\subsubsection{SQB Performance Evaluation}
\label{subsubsec:SGTPerfEval}

\input{Tables/Table5.tex}
\input{Tables/Table6.tex}

Since the very-large scale Waterloo Exploration-II database does not have subjective ratings, it is not possible to evaluate the performance of its SQB annotation scores directly. The reason why we concatenated the objective score vectors, belonging to the nine subject-rated datasets, of the FR methods being fused in SQB computation, with those of the Waterloo Exploration-II, DR-IQA V1, and DR-IQA V2 databases, is that it allowed us to punctuate data without subject ratings with data that does have these ratings. Thus, in the overall SQB vector, the SQB scores for the nine subject-rated datasets act as regularly distributed samples. Since these samples also have subjective scores available, this allows us to comprehensively test the performance of SQB.

The perceptual quality prediction performance, of all methods under test for the nine IQA datasets, in terms of PLCC and SRCC, is given in Table \ref{table:SGT_PLCC_SRCC}. For each IQA database, all of its constituent distortions were included for testing. The WA PLCC and WA SRCC are provided in the rightmost column of Table \ref{table:SGT_PLCC_SRCC} and are used to sort the methods in the descending order. Thus, the best performing methods are towards the top of the respective table sections for PLCC and SRCC. The names of the fused FR methods are mentioned in bold, to distinguish them from the individual FR methods. The results of statistical significance testing of SQB relative to the 17 other methods are shown in Table \ref{table:SGT_StaSig}, where a ``1'', ``--'', or ``0'' means that the perceptual quality prediction performance of SQB is better, indistinguishable, or worse, respectively, than that of the method in the row for a given database (with 95\% confidence). We preceded the statistical significance testing with a kurtosis based check for Gaussianity of prediction residuals of all methods under test on all datasets (described in Section \ref{subsubsec:EvalCrit}) and found that the assumption of Gaussianity holds in around 79\% cases.

\begin{figure}[!t]
	\centering
	\includegraphics[width=3in]{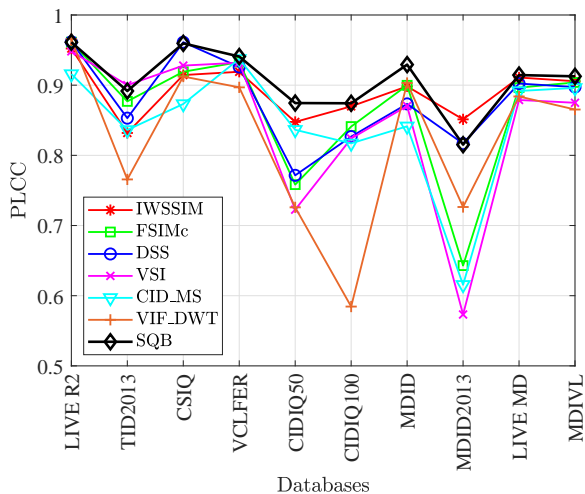}
	\vspace{-2mm}
	\caption{PLCC of SQB and selected FR methods for different IQA databases.}
	\label{fig:SGT_PLCCcomp}
	\vspace{-4mm}
\end{figure}

It can be clearly seen from Table \ref{table:SGT_PLCC_SRCC} that SQB is the top performer in terms of both WA PLCC and WA SRCC. From Table \ref{table:SGT_StaSig}, it can be observed that for the 170 method-database combinations, SQB performs statistically better than other methods in around 74\% cases, while its performance is statistically indistinguishable or inferior than other methods in around 19\% and 6\% cases, respectively. This is no small achievement given that all other methods included in the comparison, apart from PSNR, are considered state-of-the-art in FR and fused FR IQA. While RAS6 \cite{eval_Waterloo,fusion_bliss} was the top performer in our earlier comprehensive performance evaluation study \cite{eval_Waterloo}, it did not perform as well on the TID2013 database \cite{db_tid2013}, as can be seen from its PLCC and SRCC values in Table \ref{table:SGT_PLCC_SRCC}. With 3,000 distorted images and as many as 24 different distortion types, TID2013 can be considered as one of the largest and most diverse subject-rated IQA databases, making it quite challenging. It is clear from Table \ref{table:SGT_PLCC_SRCC} that SQB performs quite well on the TID2013 database, when compared to other methods. From Table \ref{table:SGT_StaSig}, it can be seen that on the TID2013 database SQB is outperformed only by the fused FR method MMF \cite{fusion_mmf_tip} and the FR method VSI \cite{fr_vsi}. MMF is a learning-based fusion method and we trained it on the TID2013 database. Thus, it is unfair to compare other methods with MMF on TID2013. While VSI outperforms SQB on TID2013, SQB performs better than VSI on almost all other datasets. We believe that the performance gain of SQB on TID2013, especially when compared to RAS6, is explained by the way we have selected the constant $k$ in the RRF \cite{fusion_rrf} computation (Equation \ref{eq:RRF}), as explained in Section \ref{subsec:SGT_Gen}. From Table \ref{table:SGT_StaSig}, we can see that SQB is outperformed by RAS6 on the LIVE R2 \cite{stateval_db_liveR2} and CSIQ \cite{fr_mad_db_csiq} databases. Table \ref{table:SGT_PLCC_SRCC} shows that for these datasets RAS6 outperforms SQB only slightly in terms of PLCC and SRCC. Given that the TID2013 dataset is much more diverse, in terms of distortions, when compared to LIVE R2 and CSIQ, we believe that this performance compromise is justified. From Table \ref{table:SGT_StaSig}, it can be seen that some individual FR methods statistically outperform SQB on at most a single dataset, but are outperformed by SQB on almost all other datasets. For example, while IWSSIM \cite{fr_iwssim} outperforms SQB on MDID2013 \cite{md_sisblim_db_mdid2013}, it is outperformed by SQB on six other datasets (Table \ref{table:SGT_StaSig}), sometimes quite significantly, such as on TID2013 \cite{db_tid2013} (Table \ref{table:SGT_PLCC_SRCC}). In fact, all state-of-the-art individual FR methods perform inconsistently across different IQA datasets, where they perform well on some datasets but not on others. This behavior can be seen in Fig. \ref{fig:SGT_PLCCcomp}, where the PLCC of SQB and some state-of-the-art FR methods is plotted for different datasets. These FR methods include IWSSIM \cite{fr_iwssim}, FSIMc \cite{fr_fsim}, DSS \cite{fr_dss}, and VSI \cite{fr_vsi}, which were found to be the top performers in our earlier comprehensive study \cite{eval_Waterloo}, where we evaluated 43 FR methods. We have also included the FR methods CID\_MS \cite{fr_cid} and VIF\_DWT \cite{fr_vifdwt_ssimdwt} in Fig. \ref{fig:SGT_PLCCcomp} as they, together with IWSSIM and DSS, are fused together in SQB. From Fig. \ref{fig:SGT_PLCCcomp}, it can be seen that the six individual FR methods encounter wide swings in performance across different datasets which differ from each other in terms of their constituent distortions and visual content. Thus, the performance of these FR methods cannot be regarded as stable, which goes against their use as alternatives to human annotations for large-scale datasets. However, Fig. \ref{fig:SGT_PLCCcomp} also shows that the performance variations are much less pronounced for SQB across all datasets. Hence, the performance of SQB can be regarded as relatively stable regardless of the distortions that afflict the images that it evaluates, thereby making it a much more suitable candidate to replace human annotations for labeling large-scale IQA datasets. We believe that SQB displays this superior performance relative to individual state-of-the-art FR methods because: 1) It uses a rank aggregation based fusion approach, RRF \cite{fusion_rrf}, that is unsupervised and training-free, which makes it robust to unseen data, and 2) The deficiencies of some FR methods being fused through RRF, for particular distortions, are supplemented by the strengths of other FR methods for those distortions, and thus the fused combination achieves stable performance for all distortions that usually afflict visual content. It is pertinent to again mention that the four FR methods being fused in SQB have not been randomly selected, but through an exhaustive search that included evaluating 737,280 FR fusion combinations in our earlier work \cite{eval_Waterloo}.

\begin{figure*}[!t]
	\centering
	\includegraphics[width=5in]{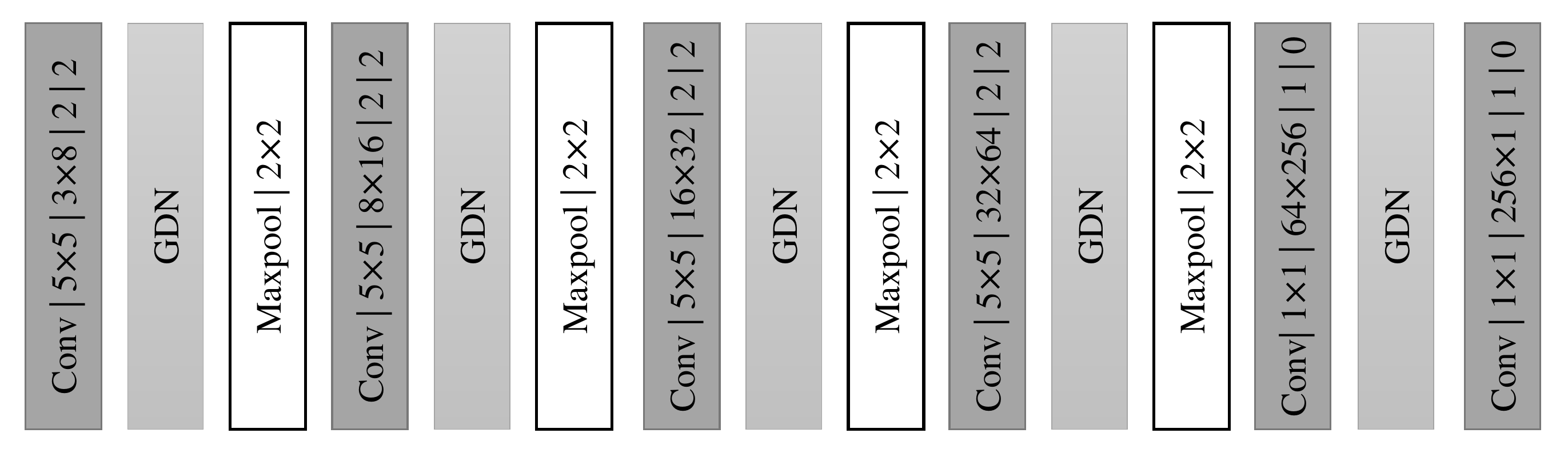}
	\vspace{-2mm}
	\caption{The architecture of the EONSS network for the BIQA task. We adopt the style and convention of \cite{E2E_ImComp} and denote the parameterization of the convolutional layer as: $Conv\ |\ kernel\ height \times kernel\ width\ |\ input\ channel \times output\ channel\ |\ stride\ |\ padding$.}
	\label{fig:EONSS_Network}
	\vspace{-4mm}
\end{figure*}

\section{The Impact of Data on DNN-based BIQA}
\label{sec:EONSS}

To investigate the impact of training data on DNN-based BIQA models and to validate our SQB approach to synthetically annotate large-scale IQA datasets, we build a model called End-to-end Optimized deep neural Network using Synthetic Scores (EONSS), which aims to provide a transparent common testing platform and thus special designs of DNN architectures and training strategies are purposely avoided. We believe that sophisticated network architectures and training strategies will certainly lead to improved performance in future work, but at this point they will make it difficult to single out the contribution of training data. While we train EONSS on the Waterloo Exploration-II database, we test it on subject-rated IQA datasets. This allows us to not only rigorously test EONSS on unseen human-rated data but also to make comparisons with other BIQA methods.

\subsection{Network Architecture and Implementation Details}
\label{subsec:EONSS_NetwArch}

The architecture of the EONSS network is illustrated in Fig. \ref{fig:EONSS_Network}. The network takes a $235 \times 235 \times 3$ RGB color image patch as input and predicts its quality in terms of a scalar value. With a few exceptions, most DNN based BIQA methods, that have been proposed so far, use a smaller input patch size of $32 \times 32$, as discussed in Section \ref{subsubsec:CurrStrat}. On the other hand, patches of larger size, such as $235 \times 235$, contain visually more meaningful content than smaller patches and can better represent the parent image, thereby reducing the label noise problem. Since earlier DNN based BIQA methods train on small-scale subject-rated datasets, they are constrained to use a smaller input patch size, as a larger patch size would dramatically reduce the overall number of patches available for training. However, our model does not suffer from this issue since the Waterloo Exploration-II database has a sufficiently large number of training images. As can be seen from Fig. \ref{fig:EONSS_Network}, the EONSS network consists of six stages of processing. The first four stages each contain a convolutional, a generalized divisive normalization (GDN) \cite{GDN}, and a max-pooling layer. The purpose of these four stages is to map the $235 \times 235 \times 3$ raw pixels from the image space to a lower-dimensional feature space where the impact of distortions on image quality can be more easily quantified in a perceptually aware manner. In the first four stages, the network reduces the spatial dimension through the use of convolution with stride $2 \times 2$ and employs $2 \times 2$ max-pooling after each GDN layer to select neurons with the highest local response. The last two stages, which consist of two fully connected layers and a GDN transform layer in between, map the extracted features to a single quality score. Before being sent to the last two fully connected layers, the spatial size of the features is reduced to $1 \times 1$, so that the number of weights in the fully connected layers is considerably reduced. Instead of using ReLU \cite{ReLU}, we use GDN \cite{GDN} as the activation function after the convolution layers in the first five stages of the network to add non-linearity to the model. While ReLU \cite{ReLU} is widely used as the activation function in CNNs, it suffers from strong higher-order dependencies, thus requiring a much larger network to achieve good performance for a given task \cite{E2E_ImComp}. We utilize a bio-inspired normalization transform, GDN \cite{GDN}, as the activation function because it helps decorrelate the high-dimensional features by using a joint nonlinear gain control mechanism, thereby enabling a much smaller network to achieve competitive performance. The GDN transform has been previously used effectively in image compression \cite{E2E_ImComp} and has also been used in a DNN based BIQA model \cite{nr_meon}. We define the loss function as the negative of PLCC. The advantages of choosing PLCC over mean absolute error (MAE) or mean squared error (MSE) are \cite{vmeon}: 1) The range of predictions is no longer restricted to the range of targets, since it is known that the range of subjective quality scores could be set arbitrarily and does not have any physical meaning. 2) It automatically normalizes the loss to the range [-1, 1], which gives more stability and flexibility to the training process. 3) PLCC is differentiable and is a frequently used evaluation criteria in the area of perceptual IQA. To empirically verify that our choice of PLCC as the loss function is valid, we also trained EONSS with MSE as the loss function. By using the nine subject-rated datasets (all distortions) mentioned in Section \ref{subsubsec:DBmethods} and the WA SRCC evaluation criteria mentioned in Section \ref{subsubsec:EvalCrit}, we found that the WA SRCC of EONSS relative to subjective data is 0.6183 and 0.6509 when using MSE and PLCC as loss functions, respectively. This clearly demonstrates the superiority of using PLCC as the loss function for EONSS.

To train the EONSS model, we randomly split the Waterloo Exploration-II database into training, validation and testing sets that consist of 60\%, 20\% and 20\% of the dataset, respectively. Since the network accepts images of size $235 \times 235 \times 3$, for the sake of speeding up the training phase, we randomly sample one $235 \times 235$ patch from each training image if its dimensions are larger. This does not prevent us from creating a sufficiently large pool of training data given the very large-scale nature of the Waterloo Exploration-II database. This also allows us to obtain a batch of image patches that have greater diversity, thereby helping to prevent model overfitting. Since the $235 \times 235$ patch size can cover a relatively large area of the original image, thereby containing perceptually meaningful content, we assign the SQB quality score of the original image as the image quality label of the sampled patch. During the validation and testing phases, we consider the entire image instead of just one patch from it. Thus, for images with larger dimensions, we sample $235 \times 235$ patches from the original image with a stride of $128 \times 128$ in an overlapping manner, and consider the average of the predicted quality scores of all patches as the predicted quality of the original image. This ensures a more rigorous validation process and also that all parts of an image are considered while testing. We initialize the weights of the convolution layers by following the approach in \cite{conv_weights_init} and use Adam \cite{Adam} for optimization. The training batch size is chosen to be $50$ and the image patches in each batch are randomly sampled from the training set only. We start with a learning rate of $0.001$ which is decreased by a factor of $10$ after every two epochs. Other parameters of Adam \cite{Adam} are set as default. The model performance, in terms of PLCC and SRCC, is tested on the validation set after each epoch and we stop training after $10$ epochs when the performance on the validation set reaches a plateau. Finally, the model after $10$ epochs of training, is applied to the testing set.

\subsection{Performance Evaluation and Comparison}
\label{subsec:EONSS_PerfEval}

To analyze the performance of EONSS and other BIQA methods, we use the same set of test datasets as mentioned in Section \ref{subsubsec:DBmethods}, which includes five singly and four multiply distorted subject-rated IQA databases. We also use the same evaluation criteria as described in Section \ref{subsubsec:EvalCrit}. However, we compute the evaluation metrics for two categories of data: 1) The \textit{all distortions category} includes all distorted images within each test dataset, that is, all distortion types are considered while computing PLCC, SRCC, and performing statistical significance testing. 2) The \textit{subset distortions category} includes a subset of distortion types in each dataset for which evaluation metrics are calculated. For singly distorted IQA datasets (LIVE R2 \cite{stateval_db_liveR2}, TID2013 \cite{db_tid2013}, CSIQ \cite{fr_mad_db_csiq}, VCLFER \cite{db_vclfer}, and CIDIQ \cite{db_cidiq}), images belonging to the following four common distortion types form the subset: 1) Gaussian (or Poisson) noise, 2) Gaussian Blur, 3) JPEG compression, and 4) JPEG2000 compression. For multiply distorted datasets, subsets of distorted images are formed by separately considering individual distortion combinations (if possible). Thus, we separately consider the Blur-JPEG and Blur-Noise combinations in LIVE MD \cite{db_livemd}, and the Blur-JPEG and Noise-JPEG combinations in MDIVL \cite{db_mdivl}. Since images in the MDID \cite{db_mdid} and MDID2013 \cite{md_sisblim_db_mdid2013} databases cannot be split into subsets, the entire datasets are considered for the subset case as well. The motivation for conducting performance evaluation on the \textit{subset distortions category} of IQA datasets, especially for singly distorted datasets, stems from the fact that most training-based opinion-aware (OA) BIQA methods are trained for the above-mentioned common distortion types that are present in almost all singly distorted datasets. Therefore, these subsets of distortions provide a fair ground for comparison. However, we also consider the \textit{all distortions category} of IQA datasets, to more rigorously test BIQA methods, as the ultimate goal of NR or \textit{blind} IQA methods is to be robust to \textit{unseen} data. Any gap in performance of BIQA methods for these two categories of test data would highlight directions for future research. We do not retrain BIQA methods on individual datasets but use the original versions, that is EONSS trained on the Waterloo Exploration-II database and author-trained versions of other BIQA methods, again to ensure rigorous testing.

\input{Tables/Table7.tex}
\input{Tables/Table8.tex}
\input{Tables/Table9.tex}
\input{Tables/Table10.tex}

For comparison we test 14 state-of-the-art BIQA methods to situate EONSS relative to the best in the field. Among them, eight methods belong to the OA BIQA category and include BIQI \cite{nr_biqi}, BRISQUE \cite{nr_brisque}, CORNIA \cite{nr_cornia}, GWHGLBP \cite{md_gwhglbp}, HOSA \cite{nr_hosa}, MEON \cite{nr_meon}, NRSL \cite{nr_nrsl}, and WaDIQaM-NR \cite{nr_fr_deepIQA}, while six methods belong to the opinion-unaware (OU) category and include dipIQ \cite{nr_dipiq}, ILNIQE \cite{nr_ilniqe}, LPSI \cite{nr_lpsi}, NIQE \cite{nr_niqe}, QAC \cite{nr_qac}, and SISBLIM \cite{md_sisblim_db_mdid2013}. It should be noted that among these methods, MEON \cite{nr_meon} and WaDIQaM-NR \cite{nr_fr_deepIQA} are DNN based BIQA methods. While a number of other deep learning based BIQA methods have recently been proposed, as discussed in Section \ref{subsubsec:CurrStrat}, we have tested the performance of MEON \cite{nr_meon} and WaDIQaM-NR \cite{nr_fr_deepIQA} as their author-trained models are publicly available. As an additional comparison point, in subsequent analysis we also include results for FR methods IWSSIM \cite{fr_iwssim} and PSNR in order to compare the performance of EONSS and other BIQA methods with a state-of-the-art (IWSSIM) and legacy (PSNR) FR method. For all datasets, the test results for the \textit{all distortions} and \textit{subset distortions categories} are given in Tables \ref{table:EONSS_PLCC_SRCC_All} and \ref{table:EONSS_PLCC_SRCC_SS}, respectively, in terms of PLCC and SRCC. In each of these tables, the WA PLCC/SRCC values are provided in the rightmost column and methods have been sorted in descending order with respect to them. The results of statistical significance testing of EONSS relative to the two FR and 14 BIQA methods for both the \textit{all distortions} and \textit{subset distortions} categories are provided in Table \ref{table:EONSS_StaSig_All_SS_Dist}, where a ``1'', ``--'', or ``0'' means that the perceptual quality prediction performance of EONSS is better, indistinguishable, or worse, respectively, than that of the method in the row for a given database (with 95\% confidence). Each entry in the table may be composed of more than one symbol, each of which represents the outcome of the test for either the \textit{all distortions} and \textit{subset distortions} categories, as explained in the table caption. We preceded the statistical significance testing with a kurtosis based check for Gaussianity of prediction residuals of all methods under test on all datasets (described in Section \ref{subsubsec:EvalCrit}) and found that the assumption of Gaussianity holds in around 89\% cases in the \textit{all distortions category} and in around 83\% cases in the \textit{subset distortions category}.

\input{Tables/Table11.tex}
\input{Tables/Table12.tex}

From the above-mentioned tables, the following observations can be made: 1) Table \ref{table:EONSS_PLCC_SRCC_All} reveals that EONSS outperforms all other state-of-the-art BIQA methods in the \textit{all distortions category}, both in terms of WA PLCC and WA SRCC. 2) Similarly, in the \textit{subset distortions category}, which can be considered a more \textit{fair} ground for comparison as stated earlier, Table \ref{table:EONSS_PLCC_SRCC_SS} shows that EONSS considerably outperforms all other BIQA methods both in terms of WA PLCC and WA SRCC. 3) Table \ref{table:EONSS_StaSig_All_SS_Dist} shows that for the 160 method-database combinations of the \textit{all distortions category}, EONSS performs statistically better than other methods in around 62\% cases, while its performance is statistically indistinguishable or inferior than other methods in around 19\% and 19\% cases, respectively. Similarly, for the 192 method-database combinations of the \textit{subset distortions category}, EONSS performs statistically better that other methods in around 67\% cases, while its performance is statistically indistinguishable or inferior than other methods in around 13\% and 20\% cases, respectively. This again demonstrates the superiority of EONSS when compared to the very state-of-the-art in the BIQA field. 4)  While considering Tables \ref{table:EONSS_PLCC_SRCC_All}, \ref{table:EONSS_PLCC_SRCC_SS}, and \ref{table:EONSS_StaSig_All_SS_Dist}, it should be noted that the OA BIQA methods BIQI \cite{nr_biqi}, BRISQUE \cite{nr_brisque}, NRSL \cite{nr_nrsl}, CORNIA \cite{nr_cornia}, HOSA \cite{nr_hosa}, WaDIQaM-NR \cite{nr_fr_deepIQA}, and MEON \cite{nr_meon} are trained on the LIVE R2 database \cite{stateval_db_liveR2}, and GWHGLBP \cite{md_gwhglbp} is trained on the LIVE MD database \cite{db_livemd}. Thus, comparing these OA BIQA methods with other approaches on these respective databases is unreliable and unfair to those other methods. Disregarding the results of these methods on the said datasets further increases the demonstrated superiority of EONSS. 5) It is pertinent to mention that the nine subject-rated IQA datasets have only been used to test EONSS, without any retraining or fine-tuning. 6) Even though EONSS has been trained on the Waterloo Exploration-II dataset, which predominantly consists of multiply distorted images, it performs well on even the singly distorted test datasets. This is explained by the wide density of distorted images in Waterloo Exploration-II, which includes 117,810 singly distorted images and a large number of multiply distorted images that have a small amount of stage-1 distortion, thereby allowing the DNN model to learn effectively for the single distortion scenario. 7) It can be clearly seen that EONSS comprehensively outperforms the two other DNN based models, MEON \cite{nr_meon} and WaDIQaM-NR \cite{nr_fr_deepIQA}, statistically and in terms of WA PLCC and WA SRCC, for both the \textit{all} and \textit{subset distortions categories}. Since, MEON \cite{nr_meon} and WaDIQaM-NR \cite{nr_fr_deepIQA}, are trained on a small-scale singly distorted dataset (LIVE R2 \cite{stateval_db_liveR2}), they do not perform well on multiply distorted datasets, which is not the case with EONSS. By using MEON \cite{nr_meon} as an example, we show in Section \ref{subsec:Impact_of_Data} that the performance of pre-existing DNN based BIQA models can indeed be elevated by retraining on the Waterloo Exploration-II database. 8) While the performance of EONSS is a considerable distance away from the state-of-the-art FR method IWSSIM \cite{fr_iwssim} in the \textit{all distortions category} (Table \ref{table:EONSS_PLCC_SRCC_All}), its performance is relatively closer to IWSSIM in the \textit{subset distortions category} (Table \ref{table:EONSS_PLCC_SRCC_SS}). Since the Waterloo Exploration-II database does not have the wide-ranging distortions of the \textit{all distortions category}, this shows that it is possible for a DNN based BIQA method to approach FR performance for distortion types for which sufficient annotated training data is available. This is no small achievement for a BIQA method, given that it has no access to the reference image.

We evaluated the computational complexity of all IQA methods under test in terms of their execution time to determine the quality of a $1024 \times 1024$ test color image on a desktop computer with a 3.5 GHz Intel Core i7-7800X processor, 16 GB of RAM, NVIDIA GeForce GTX 1050Ti GPU, and Ubuntu 18.04 operating system. The execution times of all methods are given in Table \ref{table:NR_MethodExecTime}, where methods have been sorted in ascending order with respect to execution time. Since the FR PSNR is the fastest method, we also provide the execution time relative to PSNR for ease in comparison. The time for DNN based methods, EONSS, MEON \cite{nr_meon}, and WaDIQaM-NR \cite{nr_fr_deepIQA}, was evaluated both on the GPU and CPU, while that of all other methods was evaluated on the CPU only. It should be noted that the execution time of some other well-known BIQA methods including BLIINDS2 \cite{nr_bliinds2}, DIIVINE \cite{nr_diivine}, FRIQUEE \cite{nr_friquee_jrnl}, MS-LQAF \cite{md_mslqaf}, NFERM \cite{nr_nferm}, and TCLT \cite{nr_tclt}, is even more than that of ILNIQE \cite{nr_ilniqe}, making them infeasible for large-scale or real-time use, which is why we have not included them in our analysis. It can be seen from Table \ref{table:NR_MethodExecTime} that the execution time of EONSS is approximately 20 to 30 times faster than competitive BIQA methods, such as CORNIA \cite{nr_cornia}, dipIQ \cite{nr_dipiq}, ILNIQE \cite{nr_ilniqe}, and SISBLIM \cite{md_sisblim_db_mdid2013}. Thus when Tables \ref{table:EONSS_PLCC_SRCC_All}, \ref{table:EONSS_PLCC_SRCC_SS}, and \ref{table:EONSS_StaSig_All_SS_Dist} are considered in conjunction with Table \ref{table:NR_MethodExecTime}, it becomes clear that EONSS not only outperforms the very best methods in the BIQA field in terms of perceptual quality prediction accuracy, but that it is also the fastest among them by a wide margin, making it an excellent choice for practical applications.

\subsection{The Impact of Data on DNN-based Models}
\label{subsec:Impact_of_Data}

\input{Tables/Table13.tex}
\input{Tables/Table14.tex}
\input{Tables/Table15.tex}
\input{Tables/Table16.tex}

The superior performance of EONSS, as demonstrated in Section \ref{subsec:EONSS_PerfEval}, can be directly attributed to the large-scale Waterloo Exploration-II database. We demonstrate this point more explicitly in this section by comparing the following four models: 1) EONSS trained on the Waterloo Exploration-II database, 2) We retrain the DNN architecture employed by EONSS (as described in Section \ref{subsec:EONSS_NetwArch}) on the small-scale subject-rated LIVE R2 database \cite{stateval_db_liveR2} and call this model EON\_L. 3) The original version of MEON \cite{nr_meon} that was trained on LIVE R2 \cite{stateval_db_liveR2}. 4) We retrain the MEON DNN on the Waterloo Exploration-II database and call it MEONSS. Since the publicly available version of MEON \cite{nr_meon} does not include the fast fading distortion type of LIVE R2 \cite{stateval_db_liveR2} in its training, we also exclude it from the training set of EON\_L so that it can be directly compared with MEON. Tables \ref{table:EONSS_LIVE_Train_All} and \ref{table:EONSS_LIVE_Train_SS} show the results for the \textit{all} and \textit{subset distortions categories}, respectively, both in terms of PLCC and SRCC, where we have sorted methods in the descending order with respect to their WA PLCC/SRCC values. It is clear from these tables that EONSS massively outperforms EON\_L in terms of WA PLCC and WA SRCC, both in the \textit{all} and \textit{subset distortions categories}. The only difference between EONSS and EON\_L is the training data used (both use exactly the same DNN). Thus, the enormous superiority of EONSS when compared to EON\_L can only be attributed to the large-scale training data that it utilizes, i.e., the Waterloo Exploration-II database, which allows the DNN to learn a robust quality model. From Tables \ref{table:EONSS_LIVE_Train_All} and \ref{table:EONSS_LIVE_Train_SS} it is also clear that MEONSS outperforms MEON in terms of WA PLCC and WA SRCC, both in the \textit{all} and \textit{subset distortions categories}. Again the only difference between MEONSS and MEON is the training data (both use exactly the same DNN). This again demonstrates the superiority of using the Waterloo Exploration-II database for BIQA model training.

From Tables \ref{table:EONSS_LIVE_Train_All} and \ref{table:EONSS_LIVE_Train_SS} we can also make the following three observations: 1) It is evident that the margin with which MEONSS outperforms MEON is smaller than the one with which EONSS outperforms EON\_L. 2) It is clear that although both EONSS and MEONSS are trained on the very large-scale Waterloo Exploration-II database, EONSS significantly outperforms MEONSS in terms of WA PLCC and WA SRCC, both in the \textit{all} and \textit{subset distortions categories}. 3) However, it can also be seen that although both EON\_L and MEON are trained on the small-scale LIVE R2 database \cite{stateval_db_liveR2}, MEON significantly outperforms EON\_L in terms of WA PLCC and WA SRCC, both in the \textit{all} and \textit{subset distortions categories}. This is a significant finding as it shows that the choice of DNN network architecture for the BIQA task is strongly impacted by the amount of available quality annotated training data. As we have discussed before, MEON takes a multi-task approach and utilizes two sub-networks, where sub-network 1 performs the task of distortion type identification for which a large amount of non-quality annotated training data is made available, and sub-network 2 performs quality prediction using the results from sub-network 1. On the other hand, EONSS takes a single task approach of quality prediction and hence its network is simpler compared to MEON. Our results show that the multi-task DNN model (MEON) performs better when only a small amount of quality annotated training data is available (MEON outperforms EON\_L), while the single-task DNN model (EONSS) performs much better when a very large amount of quality annotated training data is present (EONSS outperforms MEONSS). This shows that special DNN design and domain knowledge may be highly useful when there is a significant data shortage. However, even a simple single-task network architecture is able to learn a relatively effective quality model in a truly end-to-end manner given the availability of a large amount of quality annotated training data, thereby establishing the strength of large-scale datasets such as Waterloo Exploration-II.

Since the above analysis only utilizes a single subject-rated dataset, i.e., LIVE R2 \cite{stateval_db_liveR2} without one of its distortion types (fast fading), we make an additional attempt to understand the impact of data by separately training the DNN architecture of EONSS on three small-scale subject-rated datasets and use all distorted images within each dataset. These include the singly distorted LIVE R2 \cite{stateval_db_liveR2} and TID2013 \cite{db_tid2013} databases, and the multiply distorted MDID \cite{db_mdid} database. The trained models are respectively called EON\_LIVER2, EON\_TID2013, and EON\_MDID. We compare the performance of these models with that of EONSS trained on the Waterloo Exploration-II database. Tables \ref{table:EONSS_Subj_Train_All} and \ref{table:EONSS_Subj_Train_SS} show the results for the \textit{all} and \textit{subset distortions categories}, respectively, both in terms of PLCC and SRCC, where we have sorted methods in the descending order with respect to their WA PLCC/SRCC values. It can be clearly seen that EONSS substantially outperforms all other models in both the \textit{all} and \textit{subset distortions categories}. Since the weighted average calculations in Tables \ref{table:EONSS_Subj_Train_All} and \ref{table:EONSS_Subj_Train_SS} include results for the training datasets of EON\_LIVER2, EON\_TID2013, and EON\_MDID, the superiority of EONSS over these models is further increased if results for their respective training databases are disregarded. Since all four models being compared utilize the same DNN network architecture and training method, and differ only in training data, these results again provide strong evidence that the performance of DNN-based IQA models can be substantially improved with the abundance of quality training data. 

While it is difficult to determine how large the training dataset size should be to learn effective DNN based BIQA models, we try to answer this question empirically. Specifically, we consider four subsets of the Waterloo Exploration-II database which contain 1\%, 5\%, 10\%, and 20\% reference images of the original dataset along with their respective distorted versions. Next, we retrain EONSS on these dataset subsets and call the trained versions EONSS\_1, EONSS\_5, EONSS\_10, and EONSS\_20, respectively. While training, we further split each subset into training, validation, and testing sets which are composed of 60\%, 20\%, and 20\% of subset images, respectively. Tables \ref{table:EONSS_TrainSubSet_All} and \ref{table:EONSS_TrainSubSet_SS} show the results for the \textit{all} and \textit{subset distortions categories}, respectively, both in terms of PLCC and SRCC. We have repeated the results for EONSS in these tables, which utilizes 100\% of the Waterloo Exploration-II database for its training, validation, and testing. It can be observed from these tables that the model performance increases dramatically from EONSS\_1 to EONSS\_10 for both the \textit{all} and \textit{subset distortions categories}. Substantial performance increase is further seen from EONSS\_10 to EONSS\_20 for the \textit{subset distortions category}, which we believe is a more accurate category to consider for these experiments given that training and testing distortion types are more closely aligned. For both the \textit{all} and \textit{subset distortions categories} further performance gain can be seen from EONSS\_20 to EONSS (that uses the entire dataset for training, validation, and testing), but not by a wide margin, suggesting a performance saturation after 20\% of the Waterloo Exploration-II database. It should be noted that 20\% of the dataset still includes a substantial amount of annotated data (691,152 distorted images).

\section{Conclusion}
\label{sec:Conc}

The great potential of DNN-based BIQA has been largely constrained by the lack of large-scale annotated training data. Researchers have tried to address this issue by data augmentation, smart design of DNN architectures, sophisticated training strategies, or incorporating domain knowledge, but have achieved only limited success. This work suggests that the quality and quantity of annotated training data plays a dominating role in the success of DNN-based IQA model development.

To address the data challenge, we have developed the largest IQA dataset to-date, called the Waterloo Exploration-II database, which has 3,570 pristine reference and around 3.45 million, singly and multiply, distorted images. We have developed a novel mechanism to synthetically assign quality labels to the images of this dataset. Extensive tests on subject-rated datasets, reveal that these synthetic quality benchmark labels are highly accurate in perceptual quality prediction and perform better than the very best of FR IQA methods. To investigate the impact of data on DNN-based BIQA, we have developed the EONSS model trained on the Waterloo Exploration-II database and tested on nine subject-rated IQA datasets without any retraining or fine-tuning. We have comprehensively demonstrated that EONSS outperforms the very state-of-the-art BIQA methods, in both perceptual quality prediction and speed, by a wide margin. Compared to previous DNN-based BIQA models, EONSS has a relatively simple network architecture, therefore, its success is attributed to the synthetically labeled Waterloo Exploration-II database, whose enormity and content-diversity has provided sufficient data to the DNN to learn a robust BIQA model in a truly end-to-end manner. It is worth noting that there is still a significant gap between EONSS and top-performing FR IQA models such as IWSSIM \cite{fr_iwssim}, suggesting space for further improvement.

\section*{Note}
The work presented in this paper is part of an ongoing project and thus the Waterloo Exploration-II database, the SQB annotations, and the EONSS model are subject to change. Their final versions will be made publicly available once the project culminates.

\section*{Acknowledgment}
This work was supported in part by the Natural Sciences and Engineering Research Council (NSERC) of Canada.

\bibliographystyle{IEEEtran}
\bibliography{IEEEabrv,bib_allother,bib_database,bib_DNN,bib_friqa,bib_mdiqa,bib_metricfusion,bib_nriqa,bib_nss,bib_stateval}

\end{document}

%% file: Tables/Table1.tex

\begin{table*}[t!]
	\scriptsize
	\caption{Summary of contemporary subject-rated IQA databases.}
	\vspace{-3mm}
	\centering
	\begin{tabular}{ l  l  l  l  l  l  l  l  l}
		\hline
		 & \multirow{3}{*}{Database} & Published & Number of & Number of & Number of & Number of & Images per & Subjective  \\ 
		Database Category & & Year & Reference & Distorted & Distortion & Distortion & Distortion & Data Type \\
		 & & & Images & Images & Types & Levels & Type & \\ \hline
		\multirowcell{11}{Simulated Distortions\\ (Singly Distorted)} & A57 \cite{fr_vsnr} & 2007 & 3 & 54 & 6 & 3 & 9 & DMOS \\
		& CIDIQ \cite{db_cidiq} & 2014 & 23 & 690 & 6 & 5 & 115 & MOS \\
		& CSIQ \cite{fr_mad_db_csiq} & 2010 & 30 & 866 & 6 & 4 to 5 & 116 to 150 & DMOS \\
		& IVC \cite{db_ivc} & 2005 & 10 & 185 & 4 & 5 & 20 to 50 & MOS \\
		& KADID-10K \cite{db_kadid10k} & 2019 & 81 & 10,125 & 25 & 5 & 405 & DMOS \\
		& LIVE R2 \cite{stateval_db_liveR2} & 2006 & 29 & 779 & 5 & Up to 5 & 145 to 175 & DMOS \\
		& MICT-Toyama \cite{db_mict} & 2008 & 14 & 168 & 2 & 6 & 84 & MOS \\
		& PDAP-HDDS \cite{db_pdaphdds} & 2018 & 250 & 12,000 & 10 & 4 to 5 & 1,000 to 1,250 & MOS \\
		& TID2008 \cite{db_tid2008} & 2008 & 25 & 1,700 & 17 & 4 & 100 & MOS \\
		& TID2013 \cite{db_tid2013} & 2013 & 25 & 3,000 & 24 & 5 & 125 & MOS \\
		& VCLFER \cite{db_vclfer} & 2012 & 23 & 552 & 4 & 6 & 138 & MOS \\ \hline
		\multirowcell{4}{Simulated Distortions\\ (Multiply Distorted)} & LIVE MD \cite{db_livemd} & 2012 & 15 & 405 & 5 & 3 & 45 to 135 & DMOS \\
		& MDID2013 \cite{md_sisblim_db_mdid2013} & 2014 & 12 & 324 & 1 & 3 & 324 & DMOS \\
		& MDID \cite{db_mdid} & 2017 & 20 & 1,600 & 1 to 4 & 4 & N/A & MOS \\
		& MDIVL \cite{db_mdivl} & 2017 & 10 & 750 & 2 & 4 to 10 & 350 to 400 & MOS \\ \hline
		\multirowcell{4}{Authentic Distortions} & BID \cite{db_bid} & 2011 & N/A & 585 & 5 & N/A & 57 to 204 & MOS \\
		& CID2013 \cite{db_cid2013} & 2015 & N/A & 480 & 12 to 14 & N/A & N/A & MOS \\
		& KonIQ-10K \cite{db_koniq10k} & 2018 & N/A & 10,073 & N/A & N/A & N/A & MOS \\
		& LIVE Challenge \cite{db_livewc} & 2016 & N/A & 1162 & N/A & N/A & N/A & MOS \\ \hline
	\end{tabular}
	\label{table:IQAdbSummary}
	\vspace{-2mm}
\end{table*}

%% file: Tables/Table2.tex

\begin{table*}[!t]
	\scriptsize
	\caption{Distortion Levels and Target SSIMplus Scores.}
	\vspace{-2mm}
	\centering
	\begin{tabular}{ l  c  c  c  c  c  c  c  c  c  c  c  c  c  c  c  c  c  c  c  c  c}
		\hline
		Distortion Level & 1 & 2 & 3 & 4 & 5 & 6 & 7 & 8 & 9 & 10 & 11 & 12 & 13 & 14 & 15 & 16 & 17 & -- & -- & -- & -- \\ \hline
		Target SSIMplus Score & 100 & 95 & 90 & 85 & 80 & 75 & 70 & 65 & 60 & 55 & 50 & 45 & 40 & 35 & 30 & 25 & 20 & 15 & 10 & 5 & 0 \\ \hline
		Perceptual Quality Category & \multicolumn{4}{c|}{Excellent} & \multicolumn{4}{|c|}{Good} & \multicolumn{4}{|c|}{Fair} & \multicolumn{4}{|c|}{Poor} & \multicolumn{5}{|c}{Bad} \\ \hline
	\end{tabular}
	\label{table:SSIMplus_DistLevels}
	\vspace{-4mm}
\end{table*}

%% file: Tables/Table3.tex

\begin{table}[!t]
	\scriptsize
	\caption{Composition of the Waterloo Exploration-II database.}
	\vspace{-2mm}
	\centering
	\begin{tabular}{ c | c  c | c  c }
		\hline
		Reference Images & \multicolumn{2}{c |}{Stage-1 Distorted Images} & \multicolumn{2}{c}{Stage-2 Distorted Images} \\
		(Pristine Quality) & \multicolumn{2}{c |}{(Singly Distorted)} & \multicolumn{2}{c}{(Multiply Distorted)} \\ \hline
		Number of & \multirow{2}{*}{Distortion} & Number of & Distortion & Number of \\ 
		Images & & Images & Combination & Images \\ \hline
		\multirow{7}{*}{3,570} & \multirow{2}{*}{Blur} & \multirow{2}{*}{39,270} & Blur-JPEG & 667,590 \\
		 & & & Blur-Noise & 667,590 \\ \cline{2-5}
		 & JPEG & 39,270 & JPEG-JPEG & 667,590 \\ \cline{2-5}
		 & \multirow{2}{*}{Noise} & \multirow{2}{*}{39,270} & Noise-JPEG & 667,590 \\
		 & & & Noise-JP2K & 667,590 \\ \cline{2-5}
		 & Total & 117,810 & Total & 3,337,950 \\ \cline{2-5}
		 & \multicolumn{4}{c}{Overall 3,455,760 Distorted Images} \\ \hline
	\end{tabular}
	\label{table:WaterlooMD_Comp}
	\vspace{-4mm}
\end{table}

%% file: Tables/Table4.tex

\begin{table}[!t]
	\scriptsize
	\caption{Individual database and concatenated column vector sizes for SQB generation.}
	\vspace{-2mm}
	\centering
	\begin{tabular}{ c | l  c  c | c }
		\hline
		S. & \multicolumn{1}{c }{\multirow{2}{*}{Database}} & Subject & Database Column & Concatenated Column \\ 
		No. & & Rated & Vector Size & Vector Size \\ \hline
		1  & DR-IQA V1 & No & 32912$\times$1 & \multirow{12}{*}{3530595x1} \\
		2  & DR IQA V2 & No & 32912$\times$1 &  \\
		3  & Waterloo Exp.-II & No & 3455760$\times$1 &  \\ \cline{1-4}
		4  & LIVE R2 \cite{stateval_db_liveR2} & Yes & 779$\times$1 &  \\
		5  & TID2013 \cite{db_tid2013} & Yes & 3000$\times$1 &  \\
		6  & CSIQ \cite{fr_mad_db_csiq} & Yes & 866$\times$1 &  \\
		7  & VCLFER \cite{db_vclfer} & Yes & 552$\times$1 &  \\
		8  & CIDIQ \cite{db_cidiq} & Yes & 690$\times$1 &  \\
		9  & MDID \cite{db_mdid} & Yes & 1600$\times$1 &  \\
		10 & MDID2013 \cite{md_sisblim_db_mdid2013} & Yes & 324$\times$1 &  \\
		11 & LIVE MD \cite{db_livemd} & Yes & 450$\times$1 &  \\
		12 & MDIVL \cite{db_mdivl} & Yes & 750$\times$1 &  \\ \hline
	\end{tabular}
	\label{table:SGT_ColVec}
	\vspace{-2mm}
\end{table}

%% file: Tables/Table5.tex

\begin{table*}[!t]
	\scriptsize
	\caption{Test results of SQB, 2 fused FR, 14 state-of-the-art FR methods, and PSNR, on nine subject-rated IQA databases in terms of PLCC and SRCC. All distortions in each dataset were considered. The WA PLCC/SRCC are provided in the rightmost column and methods are sorted in descending order with respect to them. Fused FR methods are highlighted in bold.}
	\vspace{-2mm}
	\centering
	\begin{tabular}{ c | l | c  c  c  c  c  c  c  c  c  c  | c }
		\hline
		Evaluation & \multicolumn{1}{ c |}{\multirow{2}{*}{FR Method}} & LIVE & TID & \multirow{2}{*}{CSIQ} & \multirow{2}{*}{VCLFER} & CIDIQ & CIDIQ & \multirow{2}{*}{MDID} & MDID & LIVE & \multirow{2}{*}{MDIVL} & Weighted \\
		Criteria & & R2 & 2013 & & & 50 & 100 & & 2013 & MD & & Average \\ \hline \hline		
		\multirowcell{18}{PLCC} & \textbf{SQB} (Proposed) & 0.9612 & 0.8917 & 0.9596 & 0.9408 & 0.8745 & 0.8742 & 0.9293 & 0.8152 & 0.9144 & 0.9126 & 0.9100 \\
		& \textbf{RAS6} \cite{fusion_bliss,eval_Waterloo} & 0.9682 & 0.8488 & 0.9640 & 0.9408 & 0.8832 & 0.8585 & 0.9294 & 0.8181 & 0.9150 & 0.9202 & 0.8979 \\
		& IWSSIM \cite{fr_iwssim} & 0.9522 & 0.8319 & 0.9144 & 0.9191 & 0.8476 & 0.8698 & 0.8983 & 0.8513 & 0.9109 & 0.9056 & 0.8787 \\
		& FSIMc \cite{fr_fsim} & 0.9613 & 0.8769 & 0.9191 & 0.9329 & 0.7583 & 0.8410 & 0.8998 & 0.6429 & 0.8965 & 0.9039 & 0.8786 \\
		& DSS \cite{fr_dss} & 0.9618 & 0.8530 & 0.9612 & 0.9259 & 0.7715 & 0.8267 & 0.8733 & 0.8168 & 0.9023 & 0.8973 & 0.8757 \\
		& VSI \cite{fr_vsi} & 0.9482 & 0.9000 & 0.9279 & 0.9320 & 0.7226 & 0.8240 & 0.8703 & 0.5732 & 0.8789 & 0.8749 & 0.8714 \\
		& MCSD \cite{fr_mcsd} & 0.9675 & 0.8648 & 0.9560 & 0.9217 & 0.7532 & 0.7727 & 0.8637 & 0.8281 & 0.8847 & 0.8787 & 0.8705 \\
		& GMSD \cite{fr_gmsd} & 0.9603 & 0.8590 & 0.9541 & 0.9176 & 0.7387 & 0.7585 & 0.8776 & 0.8336 & 0.8808 & 0.8685 & 0.8672 \\
		& ESSIM \cite{fr_essim} & 0.9566 & 0.8645 & 0.9224 & 0.9094 & 0.7953 & 0.8256 & 0.8451 & 0.6648 & 0.8861 & 0.9081 & 0.8664 \\
		& SFF \cite{fr_sff} & 0.9632 & 0.8706 & 0.9643 & 0.7761 & 0.7834 & 0.7721 & 0.8590 & 0.7952 & 0.8893 & 0.8904 & 0.8658 \\
		& QASD \cite{fr_qasd} & 0.9574 & 0.8897 & 0.9481 & 0.9253 & 0.7257 & 0.8116 & 0.8063 & 0.6698 & 0.8966 & 0.8827 & 0.8638 \\
		& DVICOM \cite{fr_dvicom} & 0.9734 & 0.8194 & 0.9191 & 0.9144 & 0.8035 & 0.8018 & 0.8919 & 0.8161 & 0.8873 & 0.8773 & 0.8632 \\
		& \textbf{MMF} \cite{fusion_mmf_tip} & 0.8561 & 0.9504 & 0.9262 & 0.8624 & 0.7326 & 0.7572 & 0.8185 & 0.6788 & 0.8523 & 0.8075 & 0.8600 \\
		& CID\_MS \cite{fr_cid} & 0.9159 & 0.8362 & 0.8732 & 0.9375 & 0.8364 & 0.8171 & 0.8414 & 0.6155 & 0.8917 & 0.8961 & 0.8510 \\
		& MAD \cite{fr_mad_db_csiq} & 0.9675 & 0.8267 & 0.9502 & 0.9053 & 0.7809 & 0.8541 & 0.7552 & 0.7471 & 0.8944 & 0.8985 & 0.8464 \\
		& VIF \cite{fr_vif} & 0.9604 & 0.7720 & 0.9278 & 0.8938 & 0.7267 & 0.6415 & 0.9367 & 0.8376 & 0.9030 & 0.8736 & 0.8388 \\
		& VIF\_DWT \cite{fr_vifdwt_ssimdwt} & 0.9657 & 0.7657 & 0.9123 & 0.8969 & 0.7259 & 0.5845 & 0.9031 & 0.7264 & 0.8839 & 0.8653 & 0.8211 \\
		& PSNR & 0.8723 & 0.7017 & 0.8000 & 0.8321 & 0.6302 & 0.6808 & 0.6164 & 0.5647 & 0.7398 & 0.6806 & 0.7065 \\ \hline \hline
		\multirowcell{18}{SRCC} & \textbf{SQB} (Proposed) & 0.9665 & 0.8749 & 0.9542 & 0.9421 & 0.8760 & 0.8651 & 0.9252 & 0.8045 & 0.8857 & 0.8845 & 0.8997 \\
		& \textbf{RAS6} \cite{fusion_bliss,eval_Waterloo} & 0.9680 & 0.7930 & 0.9603 & 0.9405 & 0.8840 & 0.8532 & 0.9250 & 0.8214 & 0.8867 & 0.8954 & 0.8761 \\
		& VSI \cite{fr_vsi} & 0.9524 & 0.8965 & 0.9422 & 0.9317 & 0.7213 & 0.8106 & 0.8569 & 0.5700 & 0.8414 & 0.8269 & 0.8631 \\
		& FSIMc \cite{fr_fsim} & 0.9645 & 0.8510 & 0.9309 & 0.9323 & 0.7608 & 0.8285 & 0.8904 & 0.5806 & 0.8666 & 0.8613 & 0.8628 \\
		& IWSSIM \cite{fr_iwssim} & 0.9567 & 0.7779 & 0.9212 & 0.9163 & 0.8484 & 0.8564 & 0.8911 & 0.8551 & 0.8836 & 0.8588 & 0.8559 \\
		& SFF \cite{fr_sff} & 0.9649 & 0.8513 & 0.9627 & 0.7738 & 0.7834 & 0.7689 & 0.8396 & 0.8005 & 0.8700 & 0.8535 & 0.8527 \\
		& DSS \cite{fr_dss} & 0.9616 & 0.7921 & 0.9555 & 0.9272 & 0.7755 & 0.8246 & 0.8658 & 0.8078 & 0.8714 & 0.8759 & 0.8520 \\
		& QASD \cite{fr_qasd} & 0.9629 & 0.8674 & 0.9530 & 0.9231 & 0.7307 & 0.8079 & 0.7778 & 0.6687 & 0.8766 & 0.8315 & 0.8482 \\
		& \textbf{MMF} \cite{fusion_mmf_tip} & 0.8741 & 0.9409 & 0.9043 & 0.8594 & 0.7241 & 0.7379 & 0.8084 & 0.6799 & 0.8085 & 0.7703 & 0.8479 \\
		& MCSD \cite{fr_mcsd} & 0.9668 & 0.8089 & 0.9592 & 0.9224 & 0.7562 & 0.7808 & 0.8451 & 0.8269 & 0.8517 & 0.8370 & 0.8464 \\
		& CID\_MS \cite{fr_cid} & 0.9103 & 0.8314 & 0.8789 & 0.9366 & 0.8350 & 0.8062 & 0.8330 & 0.6168 & 0.8608 & 0.8778 & 0.8445 \\
		& GMSD \cite{fr_gmsd} & 0.9603 & 0.8044 & 0.9570 & 0.9177 & 0.7427 & 0.7675 & 0.8613 & 0.8283 & 0.8448 & 0.8210 & 0.8433 \\
		& ESSIM \cite{fr_essim} & 0.9597 & 0.8035 & 0.9325 & 0.9075 & 0.7968 & 0.8253 & 0.8250 & 0.6966 & 0.8517 & 0.8682 & 0.8418 \\
		& DVICOM \cite{fr_dvicom} & 0.9750 & 0.7598 & 0.9181 & 0.9155 & 0.8034 & 0.7903 & 0.8840 & 0.8168 & 0.8672 & 0.8374 & 0.8387 \\
		& MAD \cite{fr_mad_db_csiq} & 0.9669 & 0.7807 & 0.9466 & 0.9061 & 0.7815 & 0.8391 & 0.7249 & 0.7507 & 0.8646 & 0.8643 & 0.8220 \\
		& VIF \cite{fr_vif} & 0.9636 & 0.6769 & 0.9194 & 0.8866 & 0.7203 & 0.6257 & 0.9306 & 0.8444 & 0.8823 & 0.8381 & 0.8024 \\
		& VIF\_DWT \cite{fr_vifdwt_ssimdwt} & 0.9681 & 0.6439 & 0.9020 & 0.8930 & 0.7224 & 0.5826 & 0.8943 & 0.7553 & 0.8479 & 0.8243 & 0.7768 \\
		& PSNR & 0.8756 & 0.6394 & 0.8057 & 0.8246 & 0.6254 & 0.6701 & 0.5784 & 0.5604 & 0.6771 & 0.6136 & 0.6720 \\ \hline
	\end{tabular}
	\label{table:SGT_PLCC_SRCC}
	\vspace{-4mm}
\end{table*}

%% file: Tables/Table6.tex

\begin{table}[!t]
	\scriptsize
	\caption{Statistical significance testing of SQB with respect to fused (highlighted in bold) and individual FR methods on different databases. A ``1'', ``--'', or ``0'' means that SQB performance is statistically better, indistinguishable, or worse, respectively, than the method in the row (with 95\% confidence).}
	\vspace{-2mm}
	\centering
	\begin{tabular}{ l | c  c  c  c  c  c  c  c  c  c  }
		\hline
		\multicolumn{1}{ c |}{FR Method} & \rotatebox{90}{LIVE R2} & \rotatebox{90}{TID2013} & \rotatebox{90}{CSIQ} & \rotatebox{90}{VCLFER} & \rotatebox{90}{CIDIQ50} & \rotatebox{90}{CIDIQ100} & \rotatebox{90}{MDID} & \rotatebox{90}{MDID2013} & \rotatebox{90}{LIVE MD} & \rotatebox{90}{MDIVL} \\ \hline
		CID\_MS \cite{fr_cid} & 1 & 1 & 1 & -- & 1 & 1 & 1 & 1 & 1 & 1 \\
		DSS \cite{fr_dss} & -- & 1 & -- & 1 & 1 & 1 & 1 & -- & -- & 1 \\
		DVICOM \cite{fr_dvicom} & 0 & 1 & 1 & 1 & 1 & 1 & 1 & -- & 1 & 1 \\
		ESSIM \cite{fr_essim} & -- & 1 & 1 & 1 & 1 & 1 & 1 & 1 & 1 & -- \\
		FSIMc \cite{fr_fsim} & -- & 1 & 1 & -- & 1 & 1 & 1 & 1 & 1 & -- \\
		GMSD \cite{fr_gmsd} & -- & 1 & 1 & 1 & 1 & 1 & 1 & -- & 1 & 1 \\
		IWSSIM \cite{fr_iwssim} & 1 & 1 & 1 & 1 & 1 & -- & 1 & 0 & -- & -- \\
		MAD \cite{fr_mad_db_csiq} & 0 & 1 & 1 & 1 & 1 & 1 & 1 & 1 & 1 & 1 \\
		MCSD \cite{fr_mcsd} & 0 & 1 & -- & 1 & 1 & 1 & 1 & -- & 1 & 1 \\
		\textbf{MMF} \cite{fusion_mmf_tip} & 1 & 0 & 1 & 1 & 1 & 1 & 1 & 1 & 1 & 1 \\
		PSNR & 1 & 1 & 1 & 1 & 1 & 1 & 1 & 1 & 1 & 1 \\
		QASD \cite{fr_qasd} & -- & -- & 1 & 1 & 1 & 1 & 1 & 1 & 1 & 1 \\
		\textbf{RAS6} \cite{fusion_bliss,eval_Waterloo} & 0 & 1 & 0 & -- & -- & -- & -- & -- & -- & -- \\
		SFF \cite{fr_sff} & -- & 1 & 0 & 1 & 1 & 1 & 1 & -- & 1 & 1 \\
		VIF \cite{fr_vif} & -- & 1 & 1 & 1 & 1 & 1 & 0 & -- & -- & 1 \\
		VIF\_DWT \cite{fr_vifdwt_ssimdwt} & 0 & 1 & 1 & 1 & 1 & 1 & 1 & 1 & 1 & 1 \\
		VSI \cite{fr_vsi} & 1 & 0 & 1 & -- & 1 & 1 & 1 & 1 & 1 & 1 \\ \hline
	\end{tabular}
	\label{table:SGT_StaSig}
	\vspace{-4mm}
\end{table}

%% file: Tables/Table7.tex

\begin{table*}[t!]
	\scriptsize
	\caption{Test results of EONSS in comparison with 2 FR and 14 NR methods on nine subject-rated IQA databases in terms of PLCC and SRCC. All distortions in each test dataset were considered. The WA PLCC/SRCC are provided in the rightmost column and methods are sorted in descending order with respect to them. FR methods are highlighted in bold.}
	\vspace{-2mm}
	\centering
	\begin{tabular}{ c | l | c  c  c  c  c  c  c  c  c  c  | c }
		\hline
		Evaluation & \multicolumn{1}{ c |}{\multirow{2}{*}{NR Method}} & LIVE & TID & \multirow{2}{*}{CSIQ} & \multirow{2}{*}{VCLFER} & CIDIQ & CIDIQ & \multirow{2}{*}{MDID} & MDID & LIVE & \multirow{2}{*}{MDIVL} & Weighted \\
		Criteria & & R2 & 2013 & & & 50 & 100 & & 2013 & MD & & Average \\ \hline \hline
		\multirowcell{17}{PLCC} & \textbf{IWSSIM} \cite{fr_iwssim} & 0.9522 & 0.8319 & 0.9144 & 0.9191 & 0.8476 & 0.8698 & 0.8983 & 0.8513 & 0.9109 & 0.9056 & 0.8787 \\
		& \textbf{PSNR} & 0.8723 & 0.7017 & 0.8000 & 0.8321 & 0.6302 & 0.6808 & 0.6164 & 0.5647 & 0.7398 & 0.6806 & 0.7065 \\
		& EONSS & 0.9244 & 0.5442 & 0.7660 & 0.9120 & 0.5798 & 0.4821 & 0.8374 & 0.3020 & 0.8437 & 0.8744 & 0.6933 \\
		& CORNIA \cite{nr_cornia} & 0.9665 & 0.5729 & 0.7593 & 0.8366 & 0.4496 & 0.3530 & 0.8074 & 0.6935 & 0.8679 & 0.8277 & 0.6878 \\
		& ILNIQE \cite{nr_ilniqe} & 0.9022 & 0.5883 & 0.8538 & 0.7289 & 0.3124 & 0.3390 & 0.7245 & 0.5146 & 0.8923 & 0.6303 & 0.6452 \\
		& HOSA \cite{nr_hosa} & 0.9991 & 0.5521 & 0.7560 & 0.8496 & 0.4969 & 0.3761 & 0.6590 & 0.2513 & 0.6768 & 0.7167 & 0.6328 \\
		& dipIQ \cite{nr_dipiq} & 0.9348 & 0.4774 & 0.7720 & 0.8942 & 0.5223 & 0.3889 & 0.6789 & 0.4376 & 0.7669 & 0.7627 & 0.6284 \\
		& NRSL \cite{nr_nrsl} & 0.9815 & 0.5338 & 0.7456 & 0.8905 & 0.4672 & 0.4034 & 0.6566 & 0.3088 & 0.5183 & 0.6794 & 0.6182 \\
		& SISBLIM \cite{md_sisblim_db_mdid2013} & 0.8077 & 0.4805 & 0.7378 & 0.7574 & 0.4909 & 0.4671 & 0.6321 & 0.8135 & 0.8948 & 0.5723 & 0.6077 \\
		& GWHGLBP \cite{md_gwhglbp} & 0.8079 & 0.4982 & 0.7104 & 0.6427 & 0.3653 & 0.2978 & 0.7108 & 0.7443 & 0.9655 & 0.5966 & 0.5991 \\
		& BIQI \cite{nr_biqi} & 0.9224 & 0.4678 & 0.6916 & 0.6106 & 0.3596 & 0.2661 & 0.6763 & 0.3369 & 0.7389 & 0.6215 & 0.5648 \\
		& NIQE \cite{nr_niqe} & 0.9052 & 0.4001 & 0.7188 & 0.8040 & 0.3703 & 0.2708 & 0.6728 & 0.5634 & 0.8387 & 0.5688 & 0.5646 \\
		& MEON \cite{nr_meon} & 0.9389 & 0.4919 & 0.7865 & 0.9221 & 0.4774 & 0.3854 & 0.5250 & 0.2430 & 0.2684 & 0.5722 & 0.5630 \\
		& WaDIQaM-NR \cite{nr_fr_deepIQA} & 0.9341 & 0.5712 & 0.6882 & 0.7862 & 0.4133 & 0.3481 & 0.4631 & 0.1371 & 0.2685 & 0.5214 & 0.5457 \\
		& BRISQUE \cite{nr_brisque} & 0.9671 & 0.4747 & 0.7006 & 0.8208 & 0.4155 & 0.3257 & 0.4450 & 0.1403 & 0.6045 & 0.6517 & 0.5429 \\
		& QAC \cite{nr_qac} & 0.8625 & 0.4371 & 0.7067 & 0.7615 & 0.3573 & 0.2856 & 0.6043 & 0.4240 & 0.4145 & 0.5713 & 0.5338 \\
		& LPSI \cite{nr_lpsi} & 0.8280 & 0.4892 & 0.7216 & 0.6020 & 0.4037 & 0.3981 & 0.4335 & 0.1765 & 0.5464 & 0.5715 & 0.5204 \\ \hline \hline
		\multirowcell{17}{SRCC} & \textbf{IWSSIM} \cite{fr_iwssim} & 0.9567 & 0.7779 & 0.9212 & 0.9163 & 0.8484 & 0.8564 & 0.8911 & 0.8551 & 0.8836 & 0.8588 & 0.8559 \\
		& \textbf{PSNR} & 0.8756 & 0.6394 & 0.8057 & 0.8246 & 0.6254 & 0.6701 & 0.5784 & 0.5604 & 0.6771 & 0.6136 & 0.6720 \\
		& EONSS & 0.9267 & 0.5045 & 0.6774 & 0.9063 & 0.4991 & 0.3448 & 0.8297 & 0.2874 & 0.7260 & 0.8833 & 0.6509 \\
		& CORNIA \cite{nr_cornia} & 0.9681 & 0.4288 & 0.6534 & 0.8354 & 0.3727 & 0.2071 & 0.7918 & 0.7055 & 0.8340 & 0.8336 & 0.6147 \\
		& ILNIQE \cite{nr_ilniqe} & 0.8975 & 0.4939 & 0.8144 & 0.7391 & 0.2997 & 0.3127 & 0.6900 & 0.5148 & 0.8778 & 0.6238 & 0.6031 \\
		& HOSA \cite{nr_hosa} & 0.9990 & 0.4705 & 0.5925 & 0.8574 & 0.4494 & 0.3248 & 0.6412 & 0.2993 & 0.6393 & 0.7399 & 0.5851 \\
		& dipIQ \cite{nr_dipiq} & 0.9378 & 0.4377 & 0.5266 & 0.8957 & 0.4135 & 0.2100 & 0.6612 & 0.4153 & 0.6678 & 0.7131 & 0.5620 \\
		& NRSL \cite{nr_nrsl} & 0.9796 & 0.4277 & 0.6750 & 0.8930 & 0.4249 & 0.2894 & 0.6458 & 0.4088 & 0.4145 & 0.6047 & 0.5589 \\
		& SISBLIM \cite{md_sisblim_db_mdid2013} & 0.7741 & 0.3177 & 0.6603 & 0.7622 & 0.4435 & 0.4098 & 0.6554 & 0.8089 & 0.8770 & 0.5375 & 0.5408 \\
		& GWHGLBP \cite{md_gwhglbp} & 0.7410 & 0.3844 & 0.5773 & 0.6243 & 0.3337 & 0.2412 & 0.7032 & 0.7555 & 0.9698 & 0.5841 & 0.5377 \\
		& NIQE \cite{nr_niqe} & 0.9073 & 0.3132 & 0.6271 & 0.8126 & 0.3458 & 0.2212 & 0.6523 & 0.5451 & 0.7738 & 0.5713 & 0.5181 \\
		& BIQI \cite{nr_biqi} & 0.9198 & 0.3935 & 0.6186 & 0.6170 & 0.3433 & 0.2353 & 0.6276 & 0.0077 & 0.5556 & 0.5711 & 0.5007 \\
		& MEON \cite{nr_meon} & 0.9409 & 0.3750 & 0.7248 & 0.9215 & 0.4101 & 0.2497 & 0.4861 & 0.2980 & 0.1917 & 0.5466 & 0.4969 \\
		& BRISQUE \cite{nr_brisque} & 0.9654 & 0.3672 & 0.5563 & 0.8130 & 0.3640 & 0.2496 & 0.4035 & 0.2209 & 0.5018 & 0.6647 & 0.4792 \\
		& WaDIQaM-NR \cite{nr_fr_deepIQA} & 0.9417 & 0.4393 & 0.6388 & 0.7524 & 0.3588 & 0.2235 & 0.4040 & 0.1316 & 0.2379 & 0.5614 & 0.4782 \\
		& QAC \cite{nr_qac} & 0.8683 & 0.3722 & 0.4900 & 0.7686 & 0.3196 & 0.1944 & 0.3239 & 0.2272 & 0.3579 & 0.5524 & 0.4292 \\
		& LPSI \cite{nr_lpsi} & 0.8181 & 0.3949 & 0.5303 & 0.5865 & 0.2060 & 0.1411 & 0.0306 & 0.0168 & 0.2717 & 0.5736 & 0.3558 \\ \hline	
	\end{tabular}
	\label{table:EONSS_PLCC_SRCC_All}
	\vspace{-4mm}
\end{table*}

%% file: Tables/Table8.tex

\begin{table*}[t!]
	\scriptsize
	\caption{Test results of EONSS in comparison with 2 FR and 14 NR methods on nine subject-rated IQA databases in terms of PLCC and SRCC. A subset of distortions in each test dataset were considered. The WA PLCC/SRCC are provided in the rightmost column and methods are sorted in descending order with respect to them. FR methods are highlighted in bold.}
	\vspace{-2mm}
	\centering
	\begin{tabular}{ c | l | c  c  c  c  c  c  c  c  c  c  c  c | c }
		\hline
		Evaluation & \multicolumn{1}{ c |}{\multirow{2}{*}{NR Method}} & LIVE & TID & \multirow{2}{*}{CSIQ} & VCL & CIDIQ & CIDIQ & \multirow{2}{*}{MDID} & MDID & \multicolumn{2}{|c}{LIVE MD} & \multicolumn{2}{|c|}{MDIVL} & Weighted \\
		Criteria & & R2 & 2013 & & FER & 50 & 100 & & 2013 & \multicolumn{1}{|c}{BJPG} & \multicolumn{1}{c}{BN} & \multicolumn{1}{|c}{BJPG} & NJPG & Average \\ \hline
		\multirowcell{17}{PLCC} & \textbf{IWSSIM} \cite{fr_iwssim} & 0.9556 & 0.9407 & 0.9655 & 0.9191 & 0.8745 & 0.8536 & 0.8983 & 0.8513 & 0.9164 & 0.9117 & 0.9269 & 0.9101 & 0.9116 \\
		& EONSS & 0.9462 & 0.8751 & 0.9291 & 0.9120 & 0.7973 & 0.8082 & 0.8374 & 0.3020 & 0.8622 & 0.8337 & 0.9232 & 0.8918 & 0.8430 \\
		& CORNIA \cite{nr_cornia} & 0.9715 & 0.8868 & 0.9257 & 0.8366 & 0.5898 & 0.5480 & 0.8074 & 0.6935 & 0.8774 & 0.8723 & 0.9419 & 0.7900 & 0.8145 \\
		& dipIQ \cite{nr_dipiq} & 0.9559 & 0.8879 & 0.9481 & 0.8942 & 0.7475 & 0.6706 & 0.6789 & 0.4376 & 0.8235 & 0.7895 & 0.8311 & 0.7882 & 0.7839 \\
		& HOSA \cite{nr_hosa} & 0.9992 & 0.8901 & 0.9384 & 0.8496 & 0.6774 & 0.6597 & 0.6590 & 0.2513 & 0.8968 & 0.6728 & 0.9005 & 0.7022 & 0.7600 \\
		& ILNIQE \cite{nr_ilniqe} & 0.9164 & 0.8576 & 0.9070 & 0.7289 & 0.3860 & 0.4598 & 0.7245 & 0.5146 & 0.9048 & 0.8968 & 0.8293 & 0.5759 & 0.7263 \\
		& \textbf{PSNR} & 0.8699 & 0.8912 & 0.9079 & 0.8321 & 0.6532 & 0.5560 & 0.6164 & 0.5647 & 0.7409 & 0.7751 & 0.7143 & 0.6645 & 0.7241 \\
		& NRSL \cite{nr_nrsl} & 0.9887 & 0.9153 & 0.9133 & 0.8905 & 0.6236 & 0.6145 & 0.6566 & 0.3088 & 0.3516 & 0.6263 & 0.6418 & 0.7334 & 0.7239 \\
		& SISBLIM \cite{md_sisblim_db_mdid2013} & 0.8220 & 0.7896 & 0.7967 & 0.7574 & 0.5899 & 0.6844 & 0.6321 & 0.8135 & 0.9030 & 0.8913 & 0.8056 & 0.4871 & 0.7194 \\
		& NIQE \cite{nr_niqe} & 0.9162 & 0.8091 & 0.8767 & 0.8040 & 0.4994 & 0.4712 & 0.6728 & 0.5634 & 0.9099 & 0.8481 & 0.7996 & 0.4507 & 0.7135 \\
		& GWHGLBP \cite{md_gwhglbp} & 0.8088 & 0.7675 & 0.8052 & 0.6427 & 0.5196 & 0.5347 & 0.7108 & 0.7443 & 0.9677 & 0.9684 & 0.7745 & 0.4943 & 0.7113 \\
		& BIQI \cite{nr_biqi} & 0.9534 & 0.7772 & 0.8224 & 0.6106 & 0.4957 & 0.5164 & 0.6763 & 0.3369 & 0.7743 & 0.7404 & 0.7398 & 0.6035 & 0.6827 \\
		& MEON \cite{nr_meon} & 0.9907 & 0.9053 & 0.9423 & 0.9221 & 0.6620 & 0.6510 & 0.5250 & 0.2430 & 0.2675 & 0.4927 & 0.3875 & 0.7405 & 0.6763 \\
		& QAC \cite{nr_qac} & 0.8777 & 0.8051 & 0.8736 & 0.7615 & 0.4512 & 0.5068 & 0.6043 & 0.4240 & 0.5378 & 0.6722 & 0.6765 & 0.6090 & 0.6637 \\
		& BRISQUE \cite{nr_brisque} & 0.9760 & 0.8659 & 0.9239 & 0.8208 & 0.5257 & 0.5421 & 0.4450 & 0.1403 & 0.8663 & 0.4594 & 0.8249 & 0.6511 & 0.6564 \\
		& \tiny{WaDIQaM-NR} \cite{nr_fr_deepIQA} & 0.9302 & 0.8994 & 0.8860 & 0.7862 & 0.5137 & 0.5530 & 0.4631 & 0.1371 & 0.6842 & 0.3921 & 0.6415 & 0.5231 & 0.6251 \\
		& LPSI \cite{nr_lpsi} & 0.8440 & 0.8114 & 0.8657 & 0.6020 & 0.5508 & 0.6289 & 0.4335 & 0.1765 & 0.8820 & 0.1182 & 0.7959 & 0.5075 & 0.5991 \\ \hline \hline
		\multirowcell{17}{SRCC} & \textbf{IWSSIM} \cite{fr_iwssim} & 0.9616 & 0.9262 & 0.9603 & 0.9163 & 0.8755 & 0.8374 & 0.8911 & 0.8551 & 0.8700 & 0.8933 & 0.8778 & 0.8713 & 0.9002 \\
		& EONSS & 0.9499 & 0.8446 & 0.8969 & 0.9063 & 0.7885 & 0.7553 & 0.8297 & 0.2874 & 0.7348 & 0.7331 & 0.8754 & 0.9085 & 0.8205 \\
		& CORNIA \cite{nr_cornia} & 0.9732 & 0.8727 & 0.8987 & 0.8354 & 0.5740 & 0.5053 & 0.7918 & 0.7055 & 0.8278 & 0.8523 & 0.9254 & 0.8027 & 0.8007 \\
		& dipIQ \cite{nr_dipiq} & 0.9574 & 0.8720 & 0.9290 & 0.8957 & 0.7460 & 0.6433 & 0.6612 & 0.4153 & 0.6979 & 0.7391 & 0.6512 & 0.7730 & 0.7562 \\
		& HOSA \cite{nr_hosa} & 0.9991 & 0.8681 & 0.9111 & 0.8574 & 0.6677 & 0.6236 & 0.6412 & 0.2993 & 0.8437 & 0.5357 & 0.8789 & 0.7150 & 0.7438 \\
		& ILNIQE \cite{nr_ilniqe} & 0.9153 & 0.8417 & 0.8802 & 0.7391 & 0.3669 & 0.4248 & 0.6900 & 0.5148 & 0.8915 & 0.8821 & 0.7915 & 0.5797 & 0.7078 \\
		& \textbf{PSNR} & 0.8731 & 0.9073 & 0.9218 & 0.8246 & 0.6553 & 0.5763 & 0.5784 & 0.5604 & 0.6621 & 0.7088 & 0.6572 & 0.5841 & 0.7048 \\
		& SISBLIM \cite{md_sisblim_db_mdid2013} & 0.7835 & 0.7703 & 0.8059 & 0.7622 & 0.5565 & 0.6314 & 0.6554 & 0.8089 & 0.8746 & 0.8782 & 0.7584 & 0.3320 & 0.7008 \\
		& NRSL \cite{nr_nrsl} & 0.9880 & 0.8965 & 0.8874 & 0.8930 & 0.5732 & 0.5564 & 0.6458 & 0.4088 & 0.2634 & 0.5991 & 0.4684 & 0.7125 & 0.6996 \\
		& NIQE \cite{nr_niqe} & 0.9168 & 0.7972 & 0.8710 & 0.8126 & 0.4703 & 0.4180 & 0.6523 & 0.5451 & 0.8713 & 0.7938 & 0.7625 & 0.4510 & 0.6954 \\
		& GWHGLBP \cite{md_gwhglbp} & 0.7447 & 0.6538 & 0.6728 & 0.6243 & 0.4768 & 0.4454 & 0.7032 & 0.7555 & 0.9640 & 0.9751 & 0.7584 & 0.4502 & 0.6672 \\
		& MEON \cite{nr_meon} & 0.9906 & 0.9012 & 0.9300 & 0.9215 & 0.6421 & 0.5830 & 0.4861 & 0.2980 & 0.0476 & 0.3257 & 0.3255 & 0.7397 & 0.6441 \\
		& BIQI \cite{nr_biqi} & 0.9528 & 0.7763 & 0.7972 & 0.6170 & 0.4976 & 0.4849 & 0.6276 & 0.0077 & 0.6542 & 0.4902 & 0.6591 & 0.5302 & 0.6272 \\
		& BRISQUE \cite{nr_brisque} & 0.9757 & 0.8401 & 0.8992 & 0.8130 & 0.4727 & 0.4771 & 0.4035 & 0.2209 & 0.7923 & 0.2991 & 0.7385 & 0.6612 & 0.6239 \\
		& \tiny{WaDIQaM-NR} \cite{nr_fr_deepIQA} & 0.9399 & 0.8646 & 0.8636 & 0.7524 & 0.4777 & 0.4691 & 0.4040 & 0.1316 & 0.5012 & 0.2502 & 0.6121 & 0.4830 & 0.5786 \\
		& QAC \cite{nr_qac} & 0.8857 & 0.8055 & 0.8415 & 0.7686 & 0.4450 & 0.4566 & 0.3239 & 0.2272 & 0.3959 & 0.4707 & 0.5537 & 0.5282 & 0.5529 \\
		& LPSI \cite{nr_lpsi} & 0.8333 & 0.7046 & 0.7711 & 0.5865 & 0.3382 & 0.3949 & 0.0306 & 0.0168 & 0.8387 & 0.0012 & 0.7348 & 0.4692 & 0.4254 \\ \hline
	\end{tabular}
	\label{table:EONSS_PLCC_SRCC_SS}
\end{table*}

%% file: Tables/Table9.tex

\begin{table}[!t]
	\scriptsize
	\caption{Statistical significance testing of EONSS with respect to FR (highlighted in bold) and NR methods on different databases, for \textit{ALL} and subset (\textit{SS}) distortions. The order of symbols within each entry is: LIVE R2 (All, SS), TID2013 (All, SS), CSIQ (All, SS), VCLFER (All), CIDIQ50 (All, SS), CIDIQ100 (All, SS), MDID (All), MDID2013 (All), LIVE MD (All, Blur-JPEG, Blur-Noise), MDIVL (All, Blur-JPEG, Noise-JPEG). A ``1'', ``--'', or ``0'' means that EONSS performance is statistically better, indistinguishable, or worse, respectively, than the method in the row (with 95\% confidence).}
	\vspace{-2mm}
	\centering
	\begin{tabular}{ l | c  c  c  c  c  c  c  c  c  c  }
		\hline
		\multicolumn{1}{ c |}{FR/NR Method} & \rotatebox{90}{LIVE R2} & \rotatebox{90}{TID2013} & \rotatebox{90}{CSIQ} & \rotatebox{90}{VCLFER} & \rotatebox{90}{CIDIQ50} & \rotatebox{90}{CIDIQ100} & \rotatebox{90}{MDID} & \rotatebox{90}{MDID2013} & \rotatebox{90}{LIVE MD} & \rotatebox{90}{MDIVL} \\ \hline
		BIQI \cite{nr_biqi} & --0 & 11 & 11 & 1 & 11 & 11 & 1 & -- & 111 & 111 \\
		BRISQUE \cite{nr_brisque} & 00 & 1-- & 1-- & 1 & 11 & 11 & 1 & -- & 1--1 & 111 \\
		CORNIA \cite{nr_cornia} & 00 & -- -- & -- -- & 1 & 11 & 11 & 1 & 0 & 0--0 & 101 \\
		dipIQ \cite{nr_dipiq} & 00 & 1-- & --0 & 1 & --1 & --1 & 1 & -- & 11-- & 111 \\
		GWHGLBP \cite{md_gwhglbp} & 11 & 11 & 11 & 1 & 11 & 11 & 1 & 0 & 000 & 111 \\
		HOSA \cite{nr_hosa} & 00 & -- -- & --0 & 1 & 11 & --1 & 1 & -- & 101 & 111 \\
		ILNIQE \cite{nr_ilniqe} & 11 & 0-- & 01 & 1 & 11 & 11 & 1 & 0 & 000 & 111 \\
		\textbf{IWSSIM} \cite{fr_iwssim} & 00 & 00 & 00 & -- & 00 & 00 & 0 & 0 & 000 & 0--0 \\
		LPSI \cite{nr_lpsi} & 11 & 11 & 11 & 1 & 11 & --1 & 1 & -- & 1--1 & 111 \\
		MEON \cite{nr_meon} & 00 & 10 & --0 & -- & 11 & --1 & 1 & 1 & 111 & 111 \\
		NIQE \cite{nr_niqe} & 11 & 11 & 11 & 1 & 11 & 11 & 1 & 0 & --0-- & 111 \\
		NRSL \cite{nr_nrsl} & 00 & --0 & --1 & 1 & 11 & --1 & 1 & -- & 111 & 111 \\
		\textbf{PSNR} & 11 & 0-- & 01 & 1 & --1 & 01 & 1 & 0 & 111 & 111 \\
		QAC \cite{nr_qac} & 11 & 11 & 11 & 1 & 11 & 11 & 1 & -- & 111 & 111 \\
		SISBLIM \cite{md_sisblim_db_mdid2013} & 11 & 11 & --1 & 1 & 11 & --1 & 1 & 0 & 000 & 111 \\
		{\tiny WaDIQaM-NR} \cite{nr_fr_deepIQA} & 01 & --0 & 11 & 1 & 11 & 11 & 1 & -- & 111 & 111 \\ \hline
	\end{tabular}
	\label{table:EONSS_StaSig_All_SS_Dist}
	\vspace{-4mm}
\end{table}

%% file: Tables/Table10.tex

\begin{table}[t!]
	\scriptsize
	\caption{Execution Time of NR methods on a test image. Methods are sorted in ascending order with respect to the execution time.}
	\vspace{-2mm}
	\centering
	\begin{tabular}{ l | c | c | c }
		\hline
		\multirow{2}{*}{FR/NR Method} & Processing & Execution Time & Execution Time \\
		& Unit & (Seconds) & Relative to PSNR \\ \hline
		PSNR$^{\mathrm{1}}$ & CPU & 0.0013 & 1.00    \\
		LPSI \cite{nr_lpsi} & CPU & 0.0397 & 30.54   \\
		EONSS & GPU & 0.0604 & 46.46   \\
		EONSS & CPU & 0.0817 & 62.85   \\
		MEON \cite{nr_meon} & CPU & 0.0819 & 63.00   \\
		MEON \cite{nr_meon} & GPU & 0.0876 & 67.38   \\
		HOSA \cite{nr_hosa} & CPU & 0.1309 & 100.69  \\
		QAC  \cite{nr_qac}  & CPU & 0.1357 & 104.38  \\
		NRSL$^{\mathrm{2}}$ \cite{nr_nrsl} & CPU & 0.1421 & 109.31  \\
		GWHGLBP$^{\mathrm{2}}$ \cite{md_gwhglbp} & CPU & 0.1469 & 113.00  \\
		WaDIQaM-NR \cite{nr_fr_deepIQA} & GPU & 0.1549 & 119.15  \\
		BRISQUE \cite{nr_brisque} & CPU & 0.1823 & 140.23  \\
		NIQE \cite{nr_niqe} & CPU & 0.2941 & 226.23  \\
		BIQI \cite{nr_biqi} & CPU & 0.4634 & 356.46  \\
		dipIQ \cite{nr_dipiq} & CPU & 1.6592 & 1276.31 \\
		CORNIA \cite{nr_cornia} & CPU & 2.0304 & 1561.85 \\
		SISBLIM \cite{md_sisblim_db_mdid2013} & CPU & 2.2005 & 1692.69 \\
		ILNIQE \cite{nr_ilniqe} & CPU & 2.5227 & 1940.54 \\
		WaDIQaM-NR \cite{nr_fr_deepIQA} & CPU & 6.2818 & 4832.15 \\ \hline
		\multicolumn{4}{l}{$^{\mathrm{1}}$FR method included for comparison. \quad $^{\mathrm{2}}$Feature extraction time only.} \\
	\end{tabular}
	\label{table:NR_MethodExecTime}
	\vspace{-5mm}
\end{table}

%% file: Tables/Table11.tex

\begin{table*}[t!]
	\scriptsize
	\caption{PLCC and SRCC values for EONSS, EON\_L, MEONSS, and MEON when tested on nine subject-rated IQA databases. All distortions in each test dataset were considered. The WA PLCC/SRCC are provided in the rightmost column and methods are sorted in descending order with respect to them.}
	\vspace{-2mm}
	\centering
	\begin{tabular}{ c | l | c  c  c  c  c  c  c  c  c  c | c }
		\hline
		Evaluation & \multirow{2}{*}{Method} & LIVE & TID & \multirow{2}{*}{CSIQ} & \multirow{2}{*}{VCLFER} & CIDIQ & CIDIQ & \multirow{2}{*}{MDID} & MDID & LIVE & \multirow{2}{*}{MDIVL} & Weighted \\
		Criteria & & R2 & 2013 & & & 50 & 100 & & 2013 & MD & & Average \\ \hline \hline
		\multirowcell{4}{PLCC} & EONSS & 0.9244 & 0.5442 & 0.7660 & 0.9120 & 0.5798 & 0.4821 & 0.8374 & 0.3020 & 0.8437 & 0.8744 & 0.6933 \\
		& MEONSS & 0.8975 & 0.4270 & 0.7359 & 0.9079 & 0.4459 & 0.3218 & 0.6916 & 0.2855 & 0.7885 & 0.9012 & 0.6059 \\
		& MEON & 0.9389 & 0.4919 & 0.7865 & 0.9221 & 0.4774 & 0.3854 & 0.5250 & 0.2430 & 0.2684 & 0.5722 & 0.5630 \\
		& EON\_L & 0.8586 & 0.3772 & 0.6214 & 0.7557 & 0.2758 & 0.1942 & 0.6293 & 0.1865 & 0.4963 & 0.4238 & 0.4833 \\ \hline \hline
		\multirowcell{4}{SRCC} & EONSS & 0.9267 & 0.5045 & 0.6774 & 0.9063 & 0.4991 & 0.3448 & 0.8297 & 0.2874 & 0.7260 & 0.8833 & 0.6509 \\
		& MEONSS & 0.9060 & 0.3796 & 0.6547 & 0.9087 & 0.3768 & 0.1748 & 0.6985 & 0.2709 & 0.6211 & 0.8918 & 0.5615 \\
		& MEON & 0.9409 & 0.3750 & 0.7248 & 0.9215 & 0.4101 & 0.2497 & 0.4861 & 0.2980 & 0.1917 & 0.5466 & 0.4969 \\
		& EON\_L & 0.8636 & 0.2464 & 0.5284 & 0.7456 & 0.2245 & 0.1395 & 0.5453 & 0.1343 & 0.3670 & 0.3643 & 0.4006 \\ \hline
	\end{tabular}
	\label{table:EONSS_LIVE_Train_All}
\end{table*}

%% file: Tables/Table12.tex

\begin{table*}[t!]
	\scriptsize
	\caption{PLCC and SRCC values for EONSS, EON\_L, MEONSS, and MEON when tested on nine subject-rated IQA databases. A subset of distortions in each test dataset were considered. The WA PLCC/SRCC are provided in the rightmost column and methods are sorted in descending order with respect to them.}
	\vspace{-2mm}
	\centering
	\begin{tabular}{ c | l | c  c  c  c  c  c  c  c  c  c  c  c | c }
		\hline
		Evaluation & \multirow{2}{*}{Method} & LIVE & TID & \multirow{2}{*}{CSIQ} & \multirow{2}{*}{VCLFER} & CIDIQ & CIDIQ & \multirow{2}{*}{MDID} & MDID & \multicolumn{2}{|c}{LIVE MD} & \multicolumn{2}{|c|}{MDIVL} & Weighted \\
		Criteria & & R2 & 2013 & & & 50 & 100 & & 2013 & \multicolumn{1}{|c}{BJPG} & \multicolumn{1}{c}{BN} & \multicolumn{1}{|c}{BJPG} & NJPG & Average \\ \hline \hline
		\multirowcell{4}{PLCC} & EONSS & 0.9462 & 0.8751 & 0.9291 & 0.9120 & 0.7973 & 0.8082 & 0.8374 & 0.3020 & 0.8622 & 0.8337 & 0.9232 & 0.8918 & 0.8430 \\
		& MEONSS & 0.9213 & 0.8255 & 0.9125 & 0.9079 & 0.6307 & 0.5836 & 0.6916 & 0.2855 & 0.8019 & 0.7911 & 0.9104 & 0.9333 & 0.7668 \\
		& MEON & 0.9907 & 0.9053 & 0.9423 & 0.9221 & 0.6620 & 0.6510 & 0.5250 & 0.2430 & 0.2675 & 0.4927 & 0.3875 & 0.7405 & 0.6763 \\
		& EON\_L & 0.8769 & 0.8006 & 0.7562 & 0.7557 & 0.3551 & 0.3854 & 0.6293 & 0.1865 & 0.5556 & 0.5342 & 0.5663 & 0.3602 & 0.6039 \\ \hline \hline
		\multirowcell{4}{SRCC} & EONSS & 0.9499 & 0.8446 & 0.8969 & 0.9063 & 0.7885 & 0.7553 & 0.8297 & 0.2874 & 0.7348 & 0.7331 & 0.8754 & 0.9085 & 0.8205 \\
		& MEONSS & 0.9280 & 0.8151 & 0.9095 & 0.9087 & 0.6199 & 0.5428 & 0.6985 & 0.2709 & 0.5756 & 0.6722 & 0.8437 & 0.9365 & 0.7479 \\
		& MEON & 0.9906 & 0.9012 & 0.9300 & 0.9215 & 0.6421 & 0.5830 & 0.4861 & 0.2980 & 0.0476 & 0.3257 & 0.3255 & 0.7397 & 0.6441 \\
		& EON\_L & 0.8882 & 0.7822 & 0.7706 & 0.7456 & 0.3274 & 0.3495 & 0.5453 & 0.1343 & 0.4309 & 0.3074 & 0.3928 & 0.2620 & 0.5472 \\ \hline
	\end{tabular}
	\label{table:EONSS_LIVE_Train_SS}
\end{table*}

%% file: Tables/Table13.tex

\begin{table*}[t!]
	\scriptsize
	\caption{PLCC and SRCC values for EONSS, EON\_LIVER2, EON\_TID2013, and EON\_MDID when tested on nine subject-rated IQA databases. All distortions in each test dataset were considered. The WA PLCC/SRCC are provided in the rightmost column and methods are sorted in descending order with respect to them.}
	\vspace{-2mm}
	\centering
	\begin{tabular}{ c | l | c  c  c  c  c  c  c  c  c  c | c }
		\hline
		Evaluation & \multirow{2}{*}{Method} & LIVE & TID & \multirow{2}{*}{CSIQ} & \multirow{2}{*}{VCLFER} & CIDIQ & CIDIQ & \multirow{2}{*}{MDID} & MDID & LIVE & \multirow{2}{*}{MDIVL} & Weighted \\
		Criteria & & R2 & 2013 & & & 50 & 100 & & 2013 & MD & & Average \\ \hline \hline
		\multirowcell{4}{PLCC} & EONSS & 0.9244 & 0.5442 & 0.7660 & 0.9120 & 0.5798 & 0.4821 & 0.8374 & 0.3020 & 0.8437 & 0.8744 & 0.6933 \\
		& EON\_MDID & 0.7895 & 0.5575 & 0.7624 & 0.6849 & 0.3349 & 0.2835 & 0.9807 & 0.6063 & 0.7912 & 0.5446 & 0.6476 \\
		& EON\_TID2013 & 0.7663 & 0.8898 & 0.6775 & 0.4435 & 0.2827 & 0.2863 & 0.4626 & 0.1854 & 0.5075 & 0.5127 & 0.6086 \\
		& EON\_LIVER2 & 0.8957 & 0.4849 & 0.6997 & 0.7888 & 0.2888 & 0.2335 & 0.6048 & 0.1126 & 0.5142 & 0.4424 & 0.5279 \\ \hline \hline
		\multirowcell{4}{SRCC} & EONSS & 0.9267 & 0.5045 & 0.6774 & 0.9063 & 0.4991 & 0.3448 & 0.8297 & 0.2874 & 0.7260 & 0.8833 & 0.6509 \\
		& EON\_TID2013 & 0.7749 & 0.8795 & 0.5798 & 0.4109 & 0.2319 & 0.2268 & 0.4369 & 0.0502 & 0.4558 & 0.5065 & 0.5760 \\
		& EON\_MDID & 0.7839 & 0.3777 & 0.6968 & 0.6909 & 0.2768 & 0.2615 & 0.9800 & 0.6091 & 0.7746 & 0.4880 & 0.5752 \\
		& EON\_LIVER2 & 0.9048 & 0.3644 & 0.6753 & 0.7822 & 0.2350 & 0.1583 & 0.5344 & 0.0718 & 0.3843 & 0.3423 & 0.4529 \\ \hline
	\end{tabular}
	\label{table:EONSS_Subj_Train_All}
	\vspace{-3mm}
\end{table*}

%% file: Tables/Table14.tex

\begin{table*}[t!]
	\scriptsize
	\caption{PLCC and SRCC values for EONSS, EON\_LIVER2, EON\_TID2013, and EON\_MDID when tested on nine subject-rated IQA databases. A subset of distortions in each test dataset were considered. The WA PLCC/SRCC are provided in the rightmost column and methods are sorted in descending order with respect to them.}
	\vspace{-2mm}
	\centering
	\begin{tabular}{ c | l | c  c  c  c  c  c  c  c  c  c  c  c | c }
		\hline
		Evaluation & \multirow{2}{*}{Method} & LIVE & TID & \multirow{2}{*}{CSIQ} & \multirow{2}{*}{VCLFER} & CIDIQ & CIDIQ & \multirow{2}{*}{MDID} & MDID & \multicolumn{2}{|c}{LIVE MD} & \multicolumn{2}{|c|}{MDIVL} & Weighted \\
		Criteria & & R2 & 2013 & & & 50 & 100 & & 2013 & \multicolumn{1}{|c}{BJPG} & \multicolumn{1}{c}{BN} & \multicolumn{1}{|c}{BJPG} & NJPG & Average \\ \hline \hline
		\multirowcell{4}{PLCC} & EONSS & 0.9462 & 0.8751 & 0.9291 & 0.9120 & 0.7973 & 0.8082 & 0.8374 & 0.3020 & 0.8622 & 0.8337 & 0.9232 & 0.8918 & 0.8430 \\
		& EON\_MDID & 0.7885 & 0.7549 & 0.8071 & 0.6849 & 0.3652 & 0.3943 & 0.9807 & 0.6063 & 0.8131 & 0.7768 & 0.7837 & 0.3604 & 0.7316 \\
		& EON\_LIVER2 & 0.8992 & 0.8533 & 0.7922 & 0.7888 & 0.3711 & 0.4248 & 0.6048 & 0.1126 & 0.7056 & 0.4824 & 0.5625 & 0.4237 & 0.6179 \\ 
		& EON\_TID2013 & 0.7586 & 0.9330 & 0.6463 & 0.4435 & 0.2812 & 0.3339 & 0.4626 & 0.1854 & 0.7935 & 0.3501 & 0.6720 & 0.4567 & 0.5274 \\ \hline \hline
		\multirowcell{4}{SRCC} & EONSS & 0.9499 & 0.8446 & 0.8969 & 0.9063 & 0.7885 & 0.7553 & 0.8297 & 0.2874 & 0.7348 & 0.7331 & 0.8754 & 0.9085 & 0.8205 \\
		& EON\_MDID & 0.7831 & 0.6502 & 0.7373 & 0.6909 & 0.3009 & 0.3500 & 0.9800 & 0.6091 & 0.7853 & 0.7658 & 0.7573 & 0.2566 & 0.6993 \\
		& EON\_LIVER2 & 0.9115 & 0.8334 & 0.8230 & 0.7822 & 0.3302 & 0.3639 & 0.5344 & 0.0718 & 0.5865 & 0.2092 & 0.4245 & 0.2024 & 0.5571 \\
		& EON\_TID2013 & 0.7678 & 0.9189 & 0.6241 & 0.4109 & 0.1987 & 0.2516 & 0.4369 & 0.0502 & 0.7673 & 0.2633 & 0.6178 & 0.4381 & 0.4887 \\ \hline
	\end{tabular}
	\label{table:EONSS_Subj_Train_SS}
	\vspace{-3mm}
\end{table*}

%% file: Tables/Table15.tex

\begin{table*}[t!]
	\scriptsize
	\caption{PLCC and SRCC values for various versions of EONSS trained on different subsets of the Waterloo Exploration-II database and tested on nine subject-rated IQA databases. All distortions in each test dataset were considered. The WA PLCC/SRCC are provided in the rightmost column and methods are sorted in descending order with respect to them.}
	\vspace{-2mm}
	\centering
	\begin{tabular}{ c | l | c  c  c  c  c  c  c  c  c  c | c }
		\hline
		Evaluation & \multicolumn{1}{ c |}{EONSS} & LIVE & TID & \multirow{2}{*}{CSIQ} & \multirow{2}{*}{VCLFER} & CIDIQ & CIDIQ & \multirow{2}{*}{MDID} & MDID & LIVE & \multirow{2}{*}{MDIVL} & Weighted \\
		Criteria & \multicolumn{1}{ c |}{Version} & R2 & 2013 & & & 50 & 100 & & 2013 & MD & & Average \\ \hline \hline
		\multirowcell{5}{PLCC} & EONSS & 0.9244 & 0.5442 & 0.7660 & 0.9120 & 0.5798 & 0.4821 & 0.8374 & 0.3020 & 0.8437 & 0.8744 & 0.6933 \\
		& EONSS\_20 & 0.9231 & 0.5637 & 0.7678 & 0.9050 & 0.5800 & 0.4629 & 0.8014 & 0.2346 & 0.8330 & 0.8749 & 0.6890 \\
		& EONSS\_10 & 0.9146 & 0.5788 & 0.8112 & 0.8809 & 0.4723 & 0.4081 & 0.7701 & 0.5231 & 0.8089 & 0.8538 & 0.6856 \\
		& EONSS\_5 & 0.8972 & 0.5807 & 0.7775 & 0.8839 & 0.4359 & 0.3693 & 0.7634 & 0.2867 & 0.7922 & 0.8699 & 0.6681 \\
		& EONSS\_1 & 0.7637 & 0.5026 & 0.7438 & 0.6292 & 0.2608 & 0.2608 & 0.5662 & 0.0570 & 0.6038 & 0.6988 & 0.5334 \\ \hline \hline
		\multirowcell{5}{SRCC} & EONSS & 0.9267 & 0.5045 & 0.6774 & 0.9063 & 0.4991 & 0.3448 & 0.8297 & 0.2874 & 0.7260 & 0.8833 & 0.6509 \\
		& EONSS\_20 & 0.9214 & 0.5165 & 0.6651 & 0.9053 & 0.5097 & 0.3530 & 0.7919 & 0.2163 & 0.7278 & 0.8736 & 0.6451 \\
		& EONSS\_10 & 0.9107 & 0.5232 & 0.7260 & 0.8879 & 0.4395 & 0.3314 & 0.7615 & 0.4050 & 0.6923 & 0.8495 & 0.6420 \\
		& EONSS\_5 & 0.8895 & 0.5407 & 0.6794 & 0.8899 & 0.4094 & 0.3057 & 0.7544 & 0.2622 & 0.6907 & 0.8724 & 0.6335 \\
		& EONSS\_1 & 0.7373 & 0.4377 & 0.6600 & 0.6432 & 0.2426 & 0.2164 & 0.5163 & 0.0268 & 0.5289 & 0.6899 & 0.4866 \\ \hline
	\end{tabular}
	\label{table:EONSS_TrainSubSet_All}
	\vspace{-3mm}
\end{table*}

%% file: Tables/Table16.tex

\begin{table*}[t!]
	\scriptsize
	\caption{PLCC and SRCC values for various versions of EONSS trained on different subsets of the Waterloo Exploration-II database and tested on nine subject-rated IQA databases. A subset of distortions in each test dataset were considered. The WA PLCC/SRCC are provided in the rightmost column and methods are sorted in descending order with respect to them.}
	\vspace{-2mm}
	\centering
	\begin{tabular}{ c | l | c  c  c  c  c  c  c  c  c  c  c  c | c }
		\hline
		Evaluation & \multicolumn{1}{ c |}{EONSS} & LIVE & TID & \multirow{2}{*}{CSIQ} & \multirow{2}{*}{VCLFER} & CIDIQ & CIDIQ & \multirow{2}{*}{MDID} & MDID & \multicolumn{2}{|c}{LIVE MD} & \multicolumn{2}{|c|}{MDIVL} & Weighted \\
		Criteria & \multicolumn{1}{ c |}{Version} & R2 & 2013 & & & 50 & 100 & & 2013 & \multicolumn{1}{|c}{BJPG} & \multicolumn{1}{c}{BN} & \multicolumn{1}{|c}{BJPG} & NJPG & Average \\ \hline \hline
		\multirowcell{5}{PLCC} & EONSS & 0.9462 & 0.8751 & 0.9291 & 0.9120 & 0.7973 & 0.8082 & 0.8374 & 0.3020 & 0.8622 & 0.8337 & 0.9232 & 0.8918 & 0.8430 \\
		& EONSS\_20 & 0.9403 & 0.8705 & 0.9244 & 0.9050 & 0.7899 & 0.7925 & 0.8014 & 0.2346 & 0.8753 & 0.7971 & 0.9116 & 0.8882 & 0.8250 \\
		& EONSS\_10 & 0.9385 & 0.8687 & 0.9264 & 0.8809 & 0.6121 & 0.6421 & 0.7701 & 0.5231 & 0.8301 & 0.7963 & 0.8685 & 0.8719 & 0.8007 \\
		& EONSS\_5 & 0.9117 & 0.8515 & 0.9146 & 0.8839 & 0.5641 & 0.5904 & 0.7634 & 0.2867 & 0.8297 & 0.7717 & 0.9100 & 0.8719 & 0.7762 \\
		& EONSS\_1 & 0.8038 & 0.7663 & 0.8037 & 0.6292 & 0.2450 & 0.3184 & 0.5662 & 0.0570 & 0.6429 & 0.5981 & 0.7439 & 0.6957 & 0.5883 \\ \hline \hline
		\multirowcell{5}{SRCC} & EONSS & 0.9499 & 0.8446 & 0.8969 & 0.9063 & 0.7885 & 0.7553 & 0.8297 & 0.2874 & 0.7348 & 0.7331 & 0.8754 & 0.9085 & 0.8205 \\
		& EONSS\_20 & 0.9402 & 0.8411 & 0.8875 & 0.9053 & 0.7749 & 0.7331 & 0.7919 & 0.2163 & 0.7839 & 0.6893 & 0.8655 & 0.8907 & 0.8010 \\
		& EONSS\_10 & 0.9385 & 0.8258 & 0.8871 & 0.8879 & 0.6112 & 0.6115 & 0.7615 & 0.4050 & 0.7020 & 0.6949 & 0.8257 & 0.8800 & 0.7737 \\
		& EONSS\_5 & 0.9070 & 0.8049 & 0.8635 & 0.8899 & 0.5664 & 0.5569 & 0.7544 & 0.2622 & 0.7236 & 0.6657 & 0.8711 & 0.8803 & 0.7528 \\
		& EONSS\_1 & 0.7803 & 0.6809 & 0.7306 & 0.6432 & 0.2677 & 0.2811 & 0.5163 & 0.0268 & 0.5545 & 0.5157 & 0.6809 & 0.7137 & 0.5498 \\ \hline
	\end{tabular}
	\label{table:EONSS_TrainSubSet_SS}
	\vspace{-5mm}
\end{table*}

%% file: Database_Paper.bbl
\begin{thebibliography}{100}
\providecommand{\url}[1]{#1}
\csname url@samestyle\endcsname
\providecommand{\newblock}{\relax}
\providecommand{\bibinfo}[2]{#2}
\providecommand{\BIBentrySTDinterwordspacing}{\spaceskip=0pt\relax}
\providecommand{\BIBentryALTinterwordstretchfactor}{4}
\providecommand{\BIBentryALTinterwordspacing}{\spaceskip=\fontdimen2\font plus
\BIBentryALTinterwordstretchfactor\fontdimen3\font minus
  \fontdimen4\font\relax}
\providecommand{\BIBforeignlanguage}[2]{{%
\expandafter\ifx\csname l@#1\endcsname\relax
\typeout{** WARNING: IEEEtran.bst: No hyphenation pattern has been}%
\typeout{** loaded for the language `#1'. Using the pattern for}%
\typeout{** the default language instead.}%
\else
\language=\csname l@#1\endcsname
\fi
#2}}
\providecommand{\BIBdecl}{\relax}
\BIBdecl

\bibitem{eonss_iciar}
Z.~Wang, S.~Athar, and Z.~Wang, ``{Blind Quality Assessment of Multiply
  Distorted Images Using Deep Neural Networks},'' in \emph{Proc. Int. Conf.
  Image Anal. Recognit. ({ICIAR})}, {Waterloo, ON, Canada}, Aug. 2019, pp.
  89--101.

\bibitem{cisco}
{Cisco and/or its affiliates}, ``{Cisco Visual Networking Index: Forecast and
  Trends, 2017–2022},'' \emph{White Paper, Cisco Public Information}, 2019.

\bibitem{iqa_book}
Z.~Wang and A.~C. Bovik, ``{Modern Image Quality Assessment},'' \emph{Synthesis
  Lectures on Image, Video, and Multimedia Processing}, vol.~2, no.~1, pp.
  1--156, 2006, {Morgan \& Claypool Publishers}.

\bibitem{rrnrSPM}
{Z. Wang and A. C. Bovik}, ``{Reduced- and No-Reference Image Quality
  Assessment},'' \emph{{IEEE} Signal Process. Mag.}, vol.~28, no.~6, pp.
  29--40, Nov. 2011.

\bibitem{fr_iwssim}
Z.~Wang and Q.~Li, ``{Information Content Weighting for Perceptual Image
  Quality Assessment},'' \emph{{IEEE} Trans. Image Process.}, vol.~20, no.~5,
  pp. 1185--1198, May 2011.

\bibitem{fr_fsim}
L.~Zhang, L.~Zhang, X.~Mou, and D.~Zhang, ``{FSIM: A Feature Similarity Index
  for Image Quality Assessment},'' \emph{{IEEE} Trans. Image Process.},
  vol.~20, no.~8, pp. 2378--2386, Aug. 2011.

\bibitem{fr_vsi}
L.~Zhang, Y.~Shen, and H.~Li, ``{VSI: A Visual Saliency-Induced Index for
  Perceptual Image Quality Assessment},'' \emph{{IEEE} Trans. Image Process.},
  vol.~23, no.~10, pp. 4270--4281, Oct. 2014.

\bibitem{fr_dss}
A.~Balanov, A.~Schwartz, Y.~Moshe, and N.~Peleg, ``{Image quality assessment
  based on DCT subband similarity},'' in \emph{Proc. {IEEE} Int. Conf. Image
  Process. ({ICIP})}, {Quebec City, QC, Canada}, Sept. 2015, pp. 2105--2109.

\bibitem{eval_fr_icip12}
L.~Zhang, L.~Zhang, X.~Mou, and D.~Zhang, ``{A comprehensive evaluation of full
  reference image quality assessment algorithms},'' in \emph{Proc. {IEEE} Int.
  Conf. Image Process. ({ICIP})}, {Orlando, FL, USA}, Sept. 2012, pp.
  1477--1480.

\bibitem{eval_fr_icip15}
M.~Pedersen, ``{Evaluation of 60 full-reference image quality metrics on the
  CID:IQ},'' in \emph{Proc. {IEEE} Int. Conf. Image Process. ({ICIP})}, {Quebec
  City, QC, Canada}, Sept. 2015, pp. 1588--1592.

\bibitem{eval_survey_latest}
Y.~Niu, Y.~Zhong, W.~Guo, Y.~Shi, and P.~Chen, ``{2D and 3D Image Quality
  Assessment: A Survey of Metrics and Challenges},'' \emph{{IEEE} Access},
  vol.~7, pp. 782--801, 2019.

\bibitem{eval_Waterloo}
S.~Athar and Z.~Wang, ``{A Comprehensive Performance Evaluation of Image
  Quality Assessment Algorithms},'' \emph{{IEEE} Access}, vol.~7, pp.
  140\,030--140\,070, Sept. 2019.

\bibitem{eval_gmad}
K.~Ma, Q.~Wu, Z.~Wang, Z.~Duanmu, H.~Yong, H.~Li, and L.~Zhang, ``{Group MAD
  Competition -- A New Methodology to Compare Objective Image Quality
  Models},'' in \emph{Proc. {IEEE} Conf. Comput. Vis. Pattern Recognit.
  ({CVPR})}, {Las Vegas, NV, USA}, June 2016, pp. 1664--1673.

\bibitem{md_sisblim_db_mdid2013}
K.~Gu, G.~Zhai, X.~Yang, and W.~Zhang, ``{Hybrid No-Reference Quality Metric
  for Singly and Multiply Distorted Images},'' \emph{{IEEE} Trans. Broadcast.},
  vol.~60, no.~3, pp. 555--567, Sept. 2014.

\bibitem{nr_lpsi}
Q.~Wu, Z.~Wang, and H.~Li, ``{A highly efficient method for blind image quality
  assessment},'' in \emph{Proc. {IEEE} Int. Conf. Image Process. ({ICIP})},
  {Quebec City, QC, Canada}, Sept. 2015, pp. 339--343.

\bibitem{nr_niqe}
A.~Mittal, R.~Soundararajan, and A.~C. Bovik, ``{Making a “Completely
  Blind” Image Quality Analyzer},'' \emph{{IEEE} Signal Process. Lett.},
  vol.~20, no.~3, pp. 209--212, Mar. 2013.

\bibitem{nr_ilniqe}
L.~Zhang, L.~Zhang, and A.~C. Bovik, ``{A Feature-Enriched Completely Blind
  Image Quality Evaluator},'' \emph{{IEEE} Trans. Image Process.}, vol.~24,
  no.~8, pp. 2579--2591, Aug. 2015.

\bibitem{SVRbook}
V.~N. Vapnik, \emph{{The Nature of Statistical Learning Theory}}.\hskip 1em
  plus 0.5em minus 0.4em\relax Springer Science \& Business Media, 2000.

\bibitem{libsvm}
C.~C. Chang and C.~J. Lin, ``{LIBSVM: A Library for Support Vector Machines},''
  \emph{ACM Trans. Intell. Syst. Technol.}, vol.~2, no.~3, pp. 27:1--27:27,
  Apr. 2011.

\bibitem{nr_biqi}
A.~K. Moorthy and A.~C. Bovik, ``{A Two-Step Framework for Constructing Blind
  Image Quality Indices},'' \emph{{IEEE} Signal Process. Lett.}, vol.~17,
  no.~5, pp. 513--516, May 2010.

\bibitem{nr_diivine}
{A. K. Moorthy and A. C. Bovik}, ``{Blind Image Quality Assessment: From
  Natural Scene Statistics to Perceptual Quality},'' \emph{{IEEE} Trans. Image
  Process.}, vol.~20, no.~12, pp. 3350--3364, Dec. 2011.

\bibitem{nr_bliinds2}
M.~A. Saad, A.~C. Bovik, and C.~Charrier, ``{Blind Image Quality Assessment: A
  Natural Scene Statistics Approach in the DCT Domain},'' \emph{{IEEE} Trans.
  Image Process.}, vol.~21, no.~8, pp. 3339--3352, Aug. 2012.

\bibitem{nr_brisque}
A.~Mittal, A.~K. Moorthy, and A.~C. Bovik, ``{No-Reference Image Quality
  Assessment in the Spatial Domain},'' \emph{{IEEE} Trans. Image Process.},
  vol.~21, no.~12, pp. 4695--4708, Dec. 2012.

\bibitem{nr_gmlog}
W.~Xue, X.~Mou, L.~Zhang, A.~C. Bovik, and X.~Feng, ``{Blind Image Quality
  Assessment Using Joint Statistics of Gradient Magnitude and Laplacian
  Features},'' \emph{{IEEE} Trans. Image Process.}, vol.~23, no.~11, pp.
  4850--4862, Nov. 2014.

\bibitem{nr_nferm}
K.~Gu, G.~Zhai, X.~Yang, and W.~Zhang, ``{Using Free Energy Principle For Blind
  Image Quality Assessment},'' \emph{{IEEE} Trans. Multimedia}, vol.~17, no.~1,
  pp. 50--63, Jan. 2015.

\bibitem{md_gwhglbp}
Q.~Li, W.~Lin, and Y.~Fang, ``{No-Reference Quality Assessment for
  Multiply-Distorted Images in Gradient Domain},'' \emph{{IEEE} Signal Process.
  Lett.}, vol.~23, no.~4, pp. 541--545, Apr. 2016.

\bibitem{nr_nrsl}
Q.~Li, W.~Lin, J.~Xu, and Y.~Fang, ``{Blind Image Quality Assessment Using
  Statistical Structural and Luminance Features},'' \emph{{IEEE} Trans.
  Multimedia}, vol.~18, no.~12, pp. 2457--2469, Dec. 2016.

\bibitem{nr_friquee_jrnl}
{D. Ghadiyaram and A. C. Bovik}, ``{Perceptual Quality Prediction on
  Authentically Distorted Images Using a Bag of Features Approach},'' \emph{J.
  Vis.}, vol.~17, no.~1, pp. 32:1--32:25, Jan. 2017.

\bibitem{nr_cornia}
P.~Ye, J.~Kumar, L.~Kang, and D.~Doermann, ``{Unsupervised Feature Learning
  Framework for No-Reference Image Quality Assessment},'' in \emph{Proc. {IEEE}
  Conf. Comput. Vis. Pattern Recognit. ({CVPR})}, {Providence, RI, USA}, June
  2012, pp. 1098--1105.

\bibitem{nr_hosa}
J.~Xu, P.~Ye, Q.~Li, H.~Du, Y.~Liu, and D.~Doermann, ``{Blind Image Quality
  Assessment Based on High Order Statistics Aggregation},'' \emph{{IEEE} Trans.
  Image Process.}, vol.~25, no.~9, pp. 4444--4457, Sept. 2016.

\bibitem{nr_dipiq}
K.~Ma, W.~Liu, T.~Liu, Z.~Wang, and D.~Tao, ``{dipIQ: Blind Image Quality
  Assessment by Learning-to-Rank Discriminable Image Pairs},'' \emph{{IEEE}
  Trans. Image Process.}, vol.~26, no.~8, pp. 3951--3964, Aug. 2017.

\bibitem{eval_dnn_latest}
X.~Yang, F.~Li, and H.~Liu, ``{A Survey of DNN Methods for Blind Image Quality
  Assessment},'' \emph{{IEEE} Access}, vol.~7, pp. 123\,788--123\,806, Sept.
  2019.

\bibitem{spm_dnn}
J.~Kim, H.~Zeng, D.~Ghadiyaram, S.~Lee, L.~Zhang, and A.~C. Bovik, ``{Deep
  Convolutional Neural Models for Picture-Quality Prediction: Challenges and
  Solutions to Data-Driven Image Quality Assessment},'' \emph{{IEEE} Signal
  Process. Mag.}, vol.~34, no.~6, pp. 130--141, Nov. 2017.

\bibitem{db_Tiny}
A.~Torralba, R.~Fergus, and W.~T. Freeman, ``{80 Million Tiny Images: A Large
  Data Set for Nonparametric Object and Scene Recognition},'' \emph{{IEEE}
  Trans. Pattern Anal. Mach. Intell.}, vol.~30, no.~11, pp. 1958--1970, Nov.
  2008.

\bibitem{db_ImageNet}
J.~{Deng}, W.~{Dong}, R.~{Socher}, L.~{Li}, K.~{Li}, and L.~{Fei-Fei},
  ``{ImageNet: A Large-Scale Hierarchical Image Database},'' in \emph{Proc.
  {IEEE} Conf. Comput. Vis. Pattern Recognit. ({CVPR})}, {Miami, FL, USA}, June
  2009, pp. 248--255.

\bibitem{db_ImageNet_source}
``{ImageNet},'' {Available: \url{http://image-net.org/index}}.

\bibitem{WordNet_Miller}
G.~A. Miller, ``{WordNet: A Lexical Database for English},'' \emph{Commun.
  ACM}, vol.~38, no.~11, pp. 39--41, Nov. 1995.

\bibitem{WordNet_Fellbaum}
C.~Fellbaum, \emph{{WordNet: An Electronic Lexical Database}}.\hskip 1em plus
  0.5em minus 0.4em\relax MIT press, 1998.

\bibitem{amazonMT}
``{Amazon Mechanical Turk},'' {\url{https://www.mturk.com/}}.

\bibitem{fr_vsnr}
D.~M. Chandler and S.~S. Hemami, ``{VSNR: A Wavelet-Based Visual
  Signal-to-Noise Ratio for Natural Images},'' \emph{{IEEE} Trans. Image
  Process.}, vol.~16, no.~9, pp. 2284--2298, Sept. 2007.

\bibitem{db_cidiq}
X.~Liu, M.~Pedersen, and J.~Y. Hardeberg, ``{CID:IQ -- A New Image Quality
  Database},'' in \emph{Proc. Int. Conf. Image, Signal Process. ({ICISP})},
  {Cherbourg, France}, July 2014, pp. 193--202.

\bibitem{fr_mad_db_csiq}
E.~C. Larson and D.~M. Chandler, ``{Most apparent distortion: full-reference
  image quality assessment and the role of strategy},'' \emph{J. Electron.
  Imag.}, vol.~19, no.~1, pp. 011\,006:1--011\,006:21, Jan. 2010.

\bibitem{db_ivc}
P.~L. Callet and F.~Autrusseau, ``{Subjective quality assessment IRCCyN/IVC
  database},'' 2005, {http://www2.irccyn.ec-nantes.fr/ivcdb/}.

\bibitem{db_kadid10k}
H.~{Lin}, V.~{Hosu}, and D.~{Saupe}, ``{KADID-10k: A Large-scale Artificially
  Distorted IQA Database},'' in \emph{Proc. Int. Conf. Qual. Multimedia Exper.
  ({QoMEX})}, {Berlin, Germany}, June 2019, pp. 1--3.

\bibitem{stateval_db_liveR2}
H.~R. Sheikh, M.~F. Sabir, and A.~C. Bovik, ``{A Statistical Evaluation of
  Recent Full Reference Image Quality Assessment Algorithms},'' \emph{{IEEE}
  Trans. Image Process.}, vol.~15, no.~11, pp. 3440--3451, Nov. 2006.

\bibitem{db_mict}
Y.~Horita, K.~Shibata, and K.~Yoshikazu, ``{MICT Image Quality Evaluation
  Database},'' http://mict.eng.u-toyama.ac.jp/mictdb.html.

\bibitem{db_pdaphdds}
T.~{Liu}, H.~{Liu}, S.~{Pei}, and K.~{Liu}, ``{A High-Definition
  Diversity-Scene Database for Image Quality Assessment},'' \emph{{IEEE}
  Access}, vol.~6, pp. 45\,427--45\,438, 2018.

\bibitem{db_tid2008}
N.~Ponomarenko, V.~Lukin, A.~Zelensky, K.~Egiazarian, J.~Astola, M.~Carli, and
  F.~Battisti, ``{TID2008--A Database for Evaluation of Full-Reference Visual
  Quality Assessment Metrics},'' \emph{Adv. Modern Radioelectron.}, vol.~10,
  no.~4, pp. 30--45, 2009.

\bibitem{db_tid2013}
N.~Ponomarenko, L.~Jin, O.~Ieremeiev, V.~Lukin, K.~Egiazarian, J.~Astola,
  B.~Vozel, K.~Chehdi, M.~Carli, F.~Battisti, and C.-C.~J. Kuo, ``{Image
  database TID2013: Peculiarities, results and perspectives},'' \emph{Signal
  Process.: Image Commun.}, vol.~30, pp. 57--77, Jan. 2015.

\bibitem{db_vclfer}
A.~Zari{\'c}, N.~Tatalovi{\'c}, N.~Brajkovi{\'c}, H.~Hlevnjak,
  M.~Lon{\v{c}}ari{\'c}, E.~Dumi{\'c}, and S.~Grgi{\'c}, ``{VCL@FER Image
  Quality Assessment Database},'' \emph{AUTOMATIKA}, vol.~53, no.~4, pp.
  344--354, 2012.

\bibitem{db_livemd}
D.~Jayaraman, A.~Mittal, A.~K. Moorthy, and A.~C. Bovik, ``{Objective quality
  assessment of multiply distorted images},'' in \emph{Proc. Asilomar Conf.
  Signals, Syst., Comput. ({ASILOMAR})}, {Pacific Grove, CA, USA}, Nov. 2012,
  pp. 1693--1697.

\bibitem{db_mdid}
W.~Sun, F.~Zhou, and Q.~Liao, ``{MDID: A multiply distorted image database for
  image quality assessment},'' \emph{Pattern Recognit.}, vol.~61, pp. 153--168,
  Jan. 2017.

\bibitem{db_mdivl}
S.~Corchs and F.~Gasparini, ``{A Multidistortion Database for Image Quality},''
  in \emph{Proc. Int. Workshop Comput. Color Imag. ({CCIW})}, {Milan, Italy},
  Mar. 2017, pp. 95--104.

\bibitem{db_bid}
A.~{Ciancio}, A.~L.~N. {Targino da Costa}, E.~A.~B. {da Silva}, A.~{Said},
  R.~{Samadani}, and P.~{Obrador}, ``{No-Reference Blur Assessment of Digital
  Pictures Based on Multifeature Classifiers},'' \emph{{IEEE} Trans. Image
  Process.}, vol.~20, no.~1, pp. 64--75, Jan. 2011.

\bibitem{db_cid2013}
T.~{Virtanen}, M.~{Nuutinen}, M.~{Vaahteranoksa}, P.~{Oittinen}, and
  J.~{Häkkinen}, ``{CID2013: A Database for Evaluating No-Reference Image
  Quality Assessment Algorithms},'' \emph{{IEEE} Trans. Image Process.},
  vol.~24, no.~1, pp. 390--402, Jan. 2015.

\bibitem{db_koniq10k}
H.~Lin, V.~Hosu, and D.~Saupe, ``{KonIQ-10K: Towards an ecologically valid and
  large-scale IQA database},'' \emph{arXiv preprint arXiv:1803.08489}, 2018.

\bibitem{db_livewc}
D.~Ghadiyaram and A.~C. Bovik, ``{Massive Online Crowdsourced Study of
  Subjective and Objective Picture Quality},'' \emph{{IEEE} Trans. Image
  Process.}, vol.~25, no.~1, pp. 372--387, Jan. 2016.

\bibitem{db_ILSVRC}
O.~Russakovsky, J.~Deng, H.~Su, J.~Krause, S.~Satheesh, S.~Ma, Z.~Huang,
  A.~Karpathy, A.~Khosla, M.~Bernstein, A.~C. Berg, and L.~Fei-Fei, ``{ImageNet
  Large Scale Visual Recognition Challenge},'' \emph{Int. J. Comput. Vis.},
  vol. 115, no.~3, pp. 211--252, Dec. 2015.

\bibitem{NIPS_DNN}
A.~Krizhevsky, I.~Sutskever, and G.~E. Hinton, ``{ImageNet Classification with
  Deep Convolutional Neural Networks},'' in \emph{Proc. Adv. Neural Inf.
  Process. Syst. ({NIPS})}, 2012, pp. 1097--1105.

\bibitem{Clarifai_uses_ImageNet}
M.~D. Zeiler and R.~Fergus, ``{Visualizing and Understanding Convolutional
  Networks},'' in \emph{Proc. Eur. Conf. Comput. Vis. ({ECCV})}, Zurich,
  Switzerland, Sept. 2014, pp. 818--833.

\bibitem{OverFeat_uses_ImageNet}
P.~Sermanet, D.~Eigen, X.~Zhang, M.~Mathieu, R.~Fergus, and Y.~LeCun,
  ``{OverFeat: Integrated Recognition, Localization and Detection using
  Convolutional Networks},'' in \emph{Proc. Int. Conf. Learn. Representations
  ({ICLR})}, Banff, AB, Canada, Apr. 2014.

\bibitem{Caffe_uses_ImageNet}
Y.~Jia, E.~Shelhamer, J.~Donahue, S.~Karayev, J.~Long, R.~Girshick,
  S.~Guadarrama, and T.~Darrell, ``{Caffe: Convolutional Architecture for Fast
  Feature Embedding},'' in \emph{Proc. {ACM} Int. Conf. Multimedia ({MM})},
  Orlando, FL, USA, Nov. 2014, pp. 675--678.

\bibitem{VGG_uses_ImageNet}
K.~Simonyan and A.~Zisserman, ``{Very Deep Convolutional Networks for
  Large-Scale Image Recognition},'' in \emph{Proc. Int. Conf. Learn. Represent.
  ({ICLR})}, {San Diego, CA, USA}, May 2015.

\bibitem{GoogLeNet_uses_ImageNet}
C.~Szegedy, W.~Liu, Y.~Jia, P.~Sermanet, S.~Reed, D.~Anguelov, D.~Erhan,
  V.~Vanhoucke, and A.~Rabinovich, ``{Going Deeper with Convolutions},'' in
  \emph{Proc. {IEEE} Conf. Comput. Vis. Pattern Recognit. ({CVPR})}, {Boston,
  MA, USA}, June 2015, pp. 1--9.

\bibitem{db_pascal_voc}
M.~Everingham, S.~M.~A. Eslami, L.~V. Gool, C.~K.~I. Williams, J.~Winn, and
  A.~Zisserman, ``{The Pascal Visual Object Classes Challenge: A
  Retrospective},'' \emph{Int. J. Comput. Vis.}, vol. 111, no.~1, pp. 98--136,
  Jan. 2015.

\bibitem{db_caltech101}
L.~Fei-Fei, R.~Fergus, and P.~Perona, ``{Learning Generative Visual Models from
  Few Training Examples: An Incremental Bayesian Approach Tested on 101 Object
  Categories},'' in \emph{Proc. Conf. Comput. Vis. Pattern Recognit. Workshop},
  {Washington, DC, USA}, June 2004, pp. 178--178.

\bibitem{db_caltech256}
G.~Griffin, A.~Holub, and P.~Perona, ``{Caltech-256 Object Category Dataset},''
  California Inst. Technol., Pasadena, CA, USA, Tech. Rep. 7694, Apr. 2007.

\bibitem{bt500}
{Rec. ITU-R BT.500-13}, ``{Methodology for the subjective assessment of the
  quality of television pictures},'' Jan. 2012.

\bibitem{db_waterlooed}
K.~Ma, Z.~Duanmu, Q.~Wu, Z.~Wang, H.~Yong, H.~Li, and L.~Zhang, ``{Waterloo
  Exploration Database: New Challenges for Image Quality Assessment Models},''
  \emph{{IEEE} Trans. Image Process.}, vol.~26, no.~2, pp. 1004--1016, Feb.
  2017.

\bibitem{nr_cnn}
L.~Kang, P.~Ye, Y.~Li, and D.~Doermann, ``{Convolutional Neural Networks for
  No-Reference Image Quality Assessment},'' in \emph{Proc. {IEEE} Conf. Comput.
  Vis. Pattern Recognit. ({CVPR})}, {Columbus, OH, USA}, June 2014, pp.
  1733--1740.

\bibitem{nr_cnn_plusplus}
{L. Kang, P. Ye, Y. Li, and D. Doermann}, ``{Simultaneous estimation of image
  quality and distortion via multi-task convolutional neural networks},'' in
  \emph{Proc. {IEEE} Int. Conf. Image Process. ({ICIP})}, {Quebec City, QC,
  Canada}, Sept. 2015, pp. 2791--2795.

\bibitem{md_cnn_JieFu}
J.~Fu, H.~Wang, and L.~Zuo, ``{Blind image quality assessment for multiply
  distorted images via convolutional neural networks},'' in \emph{Proc. {IEEE}
  Int. Conf. Acoust., Speech, Signal Process. ({ICASSP})}, {Shanghai, China},
  Mar. 2016, pp. 1075--1079.

\bibitem{nr_cnn_prewitt}
J.~Li, L.~Zou, J.~Yan, D.~Deng, T.~Qu, and G.~Xie, ``{No-reference image
  quality assessment using Prewitt magnitude based on convolutional neural
  networks},'' \emph{Signal, Image, Video Process.}, vol.~10, no.~4, pp.
  609--616, Apr. 2016.

\bibitem{nr_biecon_C}
J.~Kim and S.~Lee, ``{Deep blind image quality assessment by employing
  FR-IQA},'' in \emph{Proc. {IEEE} Int. Conf. Image Process. ({ICIP})},
  {Beijing, China}, Sept. 2017, pp. 3180--3184.

\bibitem{nr_biecon_J}
{J. Kim and S. Lee}, ``{Fully Deep Blind Image Quality Predictor},''
  \emph{{IEEE} J. Sel. Topics Signal Process.}, vol.~11, no.~1, pp. 206--220,
  Feb. 2017.

\bibitem{nr_cnn_svr}
J.~Li, J.~Yan, D.~Deng, W.~Shi, and S.~Deng, ``{No-reference image quality
  assessment based on hybrid model},'' \emph{Signal, Image, Video Process.},
  vol.~11, no.~6, pp. 985--992, Sept. 2017.

\bibitem{nr_deepIQA_conf}
S.~Bosse, D.~Maniry, T.~Wiegand, and W.~Samek, ``{A deep neural network for
  image quality assessment},'' in \emph{Proc. {IEEE} Int. Conf. Image Process.
  ({ICIP})}, {Phoenix, AZ, USA}, Sept. 2016, pp. 3773--3777.

\bibitem{nr_fr_deepIQA}
S.~Bosse, D.~Maniry, K.-R. Müller, T.~Wiegand, and W.~Samek, ``{Deep Neural
  Networks for No-Reference and Full-Reference Image Quality Assessment},''
  \emph{{IEEE} Trans. Image Process.}, vol.~27, no.~1, pp. 206--219, Jan. 2018.

\bibitem{nr_deepBIQ}
S.~Bianco, L.~Celona, P.~Napoletano, and R.~Schettini, ``{On the use of deep
  learning for blind image quality assessment},'' \emph{Signal, Image, Video
  Process. (SIViP)}, vol.~12, no.~2, pp. 355--362, Feb. 2018.

\bibitem{nr_cnn_pqr}
H.~Zeng, L.~Zhang, and A.~C. Bovik, ``{Blind Image Quality Assessment with a
  Probabilistic Quality Representation},'' in \emph{Proc. {IEEE} Int. Conf.
  Image Process. ({ICIP})}, {Athens, Greece}, Oct. 2018, pp. 609--613.

\bibitem{nr_diqa}
J.~Kim, A.-D. Nguyen, and S.~Lee, ``{Deep CNN-Based Blind Image Quality
  Predictor},'' \emph{{IEEE} Trans. Neural Netw. Learn. Syst.}, vol.~30, no.~1,
  pp. 11--24, Jan. 2019.

\bibitem{nr_pccnn}
X.~Qin, T.~Xiang, Y.~Yang, and X.~Liao, ``{Pair-Comparing Based Convolutional
  Neural Network for Blind Image Quality Assessment},'' in \emph{Int. Symp.
  Neural Netw. ({ISNN})}, {Moscow, Russia}, Jul. 2019, pp. 460--468.

\bibitem{fr_vis_imp_pool}
A.~K. Moorthy and A.~C. Bovik, ``{Visual Importance Pooling for Image Quality
  Assessment},'' \emph{{IEEE} J. Sel. Topics Signal Process.}, vol.~3, no.~2,
  pp. 193--201, Apr. 2009.

\bibitem{fr_vis_attn_qa}
U.~Engelke, H.~Kaprykowsky, H.~Zepernick, and P.~Ndjiki-Nya, ``{Visual
  Attention in Quality Assessment},'' \emph{{IEEE} Signal Process. Mag.},
  vol.~28, no.~6, pp. 50--59, Nov. 2011.

\bibitem{nr_meon}
K.~Ma, W.~Liu, K.~Zhang, Z.~Duanmu, Z.~Wang, and W.~Zuo, ``{End-to-End Blind
  Image Quality Assessment Using Deep Neural Networks},'' \emph{{IEEE} Trans.
  Image Process.}, vol.~27, no.~3, pp. 1202--1213, Mar. 2018.

\bibitem{nr_nima}
H.~Talebi and P.~Milanfar, ``{NIMA: Neural Image Assessment},'' \emph{{IEEE}
  Trans. Image Process.}, vol.~27, no.~8, pp. 3998--4011, Aug. 2018.

\bibitem{db_places}
B.~Zhou, A.~Lapedriza, J.~Xiao, A.~Torralba, and A.~Oliva, ``{Learning Deep
  Features for Scene Recognition using Places Database},'' in \emph{Proc. Adv.
  Neural Inf. Process. Syst. ({NIPS})}, 2014, pp. 487--495.

\bibitem{nr_blinder}
F.~Gao, J.~Yu, S.~Zhu, Q.~Huang, and Q.~Tian, ``{Blind image quality prediction
  by exploiting multi-level deep representations},'' \emph{Pattern Recognit.},
  vol.~81, pp. 432 -- 442, Sept. 2018.

\bibitem{CNN_Feat_OffShelf}
A.~S. Razavian, H.~Azizpour, J.~Sullivan, and S.~Carlsson, ``{CNN Features
  Off-the-Shelf: An Astounding Baseline for Recognition},'' in \emph{Proc.
  {IEEE} Conf. Comput. Vis. Pattern Recognit. ({CVPR}) Workshops}, {Columbus,
  OH, USA}, June 2014, pp. 512--519.

\bibitem{Delv_Deep_ConvNets}
K.~Chatfield, K.~Simonyan, A.~Vedaldi, and A.~Zisserman, ``{Return of the Devil
  in the Details: Delving Deep into Convolutional Nets},'' in \emph{Proc. Brit.
  Mach. Vis. Conf. ({BMVC})}, {Nottingham, UK}, Sept. 2014, pp. 1--12.

\bibitem{nr_dbcnn}
W.~Zhang, K.~Ma, J.~Yan, D.~Deng, and Z.~Wang, ``{Blind Image Quality
  Assessment Using A Deep Bilinear Convolutional Neural Network},''
  \emph{{IEEE} Trans. Circuits Syst. Video Technol.}, vol.~30, no.~1, pp.
  36--47, Jan. 2020.

\bibitem{nr_dliqa}
W.~Hou, X.~Gao, D.~Tao, and X.~Li, ``{Blind Image Quality Assessment via Deep
  Learning},'' \emph{{IEEE} Trans. Neural Netw. Learn. Syst.}, vol.~26, no.~6,
  pp. 1275--1286, June 2015.

\bibitem{NatImgStat_NeuRep}
E.~P. Simoncelli and B.~A. Olshausen, ``{Natural Image Statistics and Neural
  Representation},'' \emph{Annu. Rev. Neurosci.}, vol.~24, no.~1, pp.
  1193--1216, Mar. 2001.

\bibitem{db_winkler_analysis}
S.~Winkler, ``{Analysis of Public Image and Video Databases for Quality
  Assessment},'' \emph{{IEEE} J. Sel. Topics Signal Process.}, vol.~6, no.~6,
  pp. 616--625, Oct. 2012.

\bibitem{db_winkler_SI}
H.~Yu and S.~Winkler, ``Image complexity and spatial information,'' in
  \emph{Proc. Int. Workshop Qual. Multimedia Exper. ({QoMEX})}, {Klagenfurt am
  W\"orthersee, Austria}, July 2013, pp. 12--17.

\bibitem{db_colorfulness}
D.~Hasler and S.~Suesstrunk, ``{Measuring colorfulness in natural images},'' in
  \emph{Proc. SPIE Electron. Imag.}, vol. 5007, {Santa Clara, CA, USA}, June
  2003, pp. 87--95.

\bibitem{fr_ssimplus}
A.~Rehman, K.~Zeng, and Z.~Wang, ``{Display Device-Adapted Video
  Quality-of-Experience Assessment},'' in \emph{Proc. SPIE Electron. Imag.},
  vol. 9394, {San Francisco, CA, USA}, Mar. 2015, pp. 939\,406:1--939\,406:11.

\bibitem{nr_qac}
W.~Xue, L.~Zhang, and X.~Mou, ``{Learning without Human Scores for Blind Image
  Quality Assessment},'' in \emph{Proc. {IEEE} Conf. Comput. Vis. Pattern
  Recognit. ({CVPR})}, {Portland, OR, USA}, June 2013, pp. 995--1002.

\bibitem{fusion_bliss}
P.~Ye, J.~Kumar, and D.~Doermann, ``{Beyond Human Opinion Scores: Blind Image
  Quality Assessment based on Synthetic Scores},'' in \emph{Proc. {IEEE} Conf.
  Comput. Vis. Pattern Recognit. ({CVPR})}, {Columbus, OH, USA}, June 2014, pp.
  4241--4248.

\bibitem{fr_gmsd}
W.~Xue, L.~Zhang, X.~Mou, and A.~C. Bovik, ``{Gradient Magnitude Similarity
  Deviation: A Highly Efficient Perceptual Image Quality Index},'' \emph{{IEEE}
  Trans. Image Process.}, vol.~23, no.~2, pp. 684--695, Feb. 2014.

\bibitem{fr_vif}
H.~R. Sheikh and A.~C. Bovik, ``{Image Information and Visual Quality},''
  \emph{{IEEE} Trans. Image Process.}, vol.~15, no.~2, pp. 430--444, Feb. 2006.

\bibitem{fr_msssim}
Z.~Wang, E.~P. Simoncelli, and A.~C. Bovik, ``{Multiscale structural similarity
  for image quality assessment},'' in \emph{Proc. Asilomar Conf. Signals,
  Syst., Comput. ({ASILOMAR})}, {Pacific Grove, CA, USA}, Nov. 2003, pp.
  1398--1402.

\bibitem{md_musique}
Y.~Zhang and D.~M. Chandler, ``{Opinion-Unaware Blind Quality Assessment of
  Multiply and Singly Distorted Images via Distortion Parameter Estimation},''
  \emph{{IEEE} Trans. Image Process.}, vol.~27, no.~11, pp. 5433--5448, Nov.
  2018.

\bibitem{fr_ssim}
Z.~Wang, A.~C. Bovik, H.~R. Sheikh, and E.~P. Simoncelli, ``{Image Quality
  Assessment: From Error Visibility to Structural Similarity},'' \emph{{IEEE}
  Trans. Image Process.}, vol.~13, no.~4, pp. 600--612, Apr. 2004.

\bibitem{md_waterloo17}
S.~Athar, A.~Rehman, and Z.~Wang, ``{Quality assessment of images undergoing
  multiple distortion stages},'' in \emph{Proc. {IEEE} Int. Conf. Image
  Process. ({ICIP})}, {Beijing, China}, Sept. 2017, pp. 3175--3179.

\bibitem{db_dataset_impact}
S.~Tourancheau, F.~Autrusseau, Z.~M.~P. Sazzad, and Y.~Horita, ``{Impact of
  subjective dataset on the performance of image quality metrics},'' in
  \emph{Proc. {IEEE} Int. Conf. Image Process. ({ICIP})}, {San Diego, CA, USA},
  Oct. 2008, pp. 365--368.

\bibitem{fusion_mmf_tip}
{T. Liu and W. Lin and C.-C. J. Kuo}, ``{Image Quality Assessment Using
  Multi-Method Fusion},'' \emph{{IEEE} Trans. Image Process.}, vol.~22, no.~5,
  pp. 1793--1807, May 2013.

\bibitem{fusion_cnnm}
V.~V. Lukin, N.~N. Ponomarenko, O.~I. Ieremeiev, K.~O. Egiazarian, and
  J.~Astola, ``{Combining full-reference image visual quality metrics by neural
  network},'' in \emph{Proc. SPIE Electron. Imag.}, vol. 9394, {San Francisco,
  CA, USA}, Mar. 2015, pp. 93\,940K:1--93\,940K:12.

\bibitem{fusion_hfsim}
K.~Okarma, ``{Hybrid Feature Similarity Approach to Full-Reference Image
  Quality Assessment},'' in \emph{Proc. Int. Conf. Comput. Vis. Graph.
  (ICCVG)}, {Warsaw, Poland}, Sept. 2012, pp. 212--219.

\bibitem{md_cm3cm4}
{K. Okarma}, ``{Quality Assessment of Images with Multiple Distortions using
  Combined Metrics},'' \emph{Elektronika Ir Elektrotechnika}, vol.~20, no.~6,
  pp. 128--131, 2014.

\bibitem{fusion_rrf}
G.~V. Cormack, C.~L.~A. Clarke, and S.~Buettcher, ``{Reciprocal Rank Fusion
  outperforms Condorcet and Individual Rank Learning Methods},'' in \emph{Proc.
  Int. ACM SIGIR Conf. Res. Develop. Inf. Retr.}, Boston, MA, USA, July 2009,
  pp. 758--759.

\bibitem{fr_cid}
I.~Lissner, J.~Preiss, P.~Urban, M.~S. Lichtenauer, and P.~Zolliker,
  ``{Image-Difference Prediction: From Grayscale to Color},'' \emph{{IEEE}
  Trans. Image Process.}, vol.~22, no.~2, pp. 435--446, Feb. 2013.

\bibitem{fr_vifdwt_ssimdwt}
S.~Rezazadeh and S.~Coulombe, ``{A novel discrete wavelet transform framework
  for full reference image quality assessment},'' \emph{Signal, Image, Video
  Process.}, vol.~7, no.~3, pp. 559--573, May 2013.

\bibitem{fr_essim}
X.~Zhang, X.~Feng, W.~Wang, and W.~Xue, ``{Edge Strength Similarity for Image
  Quality Assessment},'' \emph{{IEEE} Signal Process. Lett.}, vol.~20, no.~4,
  pp. 319--322, Apr. 2013.

\bibitem{fr_mcsd}
T.~Wang, L.~Zhang, H.~Jia, B.~Li, and H.~Shu, ``{Multiscale contrast similarity
  deviation: An effective and efficient index for perceptual image quality
  assessment},'' \emph{Signal Process.: Image Commun.}, vol.~45, pp. 1--9, July
  2016.

\bibitem{fr_qasd}
L.~Li, H.~Cai, Y.~Zhang, W.~Lin, A.~C. Kot, and X.~Sun, ``{Sparse
  Representation-Based Image Quality Index With Adaptive Sub-Dictionaries},''
  \emph{{IEEE} Trans. Image Process.}, vol.~25, no.~8, pp. 3775--3786, Aug.
  2016.

\bibitem{fr_sff}
H.~Chang, H.~Yang, Y.~Gan, and M.~Wang, ``{Sparse Feature Fidelity for
  Perceptual Image Quality Assessment},'' \emph{{IEEE} Trans. Image Process.},
  vol.~22, no.~10, pp. 4007--4018, Oct. 2013.

\bibitem{fr_dvicom}
{E. D. Di Claudio and G. Jacovitti}, ``{A Detail-Based Method for Linear Full
  Reference Image Quality Prediction},'' \emph{{IEEE} Trans. Image Process.},
  vol.~27, no.~1, pp. 179--193, Jan. 2018.

\bibitem{vqeg_report}
{Video Quality Experts Group and others}, ``{Final report from the Video
  Quality Experts Group on the validation of objective models of video quality
  assessment, Phase II},'' 2003.

\bibitem{statTest_fTest_ref}
D.~J. Sheskin, \emph{{Handbook of Parametric and Nonparametric Statistical
  Procedures}}.\hskip 1em plus 0.5em minus 0.4em\relax Chapman \& Hall/CRC,
  2011.

\bibitem{E2E_ImComp}
J.~Ball{\'e}, V.~Laparra, and E.~P. Simoncelli, ``{End-to-end optimized image
  compression},'' in \emph{Proc. Int. Conf. Learn. Represent. ({ICLR})},
  {Toulon, France}, Apr. 2017.

\bibitem{GDN}
Q.~{Li} and Z.~{Wang}, ``{Reduced-Reference Image Quality Assessment Using
  Divisive Normalization-Based Image Representation},'' \emph{{IEEE} J. Sel.
  Topics Signal Process.}, vol.~3, no.~2, pp. 202--211, Apr. 2009.

\bibitem{ReLU}
V.~Nair and G.~E. Hinton, ``{Rectified Linear Units Improve Restricted
  Boltzmann Machines},'' in \emph{Proc. Int. Conf. Mach. Learn. ({ICML})},
  {Haifa, Israel}, June 2010, pp. 807--814.

\bibitem{vmeon}
W.~Liu, Z.~Duanmu, and Z.~Wang, ``{End-to-End Blind Quality Assessment of
  Compressed Videos Using Deep Neural Networks},'' in \emph{Proc. {ACM} Int.
  Conf. Multimedia}, {Seoul, Republic of Korea}, Oct. 2018, p. 546–554.

\bibitem{conv_weights_init}
K.~He, X.~Zhang, S.~Ren, and J.~Sun, ``{Delving Deep into Rectifiers:
  Surpassing Human-Level Performance on ImageNet Classification},'' in
  \emph{Proc. {IEEE} Int. Conf. Comput. Vis. ({ICCV})}, {Santiago, Chile}, Dec.
  2015, pp. 1026--1034.

\bibitem{Adam}
D.~P. Kingma and J.~L. Ba, ``{Adam: A Method for Stochastic Optimization},'' in
  \emph{Proc. Int. Conf. Learn. Represent. ({ICLR})}, {San Diego, CA, USA}, May
  2015.

\bibitem{md_mslqaf}
C.~Li, Y.~Zhang, X.~Wu, and Y.~Zheng, ``{A Multi-Scale Learning Local Phase and
  Amplitude Blind Image Quality Assessment for Multiply Distorted Images},''
  \emph{{IEEE} Access}, vol.~6, pp. 64\,577--64\,586, 2018.

\bibitem{nr_tclt}
Q.~Wu, H.~Li, F.~Meng, K.~N. Ngan, B.~Luo, C.~Huang, and B.~Zeng, ``{Blind
  Image Quality Assessment Based on Multichannel Feature Fusion and Label
  Transfer},'' \emph{{IEEE} Trans. Circuits Syst. Video Technol.}, vol.~26,
  no.~3, pp. 425--440, Mar. 2016.

\end{thebibliography}
